\documentclass[letterpaper,numberedappendix]{emulateapj}
\newcommand{\subij}{_{i,j}}

\newcommand{\subipj}{_{i+1,j}}
\newcommand{\subijk}{_{i,j,k}}
\newcommand{\subimjk}{_{i-1,j,k}}
\newcommand{\subimjmk}{_{i-1,j-1,k}}
\newcommand{\subimjkm}{_{i-1,j,k-1}}
\newcommand{\subipjk}{_{i+1,j,k}}
\newcommand{\subipjkm}{_{i+1,j,k-1}}
\newcommand{\subipjmk}{_{i+1,j-1,k}}
\newcommand{\subijmk}{_{i,j-1,k}}
\newcommand{\subijpk}{_{i,j+1,k}}
\newcommand{\subijpkm}{_{i,j+1,k-1}}
\newcommand{\subimjpk}{_{i-1,j+1,k}}
\newcommand{\subimjkp}{_{i-1,j,k+1}}
\newcommand{\subijmkp}{_{i,j-1,k+1}}
\newcommand{\subijkm}{_{i,j,k-1}}
\newcommand{\subijkp}{_{i,j,k+1}}
\newcommand{\subi}{_{i}}
\newcommand{\subim}{_{i-1}}
\newcommand{\subip}{_{i+1}}
\newcommand{\subj}{_{j}}
\newcommand{\subjm}{_{j-1}}
\newcommand{\subjp}{_{j+1}}
\newcommand{\subk}{_{k}}
\newcommand{\subkm}{_{k-1}}
\newcommand{\subkp}{_{k+1}}
\newcommand{\supn}{^{n}}
\newcommand{\supnp}{^{n+1}}
\newcommand{\supna}{^{n+a}}
\newcommand{\supnb}{^{n+b}}
\newcommand{\supnc}{^{n+c}}
\newcommand{\supnh}{^{n+1/2}}
\newcommand{\supnt}{^{n+\theta}}
\newcommand{\onedir}{^{(1)}}
\newcommand{\twodir}{^{(2)}}
\newcommand{\thrdir}{^{(3)}}
\newcommand{\onechrp}{^{(1+)}}
\newcommand{\twochrp}{^{(2+)}}
\newcommand{\thrchrp}{^{(3+)}}
\newcommand{\onechrm}{^{(1-)}}
\newcommand{\twochrm}{^{(2-)}}
\newcommand{\thrchrm}{^{(3-)}}
\newcommand{\DbDt}{{D \over Dt}}
\newcommand{\dbdt}{{d \over dt}}
\newcommand{\pbpt}{{\partial \over \partial t}}
\newcommand{\diver}{{\bf \nabla \cdot}}
\newcommand{\grad}{{\bf \nabla}}
\newcommand{\curl}{{\bf \nabla \times}}
\newcommand{\cross}{{\bf \times}}
\newcommand{\dkpde}{{\partial\kappa_{\rm P}\over\partial\egas}}
\newcommand{\dkede}{{\partial\kappa_{\rm E}\over\partial\egas}}
\newcommand{\dbbde}{{\partial{\rm B}\over\partial\egas}}
\newcommand{\B}{{\bf B}}
\newcommand{\E}{{\bf E}}
\newcommand{\J}{{\bf J}}
\newcommand{\Q}{{\bf Q}}
\newcommand{\F}{{\bf F}}
\newcommand{\p}{{\bf P}}
\newcommand{\f}{{\bf f}}
\newcommand{\vel}{{\bf v}}
\newcommand{\velg}{{\bf v_{g}}}
\newcommand{\divf}{\diver\F}

\newcommand{\divv}{\diver\vel}
\newcommand{\divq}{\diver\Q}
\newcommand{\gradvp}{\grad\vel :\p}
\newcommand{\gradvq}{\grad\vel :\Q}
\newcommand{\chif}{\chi}
\newcommand{\kp}{{\kappa_{\rm P}}}
\newcommand{\ke}{{\kappa_{\rm E}}}
\newcommand{\flim}{\Lambda_{\rm E}}
\newcommand{\egas}{{\rm e}}
\newcommand{\erad}{{\rm E}}
\newcommand{\pfunc}{{\rm B_p}}
\newcommand{\guniv}{{\rm G}}
\newcommand{\pg}{{\rm p}}
\newcommand{\dt}{{\Delta t}}
\newcommand{\tn}{t^{n}}
\newcommand{\tnp}{t^{n+1}}
\newcommand{\atwid}{\tilde{\rm A}}
\newcommand{\dtwid}{\tilde{\rho}}
\newcommand{\etwid}{\widetilde{\left({\rm e}\over\rho\right)}}
\newcommand{\ertwid}{\widetilde{\left({\rm E}\over\rho\right)}}

\newcommand{\mdotone}{\dot{M}^{1}}
\newcommand{\mdottwo}{\dot{M}^{2}}
\newcommand{\mdotthr}{\dot{M}^{3}}
\newcommand{\bone}{b1\subijk}
\newcommand{\btwo}{b2\subijk}
\newcommand{\bthr}{b3\subijk}
\newcommand{\emfone}{{\cal E}1\subijk}
\newcommand{\emftwo}{{\cal E}2\subijk}
\newcommand{\emfthr}{{\cal E}3\subijk}
\newcommand{\vone}{v1\subijk}
\newcommand{\vtwo}{v2\subijk}
\newcommand{\vthr}{v3\subijk}
\newcommand{\vgone}{vg1\subi}
\newcommand{\vgtwo}{vg2\subj}
\newcommand{\vgthr}{vg3\subk}

\newcommand{\abun}{{\rm X(l)}}
\newcommand{\xtwid}{\widetilde{X}(l)}
\newcommand{\sone}{{S1}\subijk}
\newcommand{\stwo}{{S2}\subijk}
\newcommand{\sthr}{{S3}\subijk}
\newcommand{\sdotone}{{\dot{S}1}}
\newcommand{\sdottwo}{{\dot{S}2}}
\newcommand{\sdotthr}{{\dot{S}3}}
\newcommand{\vonestr}{v1^*}
\newcommand{\vtwostr}{v2^*}
\newcommand{\vthrstr}{v3^*}
\newcommand{\bonestr}{b1^*}
\newcommand{\bonetwostr}{b1^{2*}}
\newcommand{\bonethrstr}{b1^{3*}}
\newcommand{\btwoonestr}{b2^{1*}}
\newcommand{\btwothrstr}{b2^{3*}}
\newcommand{\bthronestr}{b3^{1*}}
\newcommand{\bthrtwostr}{b3^{2*}}
\newcommand{\btwostr}{b2^*}
\newcommand{\bthrstr}{b3^*}
\newcommand{\vonebar}{\overline{v1}}
\newcommand{\vtwobar}{\overline{v2}}
\newcommand{\vthrbar}{\overline{v3}}
\newcommand{\bonebar}{\overline{b1}}
\newcommand{\btwobar}{\overline{b2}}
\newcommand{\bthrbar}{\overline{b3}}
\newcommand{\srd}{\sqrt{\rho}}
\newcommand{\srdp}{\sqrt{\rho^+}}
\newcommand{\srdm}{\sqrt{\rho^-}}
\newcommand{\vatwo}{v_A^{(2)}}
\newcommand{\vathr}{v_A^{(3)}}
\newcommand{\vatwop}{v_A^{(2+)}}
\newcommand{\vathrp}{v_A^{(3+)}}
\newcommand{\vatwom}{v_A^{(2-)}}
\newcommand{\vathrm}{v_A^{(3-)}}
\newcommand{\lrnt}{^{\cal L}}
\newcommand{\zmp}{ZEUS-MP}
\newcommand{\ztwd}{ZEUS-2D}
\newcommand{\zthd}{ZEUS-3D}
\newcommand{\twospc}{\phn\phn}
\newcommand{\thrspc}{\phn\phn\phn}
\begin{document}

\title{Simulating Radiating and Magnetized Flows in Multi-Dimensions with
       ZEUS-MP}

\author{John C. Hayes\altaffilmark{1}, 
Michael L. Norman\altaffilmark{1,2}, 
Robert A. Fiedler\altaffilmark{3}, 
James O. Bordner\altaffilmark{1,2}, 
Pak Shing Li\altaffilmark{4}, 
Stephen E. Clark\altaffilmark{2}, 
Asif ud-Doula\altaffilmark{5,6},
and 
Mordecai-Mark MacLow\altaffilmark{7}}

\altaffiltext{1}{Center for Astrophysics and Space Sciences, University of California,
San Diego, CA  92093}
\altaffiltext{2}{Physics Department, University of California,
San Diego, CA  92093}
\altaffiltext{3}{Center for Simulation of Advanced Rockets, University
of Illinois, Urbana, IL  61801}
\altaffiltext{4}{Astronomy Department, University of California at
Berkeley, Berkeley, CA 94720}
\altaffiltext{5}{Bartol Research Institute, University of Delaware,
Newark, DE 19716}
\altaffiltext{6}{Department of Physics and Astronomy, Swarthmore College,
Swarthmore, PA 19081}
\altaffiltext{7}{Department of Astrophysics, American Museum of Natural
History, New York, NY 10024}

\begin{abstract}

This paper describes ZEUS-MP, a multi-physics, massively parallel,
message-passing implementation of the ZEUS code.  ZEUS-MP differs
significantly from the thoroughly documented ZEUS-2D code, the
completely undocumented (in peer-reviewed literature) ZEUS-3D code, and
a marginally documented ``version 1'' of ZEUS-MP first distributed
publicly in 1999.~\zmp~offers an MHD algorithm which is better suited
for multidimensional flows than the ZEUS-2D module by virtue of modifications
to the Method of Characteristics scheme first suggested by~\citet{hawley95}.
This MHD module is shown to compare quite favorably to the TVD scheme
described by~\citet{ryu98}.~\zmp~is the first publicly-available ZEUS
code to allow the advection of multiple chemical (or nuclear) species.  Radiation
hydrodynamic simulations are enabled via an implicit flux-limited radiation diffusion
(FLD) module.  The hydrodynamic, MHD, and FLD modules may be used, singly or
in concert, in one, two, or three space dimensions.  Additionally, so-called
``1.5-D'' and ``2.5-D'' grids, in which the ``half-D'' denotes a symmetry
axis along which a constant but non-zero value of velocity or magnetic field
is evolved, are supported.  Self gravity may be included either through
the assumption of a $GM/r$ potential or a solution of Poisson's equation
using one of three linear solver packages (conjugate-gradient, multigrid,
and FFT) provided for that purpose.  Point-mass potentials are also supported.

Because~\zmp~is designed for large simulations on parallel computing platforms,
considerable attention is paid to the parallel performance characteristics of
each module in the code.  Strong-scaling tests involving pure hydrodynamics (with and
without self-gravity), MHD, and RHD are performed in which large problems
(256$^3$ zones) are distributed among as many as 1024 processors of an IBM SP3.
Parallel efficiency is a strong function of the amount of communication required
between processors in a given algorithm, but all modules are shown to scale well
on up to 1024 processors for the chosen fixed problem size.

\end{abstract}

\keywords{hydrodynamics -- methods:numerical -- methods:parallel -- MHD --
          radiative transfer}

\section{Introduction}

Since their formal introduction in the literature, the ZEUS simulation
codes have enjoyed widespread use in the numerical astrophysics
community, having been applied to such topics as planetary
nebulae~\citep{garcia99}, molecular cloud turbulence~\citep{maclow99},
solar magnetic arcades~\citep{low00}, and galactic spiral arm
formation~\citep{martos04a,martos04b}.  The numerical methods used in the
axisymmetric~\ztwd~code are documented in an often-cited trio of
papers~\citep{stone92a,stone92b,stone92c} well familiar to
the computational astrophysics community.  A reasonable first question to
ask regarding this report might therefore be, ``why write another ZEUS `method' 
paper?''  The first reason is that the code we describe in this paper, \zmp,
is a significantly different code when compared to the highly-documented
\ztwd~code, the completely undocumented (in peer-reviewed literature) ZEUS-3D
code, and a marginally documented ``version 1'' of~\zmp~made publicly
available in 1999.  The new version of~\zmp~we present is the first ZEUS
code to unite 3D hydrodynamics (HD) and 3D magnetohydrodynamics (MHD) with implicit
flux-limited radiation diffusion (FLD) and self-gravity in a software framework
designed for execution on massively parallel architectures.  This last feature
anticipates a second major reason for offering a new method paper: the computing
hardware landscape in which numerical astrophysicists operate has changed
enormously since the~\ztwd~trilogy was published.  The enormous increase
in computer processor speed and available memory has brought with it a new
paradigm for computing in which large simulations are distributed across many
(often hundreds to thousands) of parallel processors; this new environment
has spawned additional figures of merit, such as parallel scalability, by which modern
numerical algorithms must be judged.  A major component of this paper is the
demonstration of the suitability of ZEUS-MP's algorithms, both new and familiar,
for parallel execution on thousands of processors.

In addition to describing the many new features and capabilities provided in
~\zmp, this paper fills significant gaps in the evolution history of
the MHD and radiation modules offered in predecessors of~\zmp.
These gaps were partially a consequence of the evolution of the ZEUS series.
The first code formally named ZEUS was developed by David Clarke
~\citep{clarke88,clarke86} for MHD simulations of radio
jets.  Thereafter, continued development of the ZEUS method proceeded
along two parallel tracks.  One track resulted in the release of
the~\ztwd~code, which
solves the equations of self-gravitating radiation magnetohydrodynamics
in two or 2.5 dimensions. (``2.5-D'' denotes a problem computed in 2 spatial
dimensions involving the 3-component of a vector quantity, such as velocity,
that is invariant along the 3-axis but variable along the 1- and 2-axes.)
The creation of~\ztwd~occasioned the development and
incorporation of several new algorithms, including (1) a covariant
formulation, allowing simulations in various coordinate geometries;
(2) a tensor artificial viscosity; (3) a new, more accurate MHD
algorithm (MOCCT) combining the Constrained Transport
algorithm~\citep{evans88} with a Method Of Characteristics treatment for
Alfv\'en waves; and (4) a variable tensor Eddington factor (VTEF)
solution for the equations of radiation hydrodynamics (RHD).  The VTEF
radiation module was described in~\citet{stone92c} but not included in
the version of~\ztwd~offered for public distribution.  An implicit FLD-based RHD
module was publicly distributed with~\ztwd~but never documented in a published
report.  (A draft of a technical report describing the 2-D FLD module is
available on the World Wide Web\footnote{http://cosmos.ucsd.edu/lca-publications/LCA013/
index.html}.)  The VTEF module described in~\citet{stone92c} was later modified
to incorporate a different matter-radiation coupling scheme, Eddington tensors
computed from a time-dependent solution to the radiation transfer equation, and
parallelization over two space dimensions and one angle cosine.  This new VTEF code
was coupled to an early version of~\zmp~and used to compare VTEF solutions to
FLD results for test problems featuring strongly beamed radiation fields.
This work was published by~\citet{hayes03}, but as before the VTEF module remained
reserved for private use.  \zmp~as described herein provides an updated FLD module 
designed for all dimensionalities and geometries in a parallel computing environment.
This paper documents our new module at a level of detail which should aid users of
both~\zmp~and the public~\ztwd~code.

The second track resulted in the release of~\zthd, the first ZEUS code
capable of three-dimensional simulations.  Written for the
Cray-2 supercomputer by David Clarke, ZEUS-3D physics options included hydrodynamics,
MHD, self-gravity, and optically thin radiative cooling. \zthd~was the
first ZEUS code with parallel capability, accomplished using Cray
Autotasking compiler directives.  The~\zthd~MHD module differed from 
that described in the~\ztwd~papers with regard to both dimensionality
and method: the MOC treatment of Alfv\'en waves was modified to incorporate
the improvements introduced by John Hawley and James Stone~\citep{hawley95}.
This modified ``HSMOCCT'' method is the basis for the MHD module adopted
in~\zmp.

Roughly speaking, \zmp~encapsulates a qualitative union of the methods provided by the
public~\ztwd~and~\zthd~codes and enlarges upon them with new capabilities
and solution techniques.  The first public
release of~\zmp~included HD, MHD, and self gravity, but was written
exclusively for 3-D simulations, excluding at one stroke a long
menu of worthy 2-D research applications and erecting an inconvenient
barrier to expedient code testing.  The new version we describe offers a
substantially extended menu of physics, algorithm, and dimensionality options.
The HD, MHD, and RHD modules accommodate simulations in 1, 1.5, 2, 2.5, and 3
dimensions. Arbitrary equations of state are supported; gamma-law and isothermal
equations of state are provided.  \zmp~is the first ZEUS code to allow multi-species
fluids to be treated; this is achieved with the addition of a
concentration field array dimensioned for an arbitrary number of fluids.
An implicit flux-limited photon diffusion module is included for RHD problems.
As noted previously, the implicit FLD solution is based upon the method adopted
for the public version of~\ztwd~but has been extended to three dimensions.
In addition, the FLD module sits atop a scalable linear system solver
using the {\it conjugate gradient} (CG)
method~\citep{barret94}.  Supplementing the FLD driver is an opacity module
offering simple density and temperature-dependent power-law expressions
for absorption and scattering coefficients; additional user-supplied
modules (such as tabular data sets) are easily accommodated.
Self-gravity is included in several ways:
(1) spherical gravity (GM/r) is adopted for one-dimensional problems
and may also be used in two dimensions; (2) two parallel Poisson solvers
are included for problems with Neumann or Dirichlet boundary conditions
in Cartesian or curvilinear coordinates, and (3) a fast Fourier
Transform (FFTw) package is provided for problems with triply-periodic
boundaries.  In addition to self-gravity, a simple point-mass external potential
may be imposed in spherical geometry.

While the ZEUS code line has evolved significantly, this process has not
occurred in a vacuum.  Indeed, the past decade has seen the emergence of several new
MHD codes based upon Godunov methods~\citep{ryu98,londrillo00,londrillo04,
balsara04,gardiner05}.  Godunov-type schemes are accurate to second order in both space
and time, are automatically conservative, and possess superior capability for
resolving shock fronts when compared to ZEUS-type schemes at identical
resolution.  These advances, coupled with ZEUS's lower-order rate of convergence
with numerical resolution~\citep{stone92d}, and further
with recent observations~\citep{falle02} of
vulnerabilities in the ZEUS-2D implementation of MOCCT might lead one to ask
whether a new ZEUS code has a strong potential for future contributions of significance
in areas of current astrophysical interest.  We argue this point in the affirmative
on two fronts.  While we acknowledge that the ZEUS method possesses (as do
all numerical methods) weaknesses which bound its scope of applicability, we
note the encouraging comparisons~\citep{stone05} of simulation results computed
with Athena, an unsplit MHD code coupling PPM hydrodynamics
to Constrained Transport, to ZEUS results in problems of supersonic MHD
turbulence in 3D, and in shearing-box simulations of the magneto-rotational
instability in two and three dimensions.  The authors confirm the reliability
of the ZEUS results and note that the numerical dissipation in Athena is equivalent
to that of ZEUS at a grid resolution of about 1.5 along each axis.  In a similar
vein, we examine a standard 2-D
MHD test problem, due to~\citet{orszag79}, in which the results from~\zmp~are
found to compare quite favorably with those from the TVD code described in~\citet{ryu98}
and with results from different upwind Godunov codes presented by~\citet{dai98}
and~\citet{londrillo00}.

A second cause for optimism of ZEUS's future arises from the versatility inherent
in ZEUS-MP's design. As this paper demonstrates, a wide variety of physics modules
are easily implemented within ZEUS-MP's design framework.
Additionally, different solution techniques for treating the same physics
(self-gravity is an excellent example) are almost trivially accommodated.
The theme of this paper is, therefore: physics, flexibility, and parallel performance.
To demonstrate these traits we organize the paper as follows: the presentation
begins by writing the fluid equations solved by~\zmp~in~
section~\ref{equations}; section~\ref{methods} provides a largely descriptive
overview of the various numerical methods used to solve these equations on a discrete
mesh.  Sections~\ref{verify} and~\ref{perf} present two groups of test
problems.  Section~\ref{verify} provides a suite of tests to verify
the correctness of the various physics modules.  Section~\ref{perf} examines
a quartet of 3-D problems which measure performance on a massively parallel
computer.

The main body of the paper is concluded with a summary in section~\ref{summary}.
Full documentation of the finite-difference formulae describing hydrodynamic,
magnetohydrodynamic, and radiation hydrodynamic evolution in three dimensions
is provided in a series of appendices.  These appendices are preceded by a
tabular ``code map'' (Table~\ref{codemap}) in appendix~\ref{mapapp} associating
the discrete equations written in appendices~\ref{hdapp}-\ref{poissonapp} with
the subroutines in ZEUS-MP that compute them.  We offer this as an aid to users
wishing to familiarize themselves with the code and to code developers who desire
to implement their own versions or improve upon what is currently offered in
ZEUS-MP.  The HD, MHD, and FLD modules are
detailed in appendices~\ref{hdapp}, \ref{mhdapp}, and~\ref{fldapp}, respectively.
Appendix~\ref{poissonapp} documents the 3-D linearized form of Poisson's
equation expressed in covariant grid coordinates, and details of our parallel
implementation techniques and strategies are given in appendix~\ref{impapp}.

\section{The Physical Equations}\label{equations}

Our description of the physical state of a fluid element
is specified by the following set of fluid equations relating the mass density
($\rho$), velocity ($\vel$), gas internal energy density ($\egas$), radiation
energy density ($\erad$), radiation flux ($\F$), and magnetic field strength ($\B$):
\begin{equation}\label{cont}
{D\rho \over Dt} \ + \ \rho\divv \ = \ 0;
\end{equation}
\begin{equation}\label{gmom}
\rho{D\vel \over Dt} \ = \ -\grad\pg \ + \ \left(\chif\over c\right)\F \ + \
{1\over 4\pi}\left(\curl\B\right)\cross\B \ - \ {\rho\grad\Phi};
\end{equation}
\begin{equation}\label{gase}
\rho\DbDt\left({\egas \over \rho}\right) \ + \ \pg\divv \ = \ 
c\ke\erad \ - \ 4\pi\kp\pfunc;
\end{equation}
\begin{equation}\label{rade}
\rho\DbDt\left({\erad \over \rho}\right) \ + \ \divf \ + \ \gradvp \ = \
4\pi\kp\pfunc \ - \ c\ke\erad;
\end{equation}
\begin{equation}\label{bdot}
{\partial\B\over\partial t} \ = \ \curl\left(\vel\cross\B\right).
\end{equation}

The Lagrangean (or comoving) derivative is given by the usual definition:
\begin{equation}
\DbDt \ \equiv \ \pbpt \ + \ \vel\cdot\nabla\label{lagder}
\end{equation}
The four terms on the RHS of the gas momentum equation (\ref{gmom}) denote forces to due
thermal pressure gradients, radiation stress, magnetic Lorentz acceleration,
and the gravitational potential, respectively.
The RHS of (\ref{gase}) gives source/sink terms due to absorption/emission
of radiant energy.  Each term serves an inverse role on the RHS of (\ref{rade}).
In (\ref{gase}), $\pfunc$ denotes the Planck function:
\begin{equation}
\pfunc \ = \ {\sigma \over \pi} {\rm T}^{4},
\end{equation}
where T is the local material temperature.
Equations (\ref{gase}) and (\ref{rade}) are also functions of flux-mean,
Planck-mean, and energy-mean opacities, which are formally defined
as
\begin{equation}
\chif \ = \ {1 \over \F} \int_{0}^{\infty} \chi(\nu)\F(\nu)d\nu,
\end{equation}
\begin{equation}
\kp \ = \ {1 \over \pfunc} \int_{0}^{\infty} \chi(\nu)\pfunc(\nu)d\nu,
\end{equation}
\begin{equation}
\ke \ = \ {1 \over E} \int_{0}^{\infty} \chi(\nu)E(\nu)d\nu.
\end{equation}
In the simple problems we discuss in this paper, the three opacities
are computed from a single expression, differing only in that $\ke$
and $\kp$ are defined at zone centers and $\chif$ is computed at zone
faces.  In general, however, independent expressions or data sets for
the three opacities are trivially accommodated in our discrete representation
of the RHD equations.

We compute the radiation flux, $\F$, according to
the diffusion approximation:
\begin{equation}\label{radm}
\F \ = \ -\left(c\flim\over\chif\right)\grad\erad,
\end{equation}
where we have introduced a flux-limiter ($\flim$) to ensure a radiation
propagation speed that remains bounded by the speed of light (c) in 
transparent media.  Attenuation of the flux is regulated by the total
extinction coefficient, $\chif$, which in general may include contributions
from both absorption and scattering processes.

Equation (\ref{rade}) also includes a term involving the radiation stress
tensor, $\p$.  In general, $\p$ is not known {\it a priori} and must be
computed from a solution to the radiative transfer equation.  In the VTEF
methods described in~\citet{stone92c} and~\citet{hayes03}, $\p$ is written
as the product of the (known) radiation energy, E, and an (unknown) tensor
quantity, $\f$:
\begin{equation}\label{stress}
\p \ = \ \f\erad.
\end{equation}
A solution for $\f$ may be derived from a formal solution to the radiative
transfer equation~\citep{mihalas78} or may be approximated analytically.
For the RHD implementation in~\zmp, we follow the latter approach, adopting
the expression for $\f$ given by equation (13) in~\citet{turner01}:
\begin{equation}\label{eddtens}
\f \ = \ {1\over 2}\left(1 - f\right){\bf I} \ + \ {1\over 2}\left(3f - 1\right)
\hat{\bf n}\hat{\bf n},
\end{equation}
where $\hat{\bf n} = \grad\erad / \mid\grad\erad\mid$, {\bf I} is 
the unit tensor, and $f$ is a scalar
``Eddington factor'' expressed as a function of the flux limiter, $\flim$,
and E as
\begin{equation}
f \ = \ \flim \ + \ \left(\flim\mid\grad\erad\mid\over{\chif\erad}\right)^2
\end{equation}
The RHD equations are accurate only to order unity in $\vel/c$~\citep{mihalas84}, 
consistent
with the radiation modules described in~\citet{stone92c}, \citet{turner01},
and~\citet{hayes03}.  The assumption of local thermodynamic equilibrium is
reflected in our adoption of the Planck function in equations (\ref{gase}) and
(\ref{rade}) evaluated at the local material temperature; our use of the FLD 
approximation gives rise to equation (\ref{radm}) and is discussed further in 
\S~\ref{fldmeth}.

Evolution of the magnetic field (\ref{bdot}) is constrained by the assumption of ideal MHD.
We are therefore, in common with previous public ZEUS codes, assuming wave modes
which involve fluid motions but no charge separation.
This equation for $\B$ also assumes zero resistivity, a valid approximation
in astrophysical environments where the degree of ionization is sufficiently
high to ensure a rapid collision rate between charged and neutral particles
and thus strong coupling between the two.  
There exist astrophysical environments where this assumption is expected
to break down (e.g. the cores of cold, dense molecular clouds), in which
case an algorithm able to distinguish the dynamics of ionized and neutral
particles is required.  \citet{stone99} published extensions to the ZEUS
MHD algorithm to treat nonideal phenomena such as Ohmic dissipation and
partially ionized plasmas where the ionic and neutral components of the
fluid are weakly coupled via a collisional drag term.  Incorporation of these
algorithmic extensions into ZEUS-MP is currently left as an exercise to the
interested user, but may be undertaken for a future public release.

The gravitational potential $\Phi$ appearing in (\ref{gmom}) is computed
from a solution to Poisson's equation:
\begin{equation}\label{poisson}
\grad{^2}\Phi \ = \ {4\pi\guniv\rho}.
\end{equation}
Our various techniques for solving (\ref{poisson}) are described in~\S\ref{gravmeth};
the linear system which arises from discretizing (\ref{poisson}) on a
covariant coordinate mesh is derived in appendix~\ref{poissonapp}.

Our fluid equations are closed with an equation of state (EOS) expressing
the thermal pressure as a function of the internal gas energy.  The dynamic
test problems considered in this paper adopt a simple ideal EOS with
$\gamma$ = 5/3 except where noted.

\section{Numerical Methods: An Overview}\label{methods}

\begin{figure}
\leavevmode
\includegraphics[width=\columnwidth]{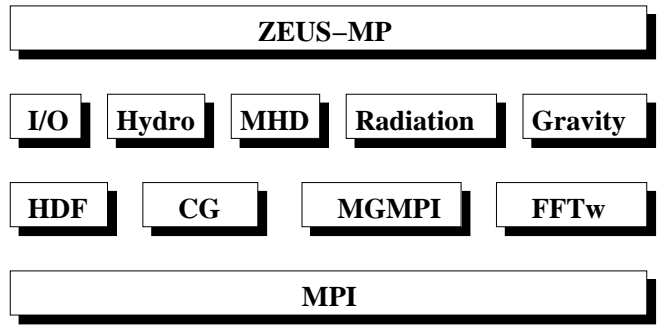}
\caption{Software implementation within~\zmp.}\label{figimp}
\end{figure}

Figure~\ref{figimp} summarizes the dependencies of ZEUS-MP's major physics and
I/O modules on underlying software libraries.  The {\it Message Passing Interface}
(MPI) software library is used to implement parallelism in ZEUS-MP and lies at
the foundation of the code.  This library is available on all NSF and DOE
supercomputing facilities and may be freely
downloaded\footnote{http://www-unix.mcs.anl.gov/mpi/mpich/}
for installation on small private clusters.  ZEUS-MP's linear system solvers and
I/O drivers access MPI functions directly and act in service of the top layer
of physics modules, which are described in the following subsections.

   \subsection{Hydrodynamics}\label{hdmeth}

Equations (\ref{cont}) through (\ref{bdot}) provide the most complete
physical description which may be invoked by~\zmp~to characterize a
problem of interest.  There exist classes of problems, however, for which
either radiation or magnetic fields (or both) are not relevant and thus
may be ignored.  In such circumstances, \zmp~evolves an appropriately
reduced subset of field variables and solves only those equations needed
to close the system.~\zmp~may therefore define systems of equations for
pure HD, MHD, RHD, or RMHD problems as necessary.  We begin the description
of our numerical methods by considering purely hydrodynamic problems in
this section; we introduce additional equations as we expand the discussion
of physical processes.

Tracing their ancestry back
to a two-dimensional, Eulerian hydrodynamics (HD) code for simulations
of rotating protostellar collapse~\citep{norman80b}, all ZEUS codes are
rooted in an HD algorithm based upon the method of finite differences
on a staggered mesh~\citep{norman80a,norman86}, which incorporates a second
order-accurate, monotonic advection scheme~\citep{vanleer77}.  The basic
elements of the ZEUS scheme arise from consideration of how the evolution
of a fluid element may be properly described on an {\it adaptive mesh}
whose grid lines move with arbitrary velocity, $\velg$.  Following the
analysis in~\citet{winkler84}, we identify three pertinent time derivatives
in an adaptive coordinate system: (1) the Eulerian time derivative 
($\partial/\partial t$),
taken with respect to coordinates fixed in the laboratory frame, (2) the 
Lagrangean derivative ($D/Dt$; cf. equation~\ref{lagder}), taken with respect to
a definite fluid element, and (3) the adaptive-mesh derivative ($d/dt$),
taken with respect to fixed values of the adaptive mesh coordinates.
Identifying $dV$ as a volume element bounded by fixed values of the 
adaptive mesh and $d{\bf S}$ as the surface bounding this element, 
one may employ the formalism of~\citet{winkler84} to split the fluid
equations into two distinct solution
steps: the {\it source step}, in which we solve
\begin{eqnarray}
\rho{\partial \vel  \over \partial t} & = & -\grad\pg \ - \ \divq \ - \ 
{\rho\grad\Phi};\label{srcvel_hd} \\
    {\partial \egas \over \partial t} & = & -\pg\divv \ - \ \gradvq;
\label{srcgas_hd}
\end{eqnarray}
and the {\it transport step}, whence
\begin{eqnarray}
\dbdt\int_{V}\rho~dV & = & -\oint_{dV}\rho~(\vel - \velg)\cdot
 d{\bf S}; \label{advrho} \\
\dbdt\int_{V}\rho\vel~dV & = & -\oint_{dV}\rho\vel~(\vel - \velg)\cdot
 d{\bf S}; \label{advvel} \\
\dbdt\int_{V}\egas~dV & = & -\oint_{dV}\egas~(\vel - \velg)\cdot
 d{\bf S}; \label{advgas}
\end{eqnarray}
where $\velg$ is the local grid velocity.  Equations (\ref{srcvel_hd}) and
(\ref{srcgas_hd}) have been further modified to include an artificial
viscous pressure, $\Q$.  \zmp~employs the method due to~\citet{vonneumann50}
to apply viscous pressure at shock fronts.  This approach is known to provide
spurious viscous heating in convergent coordinate geometries even when no
material compression is present.  \citet{tscharnuter79} describe a tensor
formalism for artificial viscosity which avoids this problem; an implementation
of this method will be provided in a future release of~\zmp.  For problems involving
very strong shocks and stagnated flows, the artificial viscosity may be augmented
with an additional term which is linear in velocity and depends upon the local
sound speed.  The precise forms of the quadratic and linear viscosity terms are
documented in Appendix~\ref{hdapp}.

   \subsection{MHD}\label{mhdmeth}

The treatment of MHD waves in~\zmp~ is by necessity more complex than that for HD
waves because MHD waves fall into two distinct families: (1) longitudinal,
compressive (fast and slow magnetosonic); and (2) transverse, non-compressive 
(Alfv\'en) waves.  The former family may be treated in the source-step portion
of the ZEUS solution in a similar fashion to their hydrodynamic analogs, but
the latter wave family couples directly to the magnetic induction equation
and therefore requires a more complex treatment.  From the algorithmic perspective,
the inclusion of MHD into the ZEUS scheme has two consequences: (1) fluid
accelerations due to {\em compressive} MHD waves introduce additional terms in equation
(\ref{srcvel_hd}); (2) fluid accelerations due to {\em transverse} MHD introduce
additional velocity acceleration terms which, owing to the coupling to the induction
equation, are computed in a separate step which follows the source step update but
precedes the ``transport'' (i.e. fluid advection) update.  In this way, the updates
due to fluid advection are deferred until all updates to the velocities are properly
recorded.  As will be extensively described in this section, the fluid accelerations
due to transverse MHD waves are combined with the evolution equation for $\B$ because
of the tight mathematical coupling between the two solutions.

We guide the following discussion by providing, in the continuum limit, the final
result. With the inclusion of MHD, the source/transport solution sequence expands
to begin with an MHD-augmented ``source'' step:
\begin{eqnarray}
\rho{\partial \vel  \over \partial t} & = & -\grad\pg - \ \grad(B^2/8\pi)
\ - \ \divq \ - \ {\rho\grad\Phi};\label{srcvel_mhd} \\
    {\partial \egas \over \partial t} & = &
\ - \pg\divv \ - \ \gradvq;
\label{srcgas_mhd}
\end{eqnarray}
This is followed by an MOCCT step, whence
\begin{eqnarray}
\rho{\partial \vel  \over \partial t}\Bigg|_{\rm final} & = & \rho{\partial\vel\over\partial t}
\Bigg|_{\rm source~step}
 \ + \ {1 \over 4\pi} \left(\B\cdot\grad\right)\B \label{transacc}; \\
\dbdt\int_{S}\B\cdot{\bf dS} & = & \vec{\epsilon}\cdot{\bf dl}, \label{induction}
\end{eqnarray}
where $\vec{\epsilon}$ is the electromotive force (EMF) and is given by
\begin{equation}
\vec{\epsilon} \ = \ (\vel-\velg)\cross \B.
\end{equation}

With velocities and $\B$ fields fully updated, we then proceed to the ``transport''
step as written in equations (\ref{advrho}) through (\ref{advgas}).

\begin{figure}
 \leavevmode
 \includegraphics[width=\columnwidth]{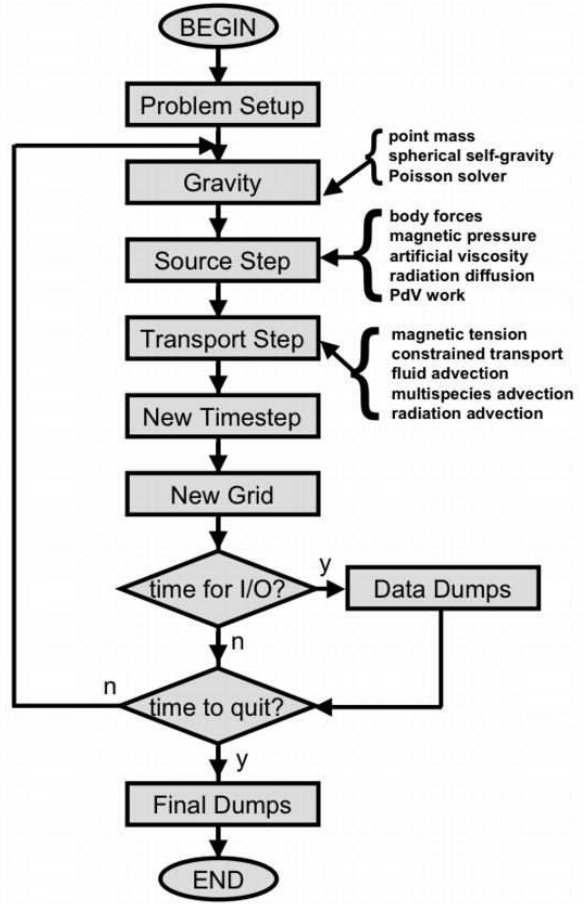}
\caption{Program control for~\zmp.}\label{figpc}
\end{figure}

Figure~\ref{figpc} shows the program control logic used to implement the
solution outlined by equations (\ref{srcvel_mhd})-(\ref{induction}) and
the previously-written expressions (\ref{advrho})-(\ref{advgas}).
Once the
problem has been initialized, Poisson's equation is solved to compute
the gravitational potential.  In the source step, updates due to longitudinal forces
(thermal and magnetic), radiation stress, artificial viscosity, 
energy exchange between
the gas and radiation field, and pdV work are performed at fixed values
of the coordinates.  Accelerations due to transverse MHD waves are then computed
and magnetic field components are updated.  Finally, velocities
updated via (\ref{srcvel_mhd}) and (\ref{transacc}) are used
to advect field variables through the moving mesh in the transport step.
Advection is performed in a series of directional sweeps which are cyclically
permuted at each time step.

The remainder of this section serves to derive the new expressions appearing in
(\ref{srcvel_mhd}), (\ref{transacc}), and (\ref{induction}), and to document
their solution.  We begin by first considering the Lorentz acceleration term
in the gas momentum equation (\ref{gmom}).  Using the vector identity
\begin{equation}\label{lorentz}
\left(\curl\B\right)\cross\B=-\grad\left(B^2/2\right)\ + \ 
\left(\B\cdot\grad\right)\B
\end{equation}
we expand the Lorentz acceleration term such that (\ref{gmom}) becomes
(ignoring radiation)
\newpage
\begin{eqnarray}\label{gmom_exp}
\rho{D\vel \over Dt} & = & -\grad\pg 
\ - \ \grad(B^2/8\pi) \ + \
{1 \over 4\pi} \left(\B\cdot\grad\right)\B  \nonumber \\
& - & {\rho\grad\Phi}.
\end{eqnarray}

The second term on the RHS of (\ref{gmom_exp}) is the gradient of the
magnetic pressure.  This term, which provides the contribution from the
compressive magnetosonic waves, is clearly a longitudinal force and is 
differenced in space and time identically to the thermal pressure term.
This expression is thus evaluated simultaneously with the other contributions to
the ``source step'' portion of the momentum equation (equation~\ref{srcvel_mhd}); 
contributions from the magnetic pressure to the discrete representation of
(\ref{srcvel_mhd}) along each axis are shown in equations (\ref{v1force}) through
(\ref{v3force}), with notational definitions provided by expressions (\ref{mpterm1})
through (\ref{mpterm12}), in appendix~\ref{hdapp}.

The third term on the RHS of (\ref{gmom_exp}) represents magnetic tension in
curved field lines and is transverse to the gradient of $\B$. This term, which
is added to the solution sequence as equation (\ref{transacc})
couples to the magnetic induction equation to produce Alfv\'en waves; the magnetic
tension force and the induction equation (\ref{bdot}) are therefore solved by a
single procedure: the Method of Characteristics + Constrained Transport (MOCCT).

\zmp~employs the MOCCT
method of ~\cite{hawley95}, which is a generalization of the algorithm 
described in ~\cite{stone92b} to 3D, with some slight modifications that
improve numerical stability. To describe MOCCT, we first derive the moving
frame induction equation. Recall that equation (\ref{bdot}) is
derived from Faraday's law
\begin{equation}\label{faraday}
{\partial\B\over\partial t} \ = \ -c\curl\E,
\end{equation}
where $\E$, $\B$ and the time derivative are measured in the Eulerian frame. 
The electric field $\E$ is specified from Ohm's law
\begin{equation}\label{ohm}
c\E = -\vel\cross\B + \J/\sigma.
\end{equation}
Equation (\ref{bdot}) results when we substitute equation \ref{ohm} 
into equation \ref{faraday}, and let the conductivity $\sigma \rightarrow \infty$.
Integrating equation \ref{faraday} over a moving surface element $S(t)$ bounded
by a moving circuit $C(t)$, the general form of Faraday's law is
\begin{equation}\label{faraday-gen}
\dbdt\int_{S}\B\cdot{\bf dS} = -c\oint_C \E'\cdot{\bf dl}
\end{equation}
where $\E'$ is the electric field measured in the moving frame. To first
order in $v/c$, $\E'=\E+(\velg\cross\B)/c$. From equation \ref{ohm}, for a perfectly 
conducting fluid $\E=-\vel\cross\B/c$. Combining these two results and
substituting into equation \ref{faraday-gen}, we get
\begin{equation}\label{induction2}
\dbdt\int_{S}\B\cdot{\bf dS} = \oint_C (\vel-\velg)\cross\B\cdot{\bf dl}.
\end{equation}

Equation (\ref{induction2}) states that the time rate of change of
the magnetic flux piercing $S$
\begin{equation}\label{flux}
\phi_S=\int_{S}\B\cdot{\bf dS}
\end{equation}
is given by the line integral of the electromotive force (EMF) $\mathbf{\epsilon=(v-v_g)\cross B}$ along $C$:
\begin{equation}\label{fdot}
{d \phi_S\over dt}=\oint_C {\epsilon}\cdot{\bf dl}.
\end{equation}
Equation (\ref{fdot}), using (\ref{flux}), is equivalent to expression (\ref{induction}) 
appearing in our
grand solution outline, and it forms, along with equation (\ref{transacc}), the target
for our MOCCT algorithm.
Equation (\ref{fdot}) is familiar from standard texts on electrodynamics, only now
$S$ and $C$ are moving with respect to the Eulerian frame. If $\velg=\vel$, we
recover the well known flux-freezing result, $d\phi_S/dt=D\phi_S/Dt=0.$

\begin{figure}
\leavevmode
\includegraphics[width=\columnwidth]{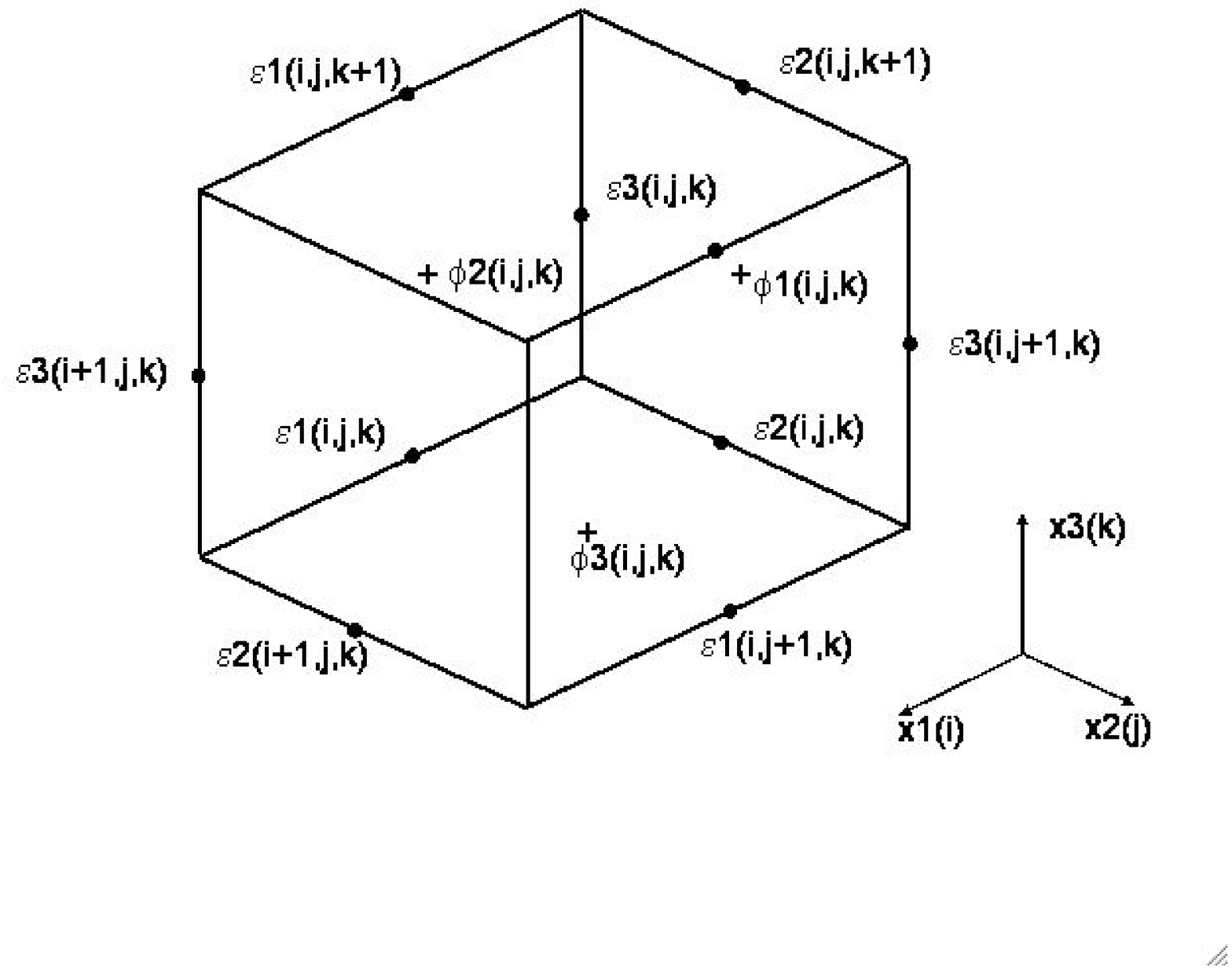}
\caption{Centering of magnetic field variables in~\zmp.}\label{b-cube}
\end{figure}

As discussed in \cite{evans88,stone92b,hawley95}, equation (\ref{fdot}) is in
the form which guarantees divergence-free magnetic field transport when finite
differenced, provided the EMFs are evaluated once and only once per time step.
Referring to the unit cell shown in Figure \ref{b-cube}, we can write the discrete
form of equation (\ref{fdot}) as
\begin{eqnarray}
{{\phi 1\subijk\supnp - \phi 1\subijk\supn}\over \dt} & = &
 \epsilon 2\subijk \Delta x2\subij + \epsilon 3\subijpk\Delta x3\subijpk \nonumber \\
& - & \epsilon 2\subijkp\Delta x2\subij - \epsilon 3\subijk \Delta x3\subijk; \label{bflux1} \\
{{\phi 2\subijk\supnp - \phi 2\subijk\supn}\over \dt} & = &
 \epsilon 1\subijkp\Delta x1\subi + \epsilon 3\subijpk\Delta x3\subijk \nonumber \\
& - & \epsilon 1\subijk \Delta x1\subi - \epsilon 3\subipjk\Delta x3\subipjk; \label{bflux2} \\
{{\phi 3\subijk\supnp - \phi 3\subijk\supn}\over \dt} & = &
 \epsilon 1\subijk \Delta x1\subi + \epsilon 2\subipjk\Delta x2\subipj \nonumber \\
& - & \epsilon 1\subijpk\Delta x1\subi - \epsilon 2\subijk \Delta x2\subij, \label{bflux3}
\end{eqnarray}
where $\phi 1, \phi 2, \phi 3$ are the face-centered 
magnetic fluxes piercing the cell
faces whose unit normals are in the $\bf{n_1, n_2, n_3}$ directions, 
respectively, $\epsilon 1, \epsilon 2, \epsilon 3$ are components
of the edge-centered EMFs, and $\Delta x1, \Delta x2, \Delta x3$
are the coordinate distances of the cell edges. The peculiar subscripting
of these line elements is made clear in Appendix \ref{mhdapp}.
It is easy to see that for any choice of EMFs, $\B\supnp$ will be divergence-free
provided $\B\supn$ is divergence-free. By Gauss's theorem, $\int_V \diver\B dV=\oint_S \B\cdot{\bf dS}=0$, where the second equality follows from the fact
$\diver\B=0.$ Analytically, the time derivative of $\diver\B$ is also zero.
Numerically, 
\begin{eqnarray}\label{divergencefree}
\dbdt\int_V \diver\B dV & = & \dbdt\oint_S \B\cdot{\bf dS}\approx\sum_{faces=1}^6 d\phi/dt
\nonumber \\
                        & = &
\sum_{faces=1}^6 \sum_{edges=1}^4 \epsilon\cdot {\bf dl}=0.
\end{eqnarray}
The last equality results from the fact that when summing over all the faces and
edges of a cell, each EMF appears twice with a change of sign, and thus cancel in
pairs. 

In principle, one could use {\em any} method to compute the EMF within
the CT formalism and still maintain divergence-free fields. In practice,
a method must be used which stably and accurately propagates both MHD
wave types: longitudinal, compressive
(fast and slow magnetosonic) waves, and the transverse, non-compressive (Alfv\'en)
waves.  As noted previously, the first wave type is straightforwardly
incorporated into the treatment of the compressive hydrodynamic waves; the
real difficulty arises in the treatment of the Alfv\'en waves.

In ideal MHD, Alfv\'en waves can exhibit discontinuities (rotational, 
transverse) at current sheets. Unlike hydrodynamical shocks, these structures
are not dissipative, which rules out the use of dissipative numerical 
algorithms to model them. In addition, Alfv\'en waves tightly couple the
evolution equations for the velocity and magnetic field components
perpendicular to the direction of propagation. This rules out operator
operator splitting these components. Finally, we need an algorithm that
can be combined with CT to give both divergence-free transport of fields
and correct local dynamics. This will be achieved if the EMFs used in the
CT scheme contain information about all wave modes, which for stability,
must be appropriately upwinded. These multiple requirements can be met using the 
Method of Characteristics (MOC) to compute the EMFs. 
The resulting hybrid scheme is MOCCT~\citep{stone92b,hawley95}. 

Schematically, the EMFs can be written as (ignoring $\velg$ for simplicity)
\begin{equation}\label{emf1}
\epsilon 1\subijk \ = \ v2^{*}\subijk~b3^{*}\subijk \ - \ v3^{*}\subijk~b2^{*}\subijk
\end{equation}
\begin{equation}\label{emf2}
\epsilon 2\subijk \ = \ v3^{*}\subijk~b1^{*}\subijk \ - \ v1^{*}\subijk~b3^{*}\subijk
\end{equation}
\begin{equation}\label{emf3}
\epsilon 3\subijk \ = \ v1^{*}\subijk~b2^{*}\subijk \ - \ v2^{*}\subijk~b1^{*}\subijk
\end{equation}
where the starred quantities represent time-centered values for these
variables resulting from the solution of the characteristic equations at
the centers of zone edges where the EMFs are located. To simplify, we apply
MOC to the Alfv\'{e}n waves only, as the longitudinal modes are adequately handled
in a previous step by finite difference methods.

Because the MOC is applied only to transverse waves, we may derive the appropriate
differential equations by considering the 1-D MHD wave equations for an 
incompressible fluid~\citep{stone92b} which reduce to
\begin{equation}\label{alfvenwave1}
{\partial v \over \partial t}=\frac{B_x}{\rho}{\partial B \over \partial x}-{\partial \over 
\partial x}(v_x v),
\end{equation}
\begin{equation}\label{alfvenwave2}
{\partial B \over \partial t}=B_x{\partial v \over \partial x}-{\partial \over \partial x}(v_x B),
\end{equation}
where we have used the divergence-free constraint in one dimension (which 
implies $\partial B_x/\partial x\equiv 0)$ and the non-compressive nature of Alfv\'{e}n waves
(which implies $\partial v_x/\partial x\equiv 0)$. 

We can rewrite the coupled equations
(\ref{alfvenwave1}) and (\ref{alfvenwave2}) in characteristic form by multiplying equation 
(\ref{alfvenwave2}) by $\rho^{-1/2}$ and then adding and subtracting 
them, yielding
\begin{equation}\label{chareqn}
{\mathcal{D}v \over \mathcal{D}t}\pm\frac{1}{\rho^{1/2}}{\mathcal{D}B \over \mathcal{D}t}=0.
\end{equation}
The plus sign denotes the characteristic equation along the forward facing 
characteristic $C^+$, while the minus sign denotes the characteristic 
equation along the backward facing characteristic $C^-$. The comoving derivative
used in equation (\ref{chareqn}) is defined as 
\begin{equation}\label{charderiv}
\mathcal{D}/\mathcal{D}t\equiv\partial/\partial t + (v_x \mp B_x/\rho^{1/2})\partial/\partial x,
\end{equation}
where the minus (plus) sign is taken for the comoving derivative along the
$C^+ (C^-)$ characteristic. Note that the coefficient of the second term 
in equation (\ref{charderiv}) is just the Alfv\'{e}n velocity in the moving fluid,
$v_x\pm v_A$. Physically, equations (\ref{chareqn}) state that along characteristics,
which are straight lines in spacetime with slopes $v_x\pm v_A$, the changes in
the velocity and magnetic field components in each direction are not independent.

The finite-difference equations used to solve the characteristic equations (\ref{chareqn})
can be generated as follows. Consider the one dimensional space-time
diagram centered at the position of one of twelve edge-centered EMFs where
we require the values $v_i^*, B_i^*$ (see Figure \ref{spacetime}). Extrapolating 
back in time along the characteristics $C^+$ and $C^-$ to time level $n$ defines
the ``footpoints". By using upwind van Leer (1977) interpolation, we can compute
the time-averaged values for these variables in each domain of dependence. 
For both the velocities and the magnetic fields the characteristic speed 
$v_x\pm v_A$ are used to compute the footpoint values $v_i^{+,n}, B_i^{+,n},
v_i^{-,n}, B_i^{-,n}$. The finite difference equations along $C^+$ and $C^-$
become
\begin{equation}\label{fdchar1}
(v_i^*-v_i^{+,n})+(B_i^*-B_i^{+,n})/(\rho_i^+)^{1/2}=0,
\end{equation}
\begin{equation}\label{fdchar2}
(v_i^*-v_i^{-,n})-(B_i^*-B_i^{-,n})/(\rho_i^-)^{1/2}=0,
\end{equation}
where the subscript $i$ refers to cell $i$, not the $i$-th component of the
vectors $\vel, \B$. For simplicity, we set $\rho_i^+ = \rho_{i-1}^n$ and
$\rho_i^- = \rho_{i}^n$. The two linear equations for the two unknowns
$v_i^*$ and $B_i^*$ are then solved algebraically.  

\begin{figure}
\leavevmode
\includegraphics[width=\columnwidth]{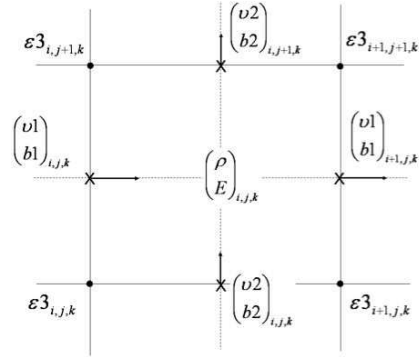}
\caption{A two dimensional ($x1-x2$) slice through the unit cell in Figure \ref{b-cube}
containing the four $\epsilon 3$'s. The computation of $\epsilon 3_{i,j,k}$ is illustrated in
Figure \ref{spacetime}.}\label{b-slice}
\end{figure}

\begin{figure}
\leavevmode
\includegraphics[width=\columnwidth]{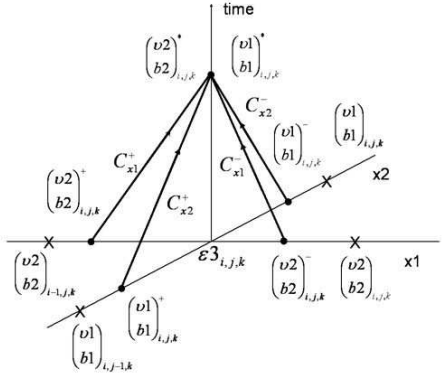}
\caption{The computation of $\epsilon 3_{i,j,k}$ involves the solution of
two 1-D characteristic equations for Alfv\'{e}n waves confined to the ($x1-x2$) plane. The
space-time diagrams for the solution of $b2^*, v2^*$ due to Alfv\'{e}n waves propagating
parallel to the $x1$ axis, and for the solution of $b1^*, v1^*$ due to Alfv\'{e}n waves
propagating parallel to the $x2$ axis are displayed.}\label{spacetime}
\end{figure}

For our multidimensional calculations, the characteristic equations are
solved in a directionally split fashion on planes passing through
the center of the cell and the cell edges where the EMFs are to be
evaluated. To illustrate, consider the calculation of
$\epsilon 3\subijk$ (Eq. \ref{emf3}).
The plane parallel to the $x1-x2$ plane passing through the four $\epsilon 3$'s 
in Figure \ref{b-cube} is shown in Figure \ref{b-slice}. Evaluating
$\epsilon 3_{i,j,k}$ requires values for 
$b1^*, v1^*, b2^*,$ and $v2^*$ at the cell corner $(x1_i, x2_j, x3_k)$. First,
as outlined above, one computes $b2^*, v2^*$ by solving the characteristic
equations for Alfv\'{e}n waves propagating in the $x1$ direction (Figure \ref{spacetime}). 
When calculating
the location of the footpoints, $v1$ and the Alfv\'{e}n
speed  are averaged to the zone corner. Then, the procedure is repeated for
$b1^*, v1^*$ by solving the characteristic
equations for Alfv\'{e}n waves propagating in the $x2$ direction, 
using $v2$ and the Alfv\'{e}n speed averaged to the zone corner.
Once all the $\epsilon 3$'s are evaluated in this way, the analogous 
procedure is followed for slices containing $\epsilon 1$ and $\epsilon 2$.
Only after all the EMFs have been evaluated over the entire grid
can the induction equation equation be updated in a divergence-free fashion.

Finally, we consider how the fluid momentum is updated due to the Lorentz Force.
The key point is that we do not want to throw away the fluid accelerations 
arising from Alfv\'{e}n waves that are implicit in the solution to the 
characteristic equations (\ref{fdchar1}) and (\ref{fdchar2}).
For example, the acceleration of $v3$ by the transverse magnetic forces is given by
\begin{equation}\label{eomforvi}
\rho{\partial v_3 \over \partial t} = -\grad_{(3)}(B^2/8\pi)+\frac{1}{4\pi}(\B\cdot\grad)B_3,
\end{equation}
or in terms of the EMFs:
\begin{eqnarray}\label{v3dottrans}
\rho\subijk (\frac{v3\subijk\supnp-v3\subijk\supn}{\Delta t}) & = &  \frac{1}{4\pi}
[\overline{b1}(\frac{b3^*_{\epsilon2(i+1,j,k)}-b3^*_{\epsilon2(i,j,k)}}{\Delta x1})
\nonumber \\  & + &
\overline{b2}(\frac{b3^*_{\epsilon1(i,j+1,k)}-b3^*_{\epsilon1(i,j,k)}}{\Delta x2})]
\end{eqnarray}
where $\overline{b1}$ and $\overline{b2}$ are four-point averages of the magnetic
field to the spatial location of $v3$, and
the $b3^*$'s are those values that enter into the EMF's referred to in the
subscripts (see Figure \ref{b-cube}.)
Similarly, the magnetic pressure calculation is
\begin{equation}\label{v3dotlong}
\rho\subijk (\frac{v3\subijk\supnp-v3\subijk\supn}{\Delta t}) = -\frac{1}{4\pi}
[\overline{b1}(\frac{\Delta b1}{\Delta x3})+
\overline{b2}(\frac{\Delta b2}{\Delta x3})].
\end{equation}

The evaluation of the EMF's outlined by equations (\ref{emf1}) through (\ref{emf3}) has
been modified according to a prescription due to~\citet{hawley95}, in which each
of the two $vB$ product terms is computed from a mix of quantities computed directly
from the characteristic equations with quantities estimated from simple advection.
Full details are provided in Appendix~\ref{mhdapp}, but the idea may illustrated
by an appropriate rewrite of (\ref{emf3}) for the evaluation of $\epsilon 3\subijk$:
\begin{eqnarray}\label{emf3hs}
\epsilon 3\subijk & = & 0.5*\left(v1^{*}\subijk~\overline{b2}\subijk +
                                  \overline{v1}\subijk~b2^{*}\subijk\right) \nonumber \\
                  & - & 0.5*\left(v2^{*}\subijk~\overline{b1}\subijk +
                                  \overline{v2}\subijk~b1^{*}\subijk\right),
\end{eqnarray}
where the starred quantities are derived from characteristic solutions and the
barred quantities arise from upwinded averages along the appropriate fluid velocity
component.  This modification (which engenders the ``HS'' in ``HSMOCCT'') introduces
a measure of diffusivity into the propagation of Alfv\'en waves which is not present
in the MOC scheme described in~\citet{stone92b}.  \citet{hawley95} note that this
change resulted in a more robust algorithm when applied to fully multidimensional
problems characterized by strong magnetic discontinuities.

   \subsection{Radiation Diffusion}\label{fldmeth}

The inclusion of radiation in the system of fluid equations to be solved
introduces four changes in the numerical solution.  First, an additional
contribution to the source-step momentum equation arises from the radiation
flux:
\begin{equation}
\rho{\partial \vel  \over \partial t} \ = \ -\grad\pg
\ - \ \divq \ + \
\left(\chif\over c\right)\F \ - \ {\rho\grad\Phi};\label{srcvel_rhd}.
\end{equation}
Second, new source/sink terms appear in the source-step gas energy equation:
\begin{equation}
    {\partial \egas \over \partial t} \ = \ c\ke\erad  \ -  4\pi\kp\pfunc
\ - \pg\divv \ - \ \gradvq.
\label{srcgas_rhd}
\end{equation}
Third, a new source-step equation is added for the radiation energy density:
\begin{equation}
    {\partial \erad \over \partial t} \ = \ 4\pi\kp\pfunc  \ - \ c\ke\erad
\ - \ \divf
 - \gradvp; \label{srcrad}
\end{equation}
and fourth, an additional advection equation for $\erad$ is added to
the transport step:
\begin{equation}
\dbdt\int_{V}\erad~dV \ = \ -\oint_{dV}\erad~(\vel - \velg)\cdot
 d{\bf S}. \label{advrad}
\end{equation}

In comparing equations (\ref{srcvel_rhd}) - (\ref{srcrad}) with either
the pure HD equations (\ref{srcvel_hd}) - (\ref{srcgas_hd}) or the
MHD analogs (\ref{srcvel_mhd}) - (\ref{induction}), it is clear that
the inclusion of radiation physics may be made to the HD or MHD systems
of equations with equal ease.  Similary, the transport step is augmented
with a solution of (\ref{advrad}) in either scenario.

\zmp~computes the evolution of radiating fluid flows through an implicit
solution to the coupled gas and radiation energy equations (\ref{srcgas_rhd}) and
(\ref{srcrad}).  Rather than solve the time-dependent radiation momentum
equation and treat the flux, $\F$, as an additional dependent variable,
we adopt the flux-limited diffusion (FLD) approximation as shown in (\ref{radm}).
This allows an algebraic substitution for $\F$ in the flux-divergence
term of the source-step equation (\ref{srcrad}) for E and avoids the need
for an additional advection equation for the flux.  The FLD approximation
is an attractive choice for multidimensional RHD applications for which
local heating/cooling approximations are inadequate.  With regard to computational
expense, FLD offers enormous economy relative to exact Boltzmann solutions
for the photon distribution function because the dimensionality of the
problem is reduced by 2 when the angular variation of the radiation field
is integrated away.  Additionally, the mathematical structure of the FLD equations
makes the solution amenable to parallel implementation.  Fundamentally, however,
the flux-limiter is a mathematical construction which interpolates
between the limiting cases of transparency and extreme opacity in a manner
that (hopefully) retains sufficient accuracy in the more difficult semi-transparent
regime.  Precisely what constitutes ``sufficient accuracy'' is dictated
by the needs of the particular application, and the techniques for meeting
that requirement may likewise depend upon the research problem.
\citet{levermore81} (LP) constructed an FLD theory which derived a form of
$\flim$ widely adopted in astrophysical applications (and in this paper).
In their work, LP use simple test problems to check the
accuracy of their FLD solution against exact transport solutions.  In simulations
of core-collapse supernovae, \citet{liebend04} compared calculations employing
energy-dependent multi-group FLD (MGFLD) calculations against those run with
an exact Boltzmann solver and found alternate forms of the limiter which better
treated the transport through the semi-transparent shocked material in the
post-bounce environment.  These calculations and others have shown that FLD
or MGFLD techniques can yield solutions that compare favorably with exact
transport, but that a ``one size fits all'' prescription for the flux limiter
is not to be expected.

In the context of applications, two other vulnerabilities of FLD bear
consideration.  \citet{hayes03} have compared FLD to VTEF solutions in
problems characterized by highly anisotropic radiation fields.  Because
the FLD equation for flux is a function of the local gradient in E, radiation
tends to flow in the direction of radiation energy gradients even when
such behavior is not physically expected, as in cases where radiation
from a distant source is shadowed by an opaque object.  In applications
where the directional dependence of the radiation is relevant (\citet{turner05}
discuss a possible example), an FLD prescription may be of limited reliability.
A second vulnerability concerns numerical resolution.  As discussed in detail
by~\citet{mihalas84}, the diffusion equation must be flux-limited because
numerically, the discrete diffusion equation advances a radiation wave
one mean-free path ($\lambda$) per time step.  Flux limiters are designed to act when
$\lambda / \Delta t$ exceeds c, which ordinarily one expects in a transparent
medium.  A problem can arise in extremely opaque media however, when the
physical mean-free path is much smaller than the width of a grid cell.  In
this case, $\lambda$ is unresolved, and the effective propagation length
scale is now determined by the much larger zone size.  Because the signal
speed is much less than c, the flux-limiter provides no constraint on the
propagation of radiation.  This can lead to unphysically
rapid heating of irradiated boundary layers which are spatially unresolved.
This problem has long bedeviled terrestrial transport applications; whether
this represents a liability for a given astrophysical application should be
carefully assessed by the user.

We consider now some basic details of our FLD module.
In our RHD prescription, matter and radiation may exchange energy through
absorption and emission processes represented by the right-hand sides of
equations (\ref{gase}) and (\ref{rade}), and the radiation stress term
on the LHS of (\ref{rade}).  The radiation energy budget is further influenced
by spatial transport, which we treat with the diffusion operator.  The
high radiation signal speed in transparent media mandates an implicit solution
to the radiation energy equation.  Coupling between the radiation
and matter is treated via an iterative scheme based upon Newton's method.
Recently, \citet{turner01} published a new FLD module for
the~\ztwd~code.  The physical assumptions underlying their method are consistent
with those identified here, but the mathematical treatment for solving the
coupled system differs from what we describe below.

Our construction of a linear system for the 3-D RHD equations begins with
expressions for the spatially and temporally discretized gas and
radiation energy equations.  Consider the gas and radiation energy densities
to be defined at discrete points along 3 orthogonal axes denoted by $i$, $j$,
and $k$; i.e. $\egas \rightarrow \egas\subijk$ and $\erad \rightarrow
\erad\subijk$.  We approximate the partial time derivative in terms of a
time-centered difference between two adjacent time levels, $\tn$ and
$\tnp$: $\dt \equiv \tnp - \tn$.  We then define two functions in $\egas\subijk$
and $\erad\subijk$:
\begin{eqnarray}
f^{(1)}\subijk & = & \erad\supnp\subijk \ - \ \erad\supn\subijk \nonumber \\
               & - & \dt\left[4\pi\kp\pfunc - c\ke\erad\supnp\subijk\right] \nonumber \\
               & - & \dt\left[\divf\subijk\supnp + \gradvp\subijk\supnp\right]; \label{discrad}\\
f^{(2)}\subijk & = & \egas\supnp\subijk \ - \ \egas\supn\subijk \nonumber \\
               & - & \dt\left[-4\pi\kp\pfunc + c\ke\erad\supnp\subijk\right] \nonumber \\
               & + & \dt\pg\divv. \label{discgas}
\end{eqnarray}
For notational economy, we have confined explicit reference to coordinate
indices and time level to the gas and radiation energy variables.  
As written above, the functions $f^{(1)}\subijk$ and $f^{(2)}\subijk$ are identically zero
for a consistent solution for $\egas\subijk$ and $\erad\subijk$.  We employ
a Newton-Raphson iteration scheme to find the roots of (\ref{discgas})
and (\ref{discrad}).  
To construct
a discrete system of equations which is linear in both energy variables, we
evaluate the E-dependent flux limiter from values of E at the previous time
level n.  The thermal pressure, Planck functions, and opacities are updated
at each iteration.  The velocities are roughly (but not formally) time-centered,
having been updated with contributions from body forces and artificial viscosity
prior to the radiation solution (cf. figure~\ref{figpc}).
We may write the linear system to be solved as
\begin{equation}\label{jacob}
J\left(x\right)\delta x \ = \ -f\left(x\right).
\end{equation}
In (\ref{jacob}), $x$ is the vector of gas and radiation energy
variables: $x \equiv \left(\erad\subijk,\egas\subijk\right)$.  Likewise,
the solution vector $\delta x$ is the set of corrections to these
variables, $\left(\delta\erad\subijk,\delta\egas\subijk\right)$. $f$
represents the vector of discrete functions $\left(f^{(1)}\subijk,f^{(2)}\subijk\right)$, 
and $J\left(x\right)$ is the Jacobian, $\partial f^i / \partial x^j$.

As written above, expression (\ref{jacob}) represents a matrix of size
(2N)x(2N), where N is the product of the numbers of mesh points along each
coordinate axis.  As will be shown in appendix~\ref{fldapp} the corrections,
$\delta\egas\subijk$, to the gas energies may be analytically expressed
as functions of the radiation energy corrections, $\delta\erad\subijk$.
This allows the solution of a reduced system of size NxN for the vector
of $\delta\erad\subijk$, from which the set of $\delta\egas\subijk$ are
immediately obtained.  These corrections are used to iteratively update
the trial values of $\egas\subijk$ and $\erad\subijk$.  We have found in
a variety of test problems that typically 2 to 10 N-R iterations are required
for a converged solution at each time step.

Each iteration in the N-R loop involves a linear system which must be solved
with matrix algebra.  Our solution for Poisson's equation also requires a
linear system solver.  We discuss the types of linear solvers
implemented in~\zmp~separately in~\S\ref{solvermeth}, with additional
details provided in appendices~\ref{fldapp}, \ref{poissonapp}, and
\ref{impapp}.

   \subsection{Self Gravity}\label{gravmeth}

\zmp~treats Newtonian gravity at three different levels of approximation:
(1) point-mass potentials, (2) spherically-symmetric gravity ($\grad\Phi = GM/r^2$),
and (3) exact solutions to Poisson's equation (\ref{poisson}).  The first two
options are trivial to implement; in this discussion we therefore focus on the
final option.  In three dimensions, the discrete Laplacian operator connects
a mesh point at coordinate (i,j,k) with both adjacent neighbors along each
axis; thus a finite-differenced form of (\ref{poisson}) takes the following
form:
\begin{eqnarray}
a_1\Phi\subijkm \ + \ a_2\Phi\subijmk \ + \ a_3\Phi\subimjk & + & \nonumber \\
a_4\Phi\subipjk \ + \ a_5\Phi\subijpk \ + \ a_6\Phi\subijkp & + & \nonumber \\
a_7\Phi\subijk & = & 4\pi G\rho\subijk.\label{discpoi}
\end{eqnarray}
If (\ref{discpoi}) is defined on a Cartesian mesh in which the zone spacing
along each axis is uniform, then the ``$a$'' coefficients
in equation~\ref{discpoi} are constant over the problem domain.  If, in addition
to uniform gridding, the problem data is characterized by spatial periodicity,
then {\it Fast Fourier Transform} (FFT) algorithms offer a highly efficient
method for solving (\ref{discpoi}).  For this class of problems (see~\citet{li04}
for a recent example), \zmp~provides an FFT module based upon the publicly
available ``FFTw'' software~\citep{frigo05}.  While FFT-based methods are
not in general restricted only to periodic problems, we emphasize that the
module implemented in~\zmp~is valid only for 3-D problems on uniform
Cartesian meshes with triply-periodic boundaries.

For multidimensional problems which do not meet all of the validity criteria
for the FFTw solver, \zmp~provides two additional solution modules.  The
most general of these accommodates 2-D and 3-D grids in Cartesian, cylindrical,
and spherical geometries and is based upon the same CG solver provided for
the FLD radiation equations. A second module currently written for 3-D Cartesian
meshes with Dirichlet or Neumann boundary conditions is based upon the
{\it multigrid} (MG) method (cf.~\S\ref{mgmeth}).

When equation (\ref{discpoi}) is formulated as a function of ZEUS covariant
grid variables, the matrix elements represented by the ``$a$'' coefficients
take on a much more complicated functional form than the constant values which
obtain for a uniform Cartesian mesh.  The form of (\ref{discpoi}) written
for the general class of 3-D covariant grids is documented in appendix
\ref{poissonapp}.  Details of all three solution techniques for Poisson's
equation are written in~\S\ref{solvermeth}.

   \subsection{Multi-species Advection}\label{madmeth}

Prior public versions of ZEUS codes have treated the gas as a single
species fluid.  In order to be forward-compatible with physical processes
such as chemistry, nuclear reactions, or lepton transport, \zmp~
offers a straightforward mechanism for identifying and tracking
separate chemical or nuclear species in a multi-species fluid mixture.
Because~\zmp~solves one set of fluid equations for the total mass
density, $\rho$, we have no facility for modeling phenomena in which
different species possess different momentum distributions and thus
move with respect to one another.  Nonetheless, a wide variety of
astrophysical applications are enabled with a mechanism for quantifying
the abundances of separate components in a mixed fluid.
Our multi-species treatment considers only the physical advection of
different species across the coordinate mesh; physics modules which
compute the local evolution of species concentrations (such as a nuclear
burning network) must be provided by the user.

Our implementation of multispecies advection proceeds by defining a
concentration array, X$_{\rm n}$, such that $\rho{\rm X_n}$ is
the fractional mass density of species n.  The advection equations in
the ZEUS transport step therefore include:
\begin{equation}
\dbdt\int_{V}\left(\rho{\rm X_n}\right)~dV \ = \ 
-\oint_{dV}\left(\rho{\rm X_n}\right)~(\vel - \velg)\cdot
 d{\bf S}. \label{advchi}
\end{equation}
This construction is evaluated such that the mass fluxes used to advect individual
species across the mesh lines are consistent with those used to advect
the other field variables defined in the application.  Discrete formulae
for the conservative advection of X$_{\rm n}$ and the other hydrodynamic
field variables are provided in appendix~\ref{hdapp}.

   \subsection{Time Step Control}\label{dtmeth}

Maintainence of stability and accuracy in a numerical calculation requires proper
management of time step evolution.  The general expression
regulating the time step in ZEUS-MP is written
\begin{eqnarray}\label{dtcon}
\dt_{\rm new} & = & C_{\rm cfl} \nonumber \\
              & \times &
          \left\{{1\over\dt_{\rm cs}^{2}} + {1\over\dt_{v1}^{2}}  +
                {1\over\dt_{v2}^{2}}+ {1\over\dt_{v3}^{2}}+\right. \nonumber \\
            &  &\mbox{}\left. {1\over\dt_{\rm al}^{2}}+{1\over\dt_{\rm av}^{2}} +
                {1\over\dt_{\rm rd}^{2}}\right\}^{1/2}, \nonumber \\
\end{eqnarray}
in which C$_{\rm cfl}$ is the Courant factor and the $\dt^2$ terms are squares of the
minimum values of the following quantities:
\begin{eqnarray}
\dt^2_{\rm cs} & = & \gamma\left(\gamma - 1\right)\cdot\left(\egas/\rho\right)~/~
                     \left(\Delta X_{\rm min}\right)^2;  \\
\dt^2_{v1}     & = & \left[\left(v1 - vg1\right)~/~dx1a\right]^2; \\
\dt^2_{v2}     & = & \left[\left(v2 - vg2\right)~/~(g2b~dx2a)\right]^2; \\
\dt^2_{v3}     & = & \left[\left(v3 - vg3\right)~/~(g31b~g32b~dx3a)\right]^2; \\
\dt^2_{\rm al} & = & \left(\bonebar^2 + \btwobar^2 + \bthrbar^2\right)~/~
                     \left[4\rho\left(\Delta X\right)^2\right]; \\
\dt^2_{\rm av} & = & \left[4~{\rm qcon}~\left|dv\over dx\right|_{\rm max}\right]^{-2}; \\
\dt^2_{\rm rd} & = & \left[{\bf ertol}\left|\erad\over{\Delta\erad}\right|\right]^2
\end{eqnarray}
These values represent, respectively, the local sound-crossing time, the local fluid
crossing time along each coordinate, the local Alfv\'en wave crossing time, the
local viscous timescale, and a radiation timescale determined by dividing the value
of $\erad$ returned from the FLD solver by the time rate of change in $\erad$ determined
by comparing the new $\erad$ to that from the previous time step.  {\bf ertol}  is
a specified tolerance which retrodictively limits the maximum fractional change in
$\erad$ allowed in a timestep.  $\Delta X_{\rm min}$ represents the minimum length of a
3-D zone edge, i.e. MIN$\left[dx1a,~g2b~dx2a,~g31b~g32b~dx3a\right]$, where each
zone length is expressed in terms of the local covariant grid coefficients
(cf. appendix~\ref{hdapp}).  As expressed in (\ref{dtcon}), $dt_{\rm new}$ represents
a trial value of the new time step, which is allowed to exceed the previous value of
the time step by no more than a preset factor; i.e. $\dt_{\rm final}~=~{\rm min}
\left[\dt_{\rm new},{\rm fac}\times\dt_{\rm old}\right]$, with ``fac'' typically
equaling 1.26.  This value allows the time step to increase by up to a factor
of 10 every 10 cycles.

   \subsection{Parallelism}\label{parmeth}

The most powerful computers available today are parallel systems with hundreds
to thousands of processors connected into a cluster.  While some systems offer
a shared-memory view to the applications programmer, others, such as Beowulf
clusters, do not. Thus, to maximize portability we have assumed ``shared
nothing'' and implemented~\zmp~as an SPMD (Single Program, Multiple Data)
parallel code using the MPI message-passing library to accomplish interprocessor
communication.  In this model, parallelism is affected via {\it domain decomposition}
\citep{foster95}, in which each CPU stores data for and performs operations
upon a unique sub-block of the problem domain.  Because finite-difference forms
of gradient, divergence, and Laplacian operators couple data at multiple
mesh points, data must be exchanged between neighboring processors when such
operations are performed along processor data boundaries.  \zmp~employs
``asynchronous'' or ``non-blocking'' communication functions which allow
interprocessor data exchange to proceed simultaneously with computational
operations.  This approach provides the attractive ability to hide a large
portion of the communication costs and thus improve parallel scalability.
Details of our method for overlapping communication and computation operations
in~\zmp~are provided in appendix~\ref{impapp}.

   \subsection{Linear Solvers}\label{solvermeth}

Our implicit formulation of the RHD equations and our solution to Poisson's
equation for self gravity require the use of an efficient linear system
solver.  Linear systems for a single unknown may involve of order 10$^6$
solution variables for a 3-D mesh at low to moderate resolution; the
number of unknowns in a high-resolution 3-D simulation can exceed 10$^9$.
In this regime, the CPU cost of the linear system solution can easily
dominate the cost of the full hydrodynamic evolution.  The choice of solution
technique, with its associated implementation requirements and performance
attributes, is therefore critically important.  Direct inversion methods such
as Gauss-Seidel are ruled out owing both to extremely high operation counts
and a spatially recursive solution which precludes parallel implementation.
As with radiation flux limiters, there is no ``best'' method, practical
choices being constrained by mathematical factors such as the matrix {\it condition
number} (cf.~\S\ref{cgmeth}), coordinate geometry, and boundary conditions, along
with performance factors such as sensitivity to problem size and ease of
parallel implementation.  This variation of suitability with problem configuration
motivated us to instrument~\zmp~with three separate linear solver packages: a
{\it preconditioned conjugate gradient} (CG) solver, a {\it multigrid} (MG)
solver, and a {\it fast Fourier transform} (FFT) solver.  We describe
each of these below.

      \subsubsection{The Conjugate Gradient Solver}\label{cgmeth}

The conjugate gradient (CG) method is one example of a broad class of
{\it non-stationary iterative} methods for sparse linear systems.  A concise
description of the theory of the CG method and a pseudo-code template for
a numerical CG module is available in~\citet{barret94}.  While a full discussion
of the CG method is beyond the scope of this paper, several key elements will
aid our discussion.  The linear systems we wish to solve may be written in the
form $Ax = b$, where $A$ is the linear system matrix, $x$ is the unknown
solution vector, and $b$ is a known RHS vector.  An associated quadratic form,
$f(x)$, may be constructed such that
\begin{equation}
f(x) \ = \ {1\over 2} x^T A x \ - \ b^T x \ + \ c,
\end{equation}
where $c$ is an arbitrary constant (the ``T'' superscript denotes a transpose).
One may show algebraically that if $A$ is symmetric ($A^T = A$) and
positive-definite ($x^T A x > 0$ for all non-zero $x$), then the vector $x$
which satisfies $Ax = b$ also satisfies the condition that $f(x)$ is
minimized, i.e.
\begin{equation}\label{cgmin}
f'(x) \ \equiv \ \left\{
\begin{array}{c}
{\partial \over x_1}f(x) \\
{\partial \over x_2}f(x) \\
          .              \\
          .              \\
          .              \\
{\partial \over x_n}f(x)
\end{array}
                 \right\} \ = \ 0.
\end{equation}
The CG method is an iterative technique for finding elements of $x$ such
that (\ref{cgmin}) is satisfied.  A key point to consider is that the convergence
rate of this approach is strongly sensitive to the {\it spectral radius} or
{\it condition number} of the matrix, given by the ratio of the largest to
smallest eigenvalues of $A$.  For matrices that are poorly conditioned, the
CG method is applied to ``pre-conditioned'' systems such that
\begin{equation}\label{pccg}
M^{-1}Ax \ = \ M^{-1}b,
\end{equation}
where the {\it preconditioner}, $M^{-1}$, is chosen such that the eigenvalues
of ($M^{-1}A$) span a smaller range in value. (Typically, one equates the
preconditioner with $M^{-1}$ rather than $M$.)

\begin{deluxetable*}{lll}
\tablecaption{Multigrid V-cycle iteration.\label{tabmgmpi}}
\tablehead{
\colhead{Keyword} & \colhead{Operation} & \colhead{Description} }
\startdata
\textsf{smooth} &
  ${\cal L}_h \bar{x}^n_h = b_h$ &
  \it{Smooth error on fine grid via stationary method} \\
\textsf{compute} &
  $r^n_h \leftarrow b_h - {\cal L}_h \bar{x}^n_h$ &
  \it{Compute residual on fine grid} \\
\textsf{restrict} &
  $r^n_{h} \rightarrow r^n_{2h}$ &
  \it{Transfer residual down to coarse grid} \\
\textsf{solve} &
  ${\cal L}_{2h} e^n_{2h} = r^n_{2h}$ &
  \it{Obtain coarse grid correction from residual equation} \\
\textsf{prolong} &
  $e^n_{h} \leftarrow e^n_{2h}$ &
  \it{Transfer coarse grid correction up to fine grid} \\
\textsf{update} &
  $x^{n+1}_h \leftarrow \bar{x}^n_h + e^n_h$ &
  \it{Update solution with coarse grid correction} \\
\enddata
\end{deluxetable*}

From (\ref{pccg}) it follows at once that the ``ideal'' preconditioner for
$A$ is simply $A^{-1}$, which of course is unknown.  However, for matrices
in which the main diagonal elements are much larger in magnitude
than the off-diagonal elements, a close approximation to $A^{-1}$ may
be constructed by defining a diagonal matrix whose elements are given by
the reciprocals of the corresponding diagonal elements of $A$.  This technique
is known as {\it diagonal preconditioning}, and we have adopted it in the
implementation of our CG solver.
The property in which the diagonal elements of $A$ strongly exceed (in absolute
value) the values of the off-diagonal elements is known as {\it diagonal
dominance}.  Diagonal dominance is a prerequisite for the profitable application
of diagonal preconditioning.  Diagonal preconditioning is an attractive technique
due to its trivial calculation, the fact that it poses no barrier to parallel
implementation, and its fairly common occurrence in linear systems.  Nonetheless,
sample calculations in~\S\ref{perf} will demonstrate cases in which diagonal
dominance breaks down, along with the associated increase in cost of the linear
solution.

      \subsubsection{The Multigrid Solver}\label{mgmeth}

Unlike the conjugate gradient method, multigrid methods~\citep{Br77}
are based on stationary iterative methods.  A key feature of multigrid
is the use of a hierarchy of nested coarse grids to dramatically
increase the rate of convergence.  Ideally, multigrid methods are
fast, capable of numerically solving elliptic PDE's with computational
cost proportional to the number of unknowns, which is optimal.  For
example, for a $k^3$ problem in three dimensions, multigrid
(specifically the
\textit{full multigrid} method, discussed below) 
requires only $O(k^3)$ operations.  Compare this to $O(k^3 \log k)$
for FFT methods, $O(k^4)$ for non-preconditioned CG, and approximately
$O(k^{3.5})$ for preconditioned CG~\citep{He97}.  Multigrid has
disadvantages as well, however; they are relatively difficult to
implement correctly, and are very sensitive to the underlying PDE and
discretization.  For example, anisotropies in the PDE coefficients or
grid spacing, discontinuities in the coefficients, or the presence of
advection terms, can all play havoc with standard multigrid's
convergence rate.

Stationary methods, on which multigrid is based, are very simple, but
also very slow to converge.  For a linear system $Ax=b$, the two main
stationary methods are Jacobi's method ($a_{ii}x_i^{n+1} = b_i -
\sum_{j \neq i} a_{ij}x_j^{n}$) and the Gauss-Seidel method
($a_{ii}x_i^{n+1} = b_i - \sum_{j < i} a_{ij}x_j^{n+1} -
\sum_{j>i}a_{ij}x_j^n$), where subscripts denote matrix and vector
components, and superscripts denote iterations.  While stationary
methods are very slow to converge to the solution (the computational
cost for both Jacobi and Gauss-Seidel methods is $O(k^5 \log k)$ for a
$k^3$ elliptic problem in $3D$), they do reduce the high-frequency
components of the error very quickly; that is, they efficiently
``smooth'' the error.  This is the first part of understanding how
multigrid works.  The second part is that a problem with a smooth
solution on a given grid can be accurately represented on a coarser
grid. This can be a very useful thing to do, because problems on
coarser grids can be solved faster.

Multigrid combines these two ideas as follows.  First, a handful of
iterations of a stationary method (frequently called a ``smoother'' in
multigrid terminology) is applied to the linear system to smooth the
error.  Next, the residual for this smoothed problem is transfered to
a coarse grid, solved there, and the resulting coarse grid correction
is used to update the solution on the original (``fine'') grid.
Table~\ref{tabmgmpi} shows the main algorithm for the
\textit{multigrid V-cycle iteration}, applied to the linear
system ${\cal L}_h x^n_h = b_h$ associated with a grid with zone
spacing $h$.

\newcommand{\codecomment}[1]{\begin{minipage}[t]{4in}\textit{#1}\end{minipage}}

Note that the coarse grid problem (keyword ``\textsf{solve}'' in 
table~\ref{tabmgmpi}) is solved recursively.  The
recursion bottoms out when the coarsest grid has a single unknown; or,
more typically, when the coarse grid is small enough to be quickly
solved using some other method, such as CG, or with a small number of
applications of the smoother.  Also, the multigrid V-cycle can
optionally have additional applications of the smoother at the end of
the iteration.  This is helpful to smooth errors introduced in the
coarse grid correction, or to symmetrize the iteration when used as a
preconditioner.

The \textit{full multigrid} method uses V-cycles in a bootstrapping
approach, first solving the problem on the the coarsest grid, then
interpolating the solution up to the next-finer grid to use as a
starting guess for a V-cycle.  Ideally, just a single V-cycle at each
successively finer grid level is required to obtain a solution whose
error is commensurate with the discretization error.

\newcommand{\mgmpi}{\textsf{MGMPI}}
Multigrid methods in~\zmp~are provided using an external
MPI-parallel C++/\linebreak[0]C/\linebreak[0]Fortran package called
\mgmpi~\citep{Bo02}.  It includes a suite of Krylov subspace methods
as well as multigrid solvers.  The user has flexible control over the
multigrid cycling strategy, boundary conditions, depth of the
multigrid mesh hierarchy, choice of multigrid components, and even
whether to use Fortran or C computational kernels.  Parallelization is
via MPI using the same domain decomposition as~\zmp.  Currently
there are limitations to grid sizes in \mgmpi: there must be $M
2^{L}-1$ zones along axes bounded by Dirichlet or Neumann boundary
conditions, and $M 2^{L}$ zones along periodic axes, where $L$ is the
number of coarse grids used, and $M$ is an arbitrary integer.  This
restriction is expected to change in future versions of \mgmpi.

\begin{figure}
\leavevmode
\includegraphics[width=\columnwidth]{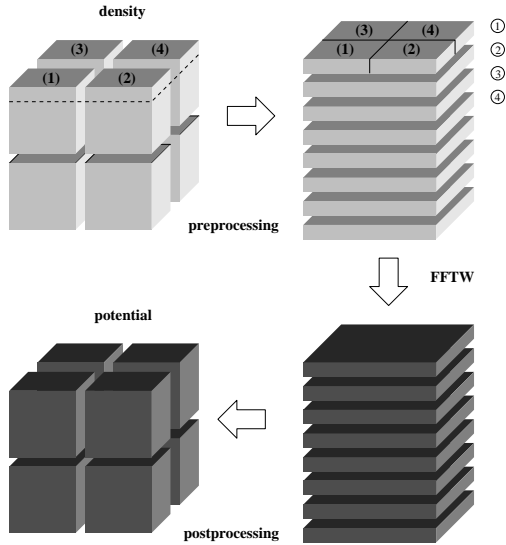}
\caption{Block-to-slab decomposition of density data before calling FFTW library and
slab-to-block decomposition of gravitational potential afterward.}\label{figfftw}
\end{figure}

      \subsubsection{The Fast Fourier Transform Solver}\label{fftmeth}

As mentioned in~\S\ref{gravmeth}, FFT algorithms offer a highly
efficient method in solving the Poisson equation.  The publicly available
``Fastest Fourier Transform in the West'' (FFTw)
algorithm~\citep{frigo05} is used as one of the gravity solvers
available in~\zmp.  Note that the parallelized version of the FFTw
library using MPI is only available in version 2.1.5 or before.  Based
on this version of MPI FFTw, the gravity solver implemented in~\zmp~
is valid only for Cartesian meshes with triply-periodic boundaries.

The transform data used by the FFTw routines is distributed, which
means that a distinct portion of data resides in each processor during
the transformation.  In particular, the data array is divided
along the first dimension of the data cube, which is sometimes
called a slab decomposition.  Users can design their data layout using
slab decomposition as the FFTw requires but it is inconvenient for
solving general problems.  Therefore, there is a preprocessing of the
density data distribution from a general block domain decomposition to a slab
decomposition before calling the FFTw routines.  After the potential
is calculated, the slab decomposition of the potential is transformed
back to block decomposition in the postprocessing stage.
Figure~\ref{figfftw} shows the idea of these two additional processes.

In figure~\ref{figfftw}, the initial density data block is first
rearranged from block decomposition into slab decomposition.  In this
example using a 2 $\times$ 2 $\times$ 2 topology, the first layer of
blocks will be divided into four slabs.  The four processors 
exchange the data slabs to ensure the density data remains organized
at the same spatial location.  Therefore, the data exchange can be
viewed as a rearrangement of the spatial location of processors.
Since data in the second layer of blocks does not overlap with the first
layer of data blocks, the non-blocking data communication among blocks
in different layers can proceed simultaneously.  After the gravitational
potential is calculated, the reverse rearrangement of potential data
from slab decomposition back to block decomposition is performed
in an analogous manner.

Because of the required slab decomposition in FFTw, the number of
processors that can be used for a given problem size is limited.  For
example, for a problem with 512$^3$ zones on 512 processors, each slab
least one cell thick.  Using more than 512 processors in this example
will not lessen the time to solution for the potential as extra processors
would simply stand idle.

\section{Verification Tests}\label{verify}

In this section we present results from a suite of test problems designed to stress
each of~\zmp's physics modules and verify the correct function of each.  We begin
with a pure HD problem which follows shock propagation in spherical geometry.  We
then examine a trio of MHD problems, two of which were considered in~\citet{stone92b},
and the final of which has become a standard multidimensional test among developers
of Godunov-based MHD codes.  The section concludes with two radiation problems,
the first treating radiation diffusion waves through static media; the second following
the evolution of radiating shock waves.  All problems with shocks use a quadratic
(von Neumann-Richtmyer) artificial viscosity coefficient qcon of 2.0.  The
Orszag-Tang vortex test uses an additional linear viscosity with a value of qlin = 0.25.

Three of the test problems discussed below which include hydrodynamic effects
are also adiabatic: these are the Sedov-Taylor HD blast wave, the MHD Riemann
problem, and the Orszag-Tang MHD vortex problem.  In these cases, the total
fluid energy integrated over the grid should remain constant.  Because~\zmp~
evolves an internal gas energy, rather than total fluid energy, equation,
the~\zmp~solution scheme is non-conservative by construction.  It therefore
behooves us to monitor and disclose errors in total energy conservation as
appropriate.  For the three adiabatic dynamics problems noted above, the
total energy conservation errors (measured relative to the initial integrated
energy) were 1.4\%, 0.8\%, and 1.6\%, respectively.

For problems involving magnetic field evolution, an additional metric of
solution fidelity is the numerical adherence to the divergence-free
constraint.  As shown analytically in~\S\ref{mhdmeth} and previously
in~\citet{hawley95,stone92b,evans88}, the Constrained Transport algorithm
is divergence-free by construction, regardless of the method chosen to
compute the EMF's.  Nonetheless, all numerical codes which evolve discrete
representations of the fluid equations are vulnerable to errors bounded
from below by machine round-off; we therefore compute $\diver\B/|\B|$
at each mesh point and record the largest positive and negative values
thus obtained.  For the Alfv\'en rotor and MHD Riemann problems, which
are computed on 1D grids, the maximum normalized divergence values at
the end of the calculations remain formally zero to machine tolerance
in double precision (15 significant figures).  For the 2D Orszag-Tang
vortex, the divergence-free constraint remains satisfied to within roughly
1 part in 10$^{12}$, consistent with machine round-off error over an
evolution of roughly 1000 timesteps.

   \subsection{Hydrodynamics: Sedov-Taylor Blast Wave}\label{verifyhydro}

\begin{figure}
\leavevmode
\includegraphics[width=\columnwidth]{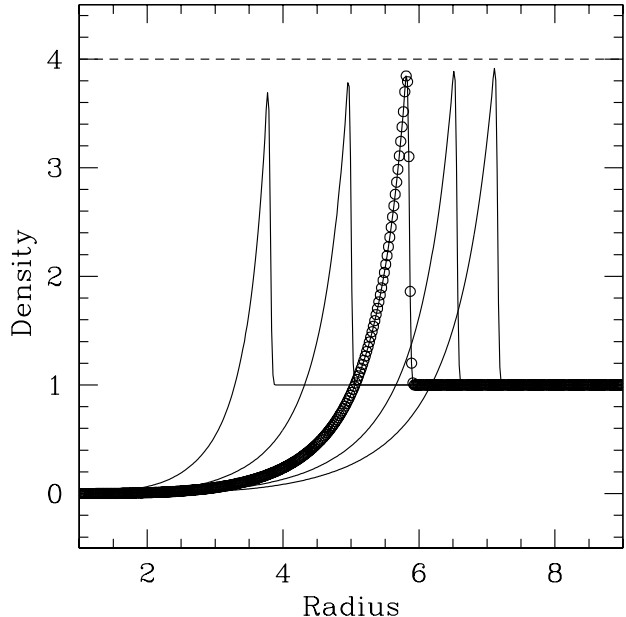}
\caption{Density vs. radius for the
Sedov-Taylor blast wave.  Density is plotted in units of $10^{-8}$ g cm$^{-3}$;
radius is plotted in units of $10^{13}$ cm.  Profiles
represent numerical results plotted in intervals of $6 \times 10^4$ seconds.
The dashed line shows the analytic value for the peak density.}
\label{strho}
\end{figure}

Our first test problem is a classic hydrodynamic test due to~\citet{sedov59}, in 
which a point explosion is induced at the center of a cold, homogeneous sphere
in the absence of gravity.  Problem parameters are chosen such that the explosion
energy is orders of magnitude larger than the total internal energy of the cloud.
In this case, the resulting shock wave evolves in a self-similar fashion in which
the shock radius and velocity evolve with time according to
\begin{equation}\label{rsh_st}
r_{sh} \ = \ \xi_{sh} \left(E_o\over\rho_o\right)^{1/5}t^{2/5},
\end{equation}
and
\begin{equation}\label{vsh_st}
v_{sh} \ = \ {2\over 5}\xi_{sh} \left(E_o\over\rho_o\right)^{1/5}t^{-3/5},
\end{equation}
where $E_o$ and $\rho_o$ are the explosion energy and the initial density,
respectively. $\xi_{sh}$ is a dimensionless constant which is equal to 1.15 for
an ideal gas with $\gamma = 5/3$.
The density, pressure, temperature, and fluid velocity at the shock front
are given by
\begin{equation}\label{d_st}
\rho_s \ = \ 4\rho_o,
\end{equation}
\begin{equation}\label{p_st}
P_s \ = \ {3\over 4}\rho_o v^2_{sh},
\end{equation}
\begin{equation}\label{t_st}
T_s \ = \ {3\over 16}{\mu m_h\over k_B} v^2_{sh},
\end{equation}
and
\begin{equation}\label{v_st}
v_s \ = \ {3\over 4} v_{sh}.
\end{equation}

\begin{figure}
\leavevmode
\includegraphics[width=\columnwidth]{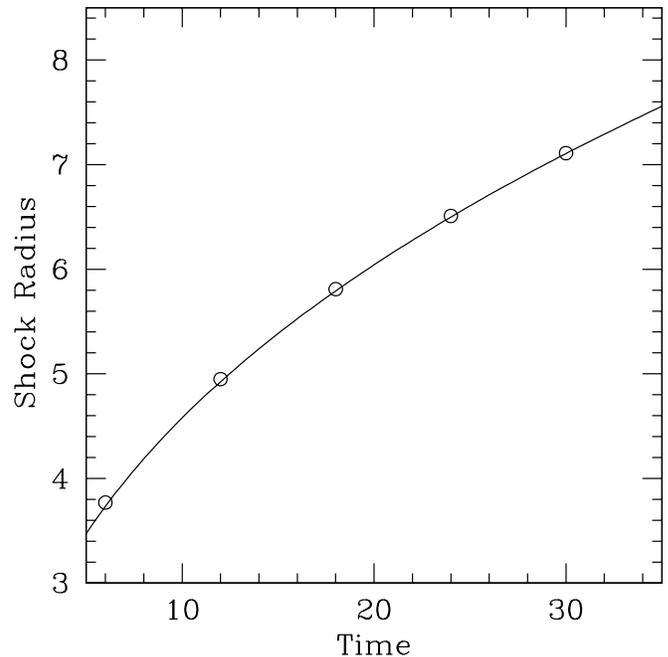}
\caption{Shock radius vs. time for the Sedov-Taylor blast wave.  Time is plotted
in units of 10$^4$ seconds in the right-hand figure.  Open circles
represent numerical results plotted at the times corresponding to profiles in
Figure~\ref{strho}.  The solid line indicates the analytic solution.}
\label{strad}
\end{figure}

Our problem was run in one dimension on a mesh of 500 zones equally spaced in radius.
We initialize a spherical cloud of radius 10$^{14}$ cm with a uniform density of
10$^{-8}$ g/cm$^3$.  The initial temperature is 50 K.  At t = 0, 10$^{50}$ ergs of
internal energy are deposited within a radius of 10$^{12}$ cm, which spreads the
blast energy over 5 zones.  Depositing the energy over a few zones within a small
region centered on the origin maintains the point-like nature of the explosion
and markedly improves the accuracy of the solution relative to that obtained if
the all of the energy is deposited in the first zone.

Figures~\ref{strho} and~\ref{strad} provide results of our Sedov-Taylor blast wave
test.  Figure~\ref{strho} shows radial plots of density separated in time by 
$6 \times 10^4$ seconds.  The density and radius are expressed in units of
$10^{-8}$ g/cm$^{-3}$ and cm, respectively.  The dashed line indicates the analytic
value for the density at the shock front.  Figure~\ref{strad} shows numerical
values (open circles) of the shock front at times identical to those used in the
Figure~\ref{strho}.  These data are superimposed upon the analytic solution (solid line)
for the shock radius given by equation~\ref{rsh_st}.

   \subsection{MHD}\label{verifymhd}

      \subsubsection{Magnetic Braking of an Aligned Rotor}

\begin{figure}
\leavevmode
\includegraphics[width=\columnwidth]{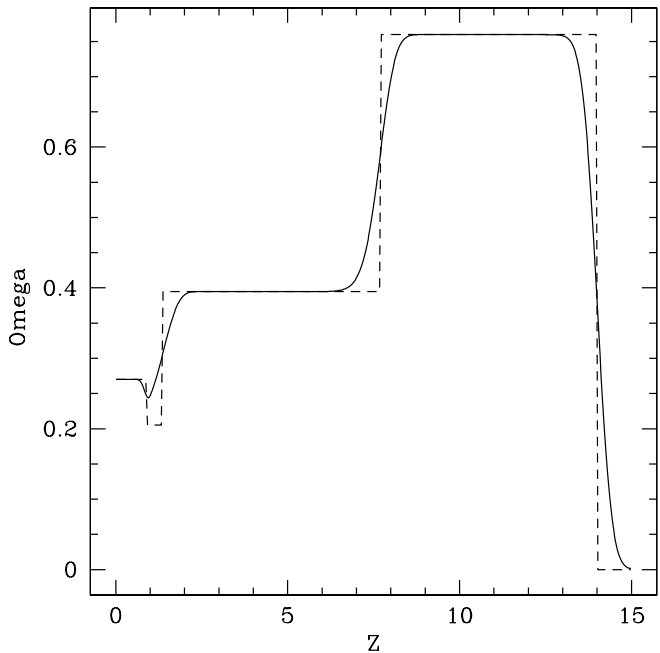}
\caption{MHD aligned rotor: angular velocity
vs. Z.  The dashed line indicates the analytic solution of \protect\citet{mouschovias80}.}
\label{rotor_dic_om}
\end{figure}

\begin{figure}
\leavevmode
\includegraphics[width=\columnwidth]{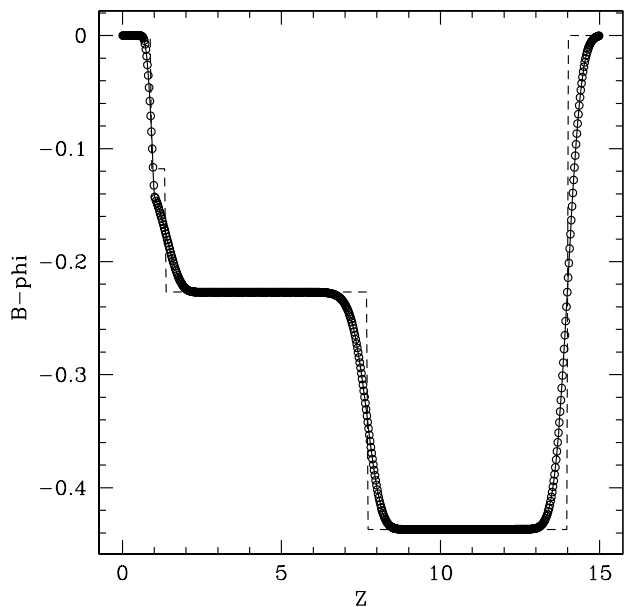}
\caption{MHD aligned rotor: $B_\phi$
vs. Z.  The dashed line indicates the analytic solution of \protect\citet{mouschovias80}.}
\label{rotor_dic_bphi}
\end{figure}

\begin{figure}
\leavevmode
\includegraphics[width=\columnwidth]{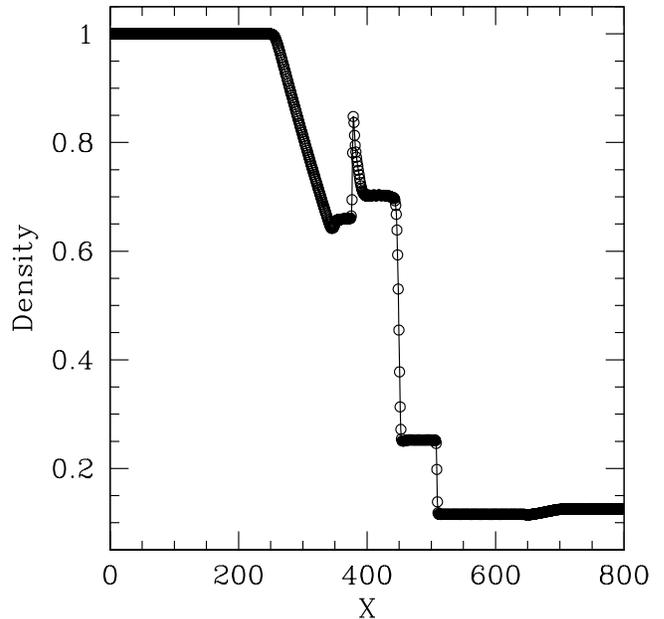}
\caption{MHD Riemann problem: density vs. X at t = 80 sec.}
\label{bwurho}
\end{figure}

\begin{figure}
\leavevmode
\includegraphics[width=\columnwidth]{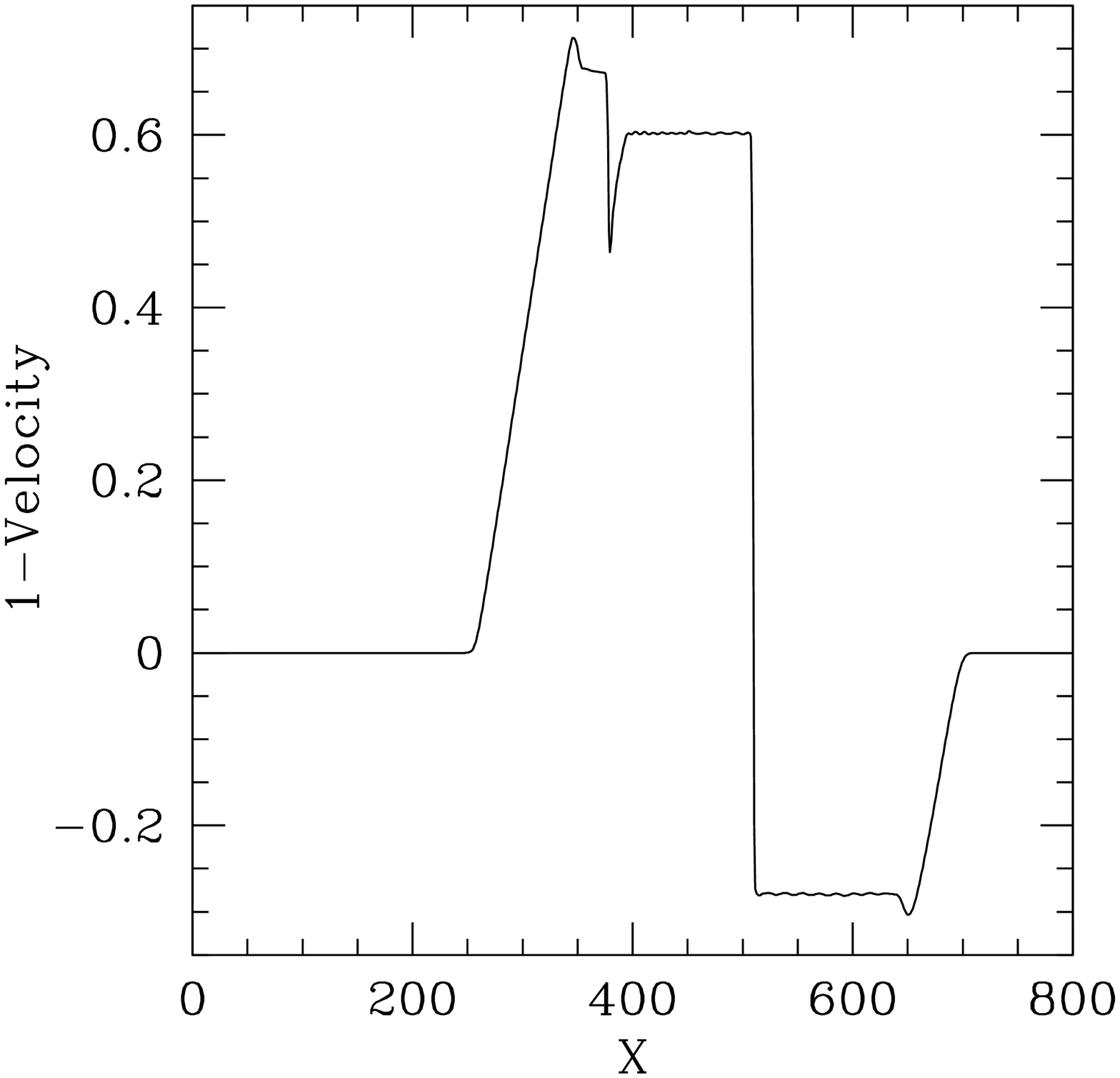}
\caption{MHD Riemann problem: 1-velocity vs. X at t = 80 sec.}
\label{bwuv1}
\end{figure}

\begin{figure}
\leavevmode
\includegraphics[width=\columnwidth]{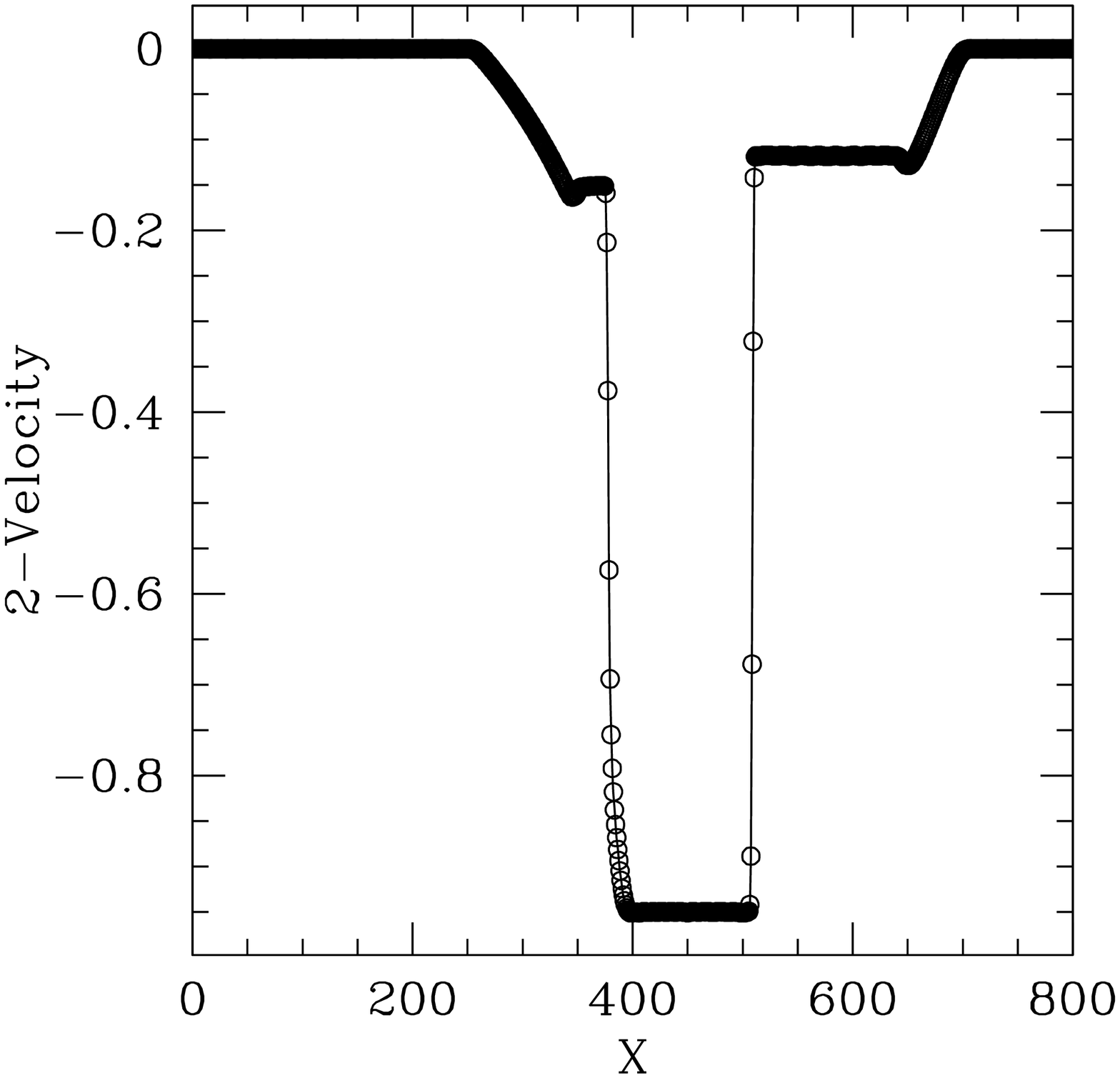}
\caption{MHD Riemann problem: 2-velocity vs. X at t = 80 sec.}
\label{bwuv2}
\end{figure}

Our first MHD test examines the propagation of torsional Alfv\'en waves generated
by an aligned rotor.  A disk of uniform density $\rho_d$, thickness $z_d$, and
angular velocity $\Omega_0$ lies in an ambient, initially static, medium with
density $\rho_m$.  The disk and medium are threaded by an initially uniform
magnetic field oriented parallel to the rotation axis of the disk.  
Considered in cylindrical geometry, rotation of
the disk produces transverse Alfv\'en waves which propagate
along the Z axis and generate non-zero $\phi$ components of velocity and
magnetic field.  Analytic solutions for $v_\phi$ and $B_\phi$ were calculated
by~\citet{mouschovias80} under the assumption that only the transverse Alfv\'en
wave modes are present; to reproduce these conditions in ZEUS-MP, compressional
wave modes due to gradients in gas and magnetic pressures are
artificially suppressed.  The utility of this restriction lies in
the fact that in more general calculations, errors in the propagation of Alfv\'en
waves may easily be masked by the effects of other wave modes in the problem.

The problem parameters as described above correspond to the case of discontinuous
initial conditions considered in~\citet{mouschovias80}.  We consider a half-plane
spanning the range $0~\leq~Z~\leq~15$, with $\rho_d$ = 10 and $\rho_m$ = 1.
Because there are no dynamical phenomena acting along the radial coordinate, we
may compute the problem on a 1-D grid of Z on which $R$- and $\phi$-invariant
values of $v_\phi$ and $B_\phi$ are computed.  Figures~\ref{rotor_dic_om} 
and~\ref{rotor_dic_bphi} show
results of the calculation at a time t = 13.  Solid curves indicate the solutions
from~\zmp; dashed lines show the analytic solution of~\citet{mouschovias80}.
These results are consistent with those obtained with~\ztwd~and shown in
\citet{stone92b}; the only salient difference between the two calculations is that
we used twice as many zones (600) as reported for the~\ztwd~calculation.  The
increased resolution is mandated by the fact that the HSMOCCT algorithm is
by construction more diffusive than the original MOCCT algorithm documented
in~\citet{stone92b}.  As noted previously in section~\ref{mhdmeth} and discussed in 
detail in~\citet{hawley95}, this added
diffusivity makes the MOCCT algorithm more robust in fully multidimensional
calculations.  The requirement within HSMOCCT of higher resolution with respect to
ZEUS-2D's older MOCCT algorithm is maximized in this test problem due to the
artificial suppression of compressive hydrodynamic waves and longitudinal MHD
waves; the true resolution requirements of HSMOCCT as implemented in ZEUS-MP
will depend in part upon the relative importance of various wave modes to the
calculation and will in general be problem dependent.

      \subsubsection{MHD Riemann Problem}

\begin{figure}
\leavevmode
\includegraphics[width=\columnwidth]{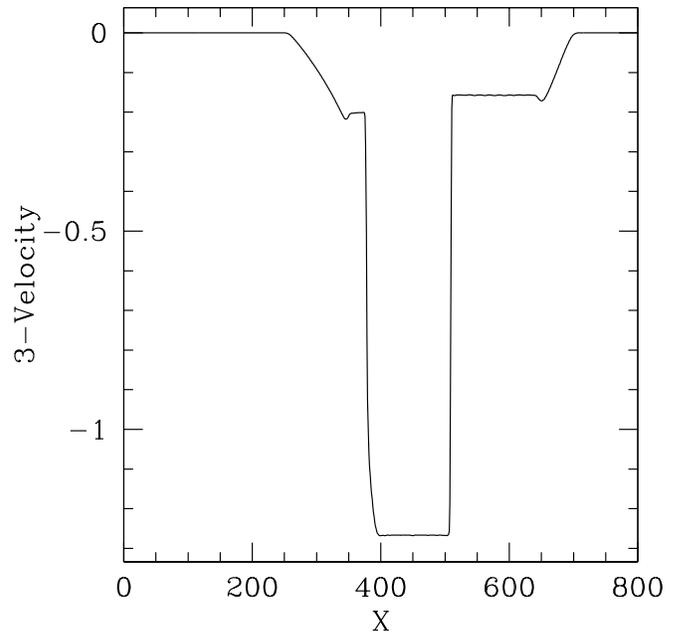}
\caption{MHD Riemann problem: 3-velocity vs. X at t = 80 sec.}
\label{bwuv3}
\end{figure}

\begin{figure}
\leavevmode
\includegraphics[width=\columnwidth]{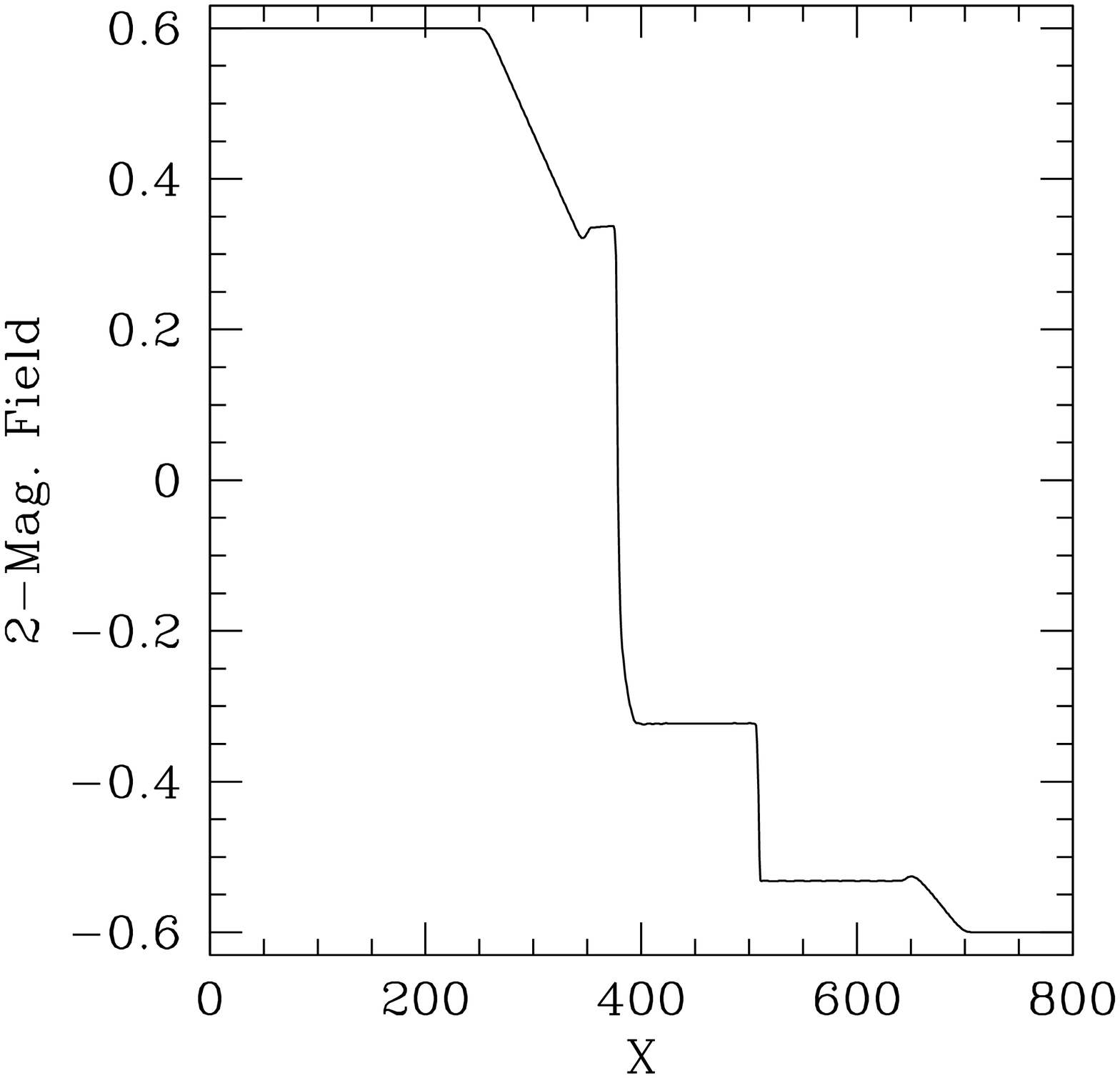}
\caption{MHD Riemann problem: 2-component of magnetic field vs. X at t = 80 sec.}
\label{bwub2}
\end{figure}

\begin{figure}
\leavevmode
\includegraphics[width=\columnwidth]{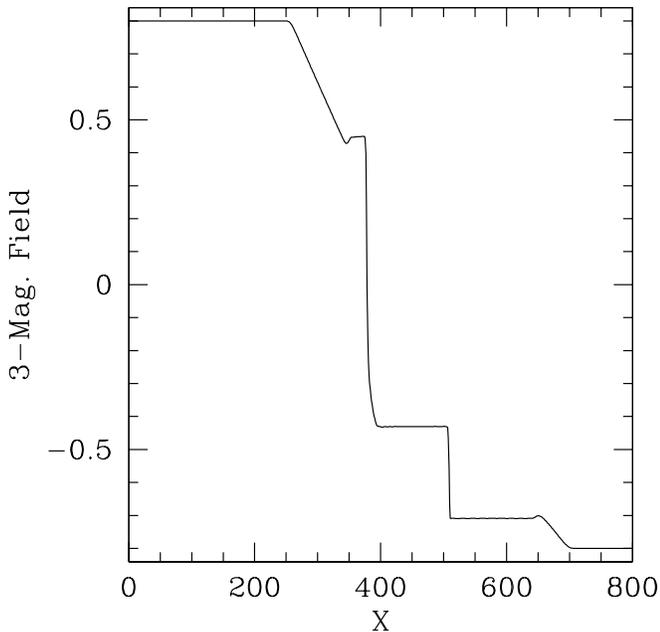}
\caption{MHD Riemann problem: 3-component of magnetic field vs. X at t = 80 sec.}
\label{bwub3}
\end{figure}

Our second MHD problem is a magnetic shock tube problem due to~\citet{brio88}.
Test results with ZEUS-2D using the van Leer advection algorithm were also
published in~\citet{stone92b}.  This test problem is described by ``left''
and ``right'' states in which the discontinuous medium is threaded by a magnetic
field which is uniform on both sides but exhibits a kink at the material interface.
Our formulation of the problem differs from that of~\citet{brio88} and~\citet{stone92b}
only in that we have oriented the transverse component of the magnetic field
to have non-zero components along both the Y and Z axes in Cartesian geometry.
At t = 0, our left state is given by $\left(\rho_l,~p_l,~B_l^x,~B_l^y,~B_l^z\right)$ =
(1.0,~1.0,~0.75,~0.6,~0.8), and the right state is given by
$\left(\rho_r,~p_r,~B_r^x,~B_r^y,~B_r^z\right)$ = (0.125,~0.1,~0.75,~-0.6,~-0.8).  All
velocities are initially zero.  The ratio of specific heats for this problem is 2.0.
As with the calculation in~\citet{stone92b}, the
problem is computed in 1-D on an 800-zone grid and run to a time of t = 80.  The
spatial domain length is 800.

Figures~\ref{bwurho} through~\ref{bwub3} show the results obtained
with~\zmp.  The 1-component of $B$ (not included in the figures) remained flat
over the domain at its initial value, as expected.
The grid resolution is identical to that used in the ZEUS-2D calculation,
and the results are evidently consistent (see Fig. 6 of~\citet{stone92b}).  While
this problem is not truly multidimensional, it does exhibit both transverse and
compressional wave modes, in contrast with the previous test problem.  In this
case, we may qualitatively match the results from ZEUS-2D without an increase
in grid resolution.

      \subsubsection{Orszag-Tang Vortex}

Our final MHD test problem is a multidimensional problem due to~\citet{orszag79}
which has been featured in a number of recent MHD method papers, such as
~\citet{dai98,ryu98} and~\citet{londrillo00,londrillo04}.  This problem follows
the evolution of a 2-D periodic box of gas with $\gamma$ = 5/3, in which fluid
velocities and magnetic field components are initialized according to
${\bf v}~=~ v_0\left[-\sin\left(2\pi y\right)\hat{\bf x} +
                      \sin\left(2\pi x\right)\hat{\bf y}\right]$
and
${\bf B}~=~ B_0\left[-\sin\left(2\pi y\right)\hat{\bf x} +
                      \sin\left(4\pi x\right)\hat{\bf y}\right]$, with $v_0$ = 1
and $B_0$ = 1/$(4\pi)^{1/2}$.  The box has length 1.0 along each side.  The density
and pressure are initially uniform with values of 25/(36$\pi$) and 5/(12$\pi$),
respectively  (these choices lead to an initial adiabatic sound speed of 1.0).
Subsequent evolution leads to a complex network
of waves, shocks, rarefactions, and stagnant flows.  \citet{ryu98} provide
greyscale snapshots of the flow field at t = 0.48; in addition, they provide
1-D cuts through the data along the line given by $y$ = 0.4277, over which the
gas and magnetic pressures are plotted as functions of $x$.  The~\citet{ryu98}
results were computed on a 256$^2$-zone Cartesian mesh.  For consistency, we also
computed the problem on a 256$^2$-zone mesh, from which comparison values of
pressure at the identical cut in $y$ may be extracted.  To explore the effect of
resolution, we also provide 2-D greyscale images from a 512$^2$-zone calculation.

\begin{figure}
\leavevmode
\includegraphics[width=\columnwidth]{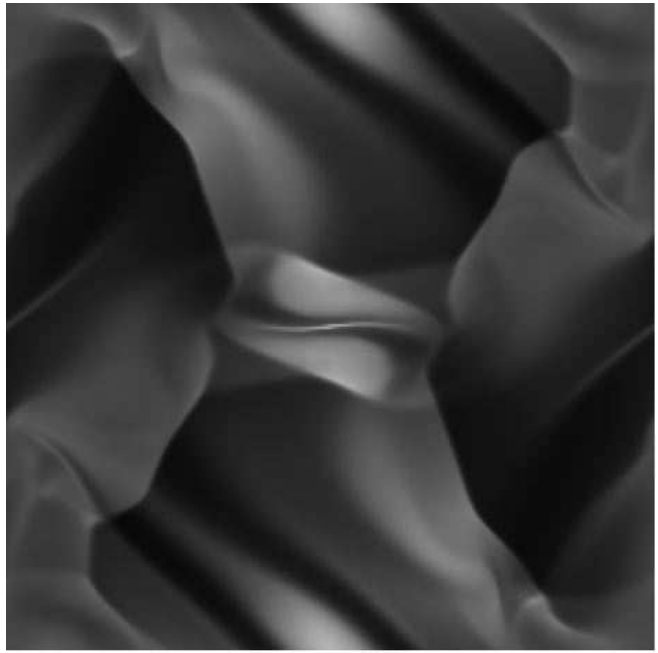}
\caption{Orszag-Tang vortex: pressure at t = 0.48 seconds. 256$^2$ zones were used.}
\label{vor_p256}
\end{figure}

\begin{figure}
\leavevmode
\includegraphics[width=\columnwidth]{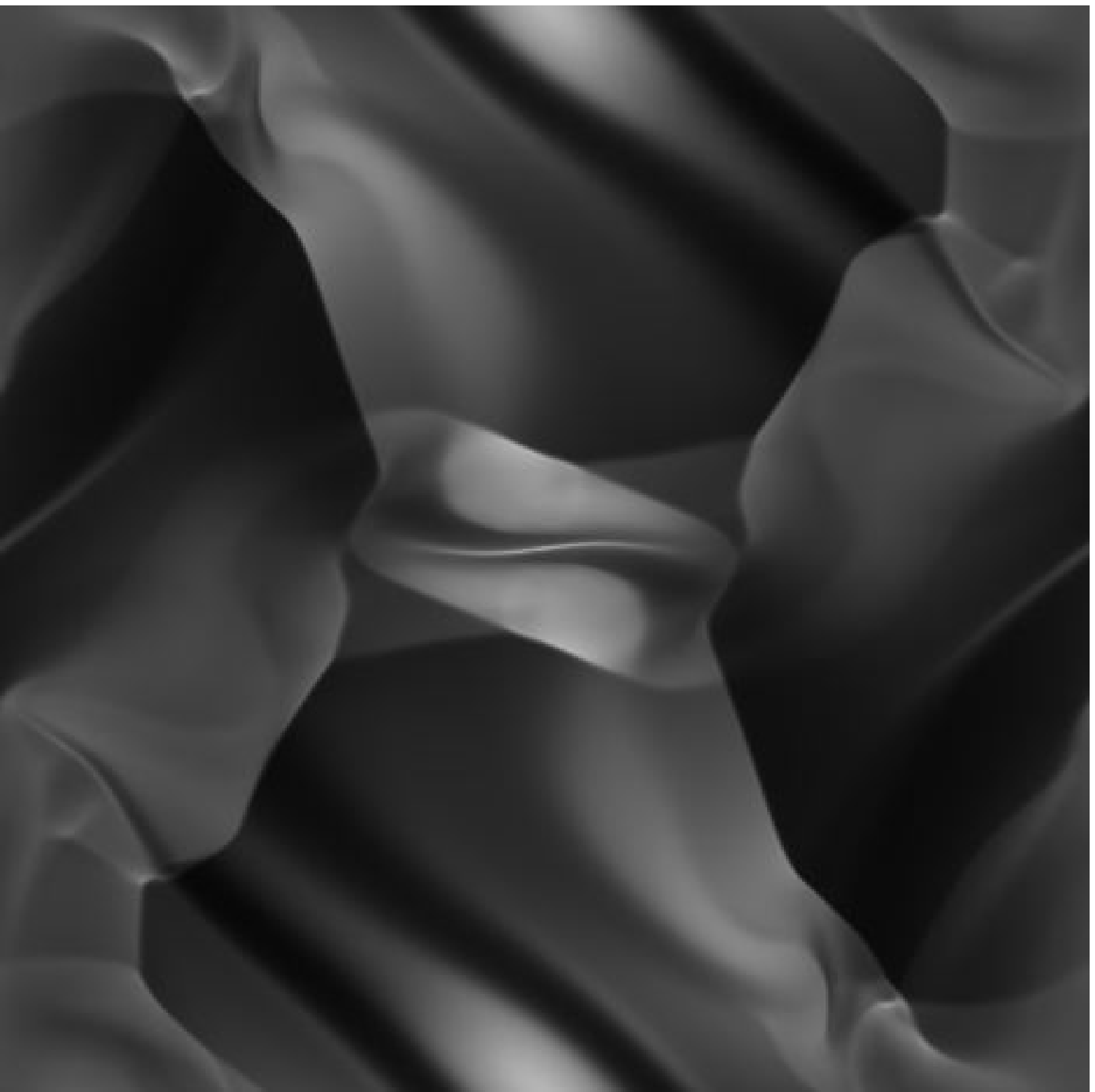}
\caption{Orszag-Tang vortex: pressure at t = 0.48 seconds. 512$^2$ zones were used.}
\label{vor_p512}
\end{figure}

\begin{figure}
\leavevmode
\includegraphics[width=\columnwidth]{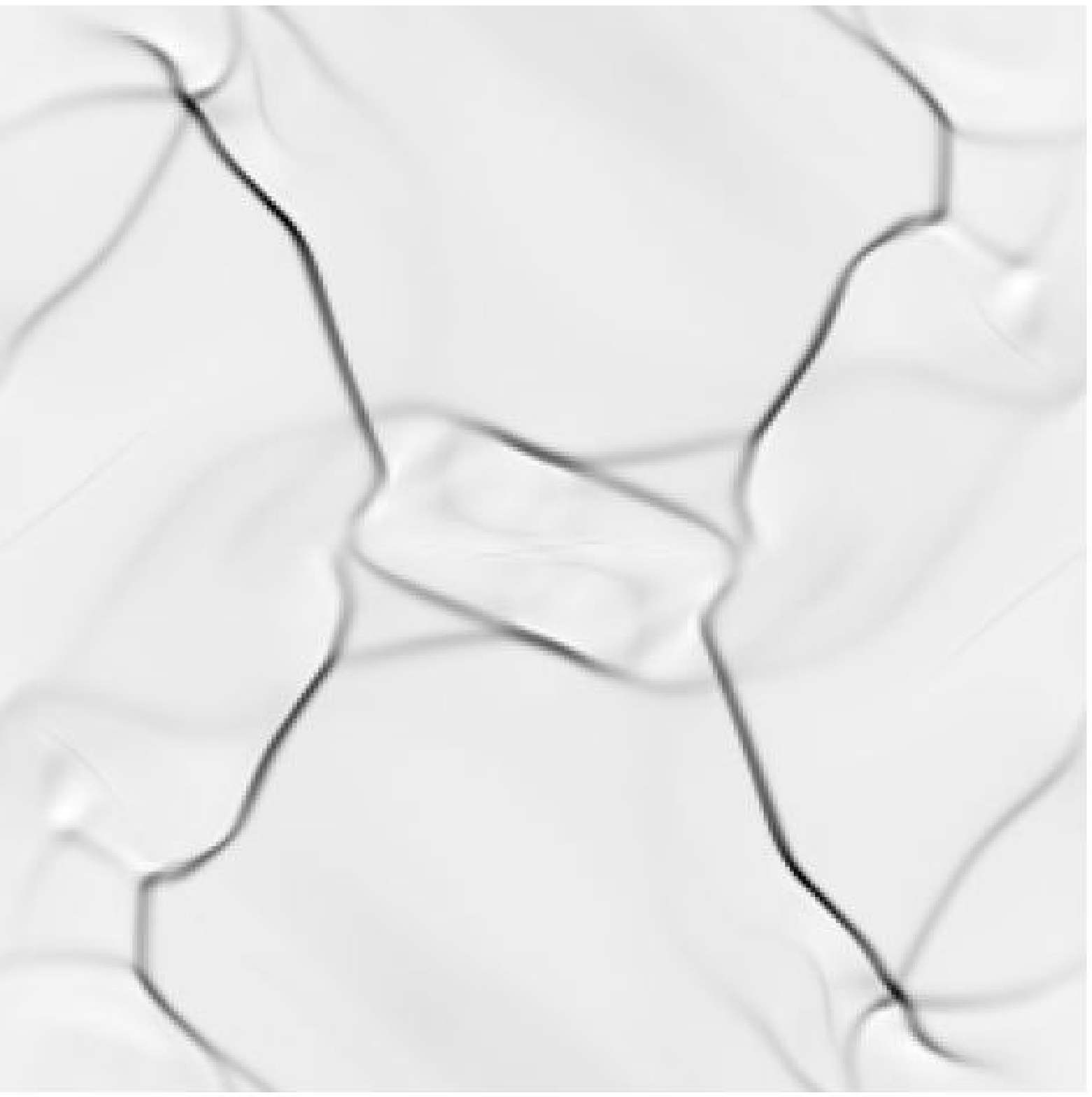}
\caption{Orszag-Tang vortex: $\nabla\cdot{\bf v}$ at t = 0.48 seconds for the
256$^2$ zone calculation.}
\label{vor_divv256}
\end{figure}

\begin{figure}
\includegraphics[width=\columnwidth]{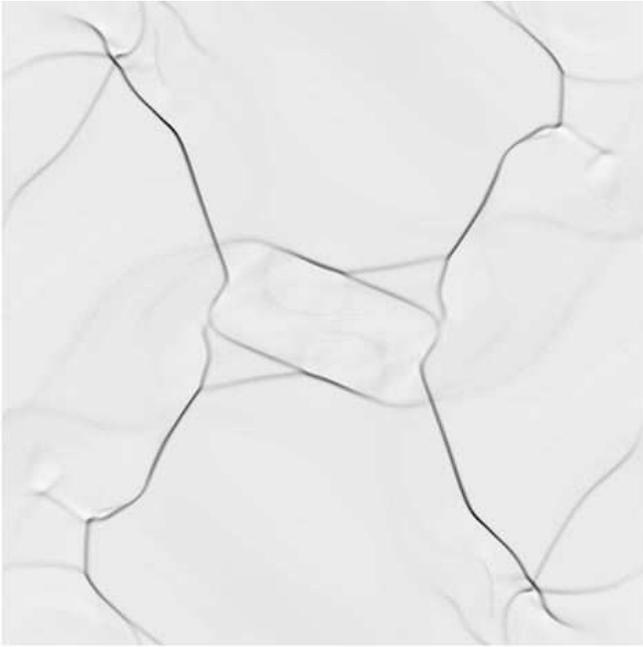}
\caption{Orszag-Tang vortex: $\nabla\cdot{\bf v}$ at t = 0.48 seconds for the
512$^2$ zone calculation.}
\label{vor_divv512}
\end{figure}

\begin{figure}
\leavevmode
\includegraphics[width=\columnwidth]{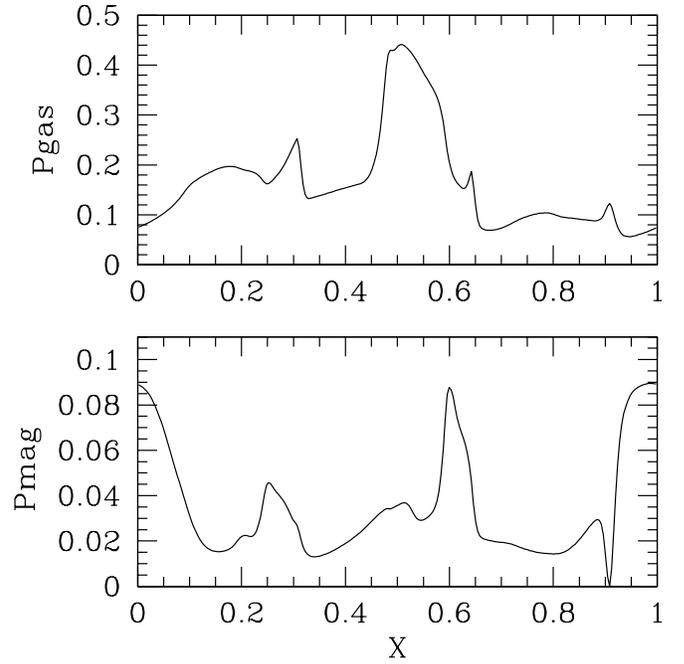}
\caption{Gas and magnetic pressures vs. $x$ along $y$ = 0.4277 at time t = 0.48 sec.}
\label{vor_pro256}
\end{figure}

Our multidimensional flow structures at t = 0.48 are given in Figures~\ref{vor_p256}
through~\ref{vor_divv512}, which are to be compared to the grey-scale panels on the
left-hand side of Figure 3 in~\citet{ryu98}.  Figure~\ref{vor_pro256} presents
line plots of gas and magnetic pressure along a line of $x$ located at $y$ = 0.4277.
Save for a very small notch in the gas pressure near x = 0.5, our pressure profiles
from the 256$^2$ calculation appear to be virtually identical to those from
~\citet{ryu98} at identical resolution.  With respect to the 2-D images, the effect
of resolution is most apparent in maps of the velocity divergence (Figures
~\ref{vor_divv256} and~\ref{vor_divv512}).
Again, the ZEUS-MP results at a grid resolution of 256$^2$ compare quite favorably
to those from~\citet{ryu98}, with subtle flow features marginally less well resolved.
Our results are likewise consistent with those computed at similar resolution by
\citet{dai98} and~\citet{londrillo00} (\citet{londrillo04} also computed the problem
but did not include figures matching the other cited works).
The 256$^2$ and 512$^2$ results clearly bracket those of~\citet{ryu98}; thus we
see that in this problem axial resolution requirements of the two codes differ by
{\it at most} a factor of 2, which we consider an agreeable result for a finite-difference
staggered-mesh code.

   \subsection{Radiation}\label{verifyradiation}

      \subsubsection{Marshak Waves}

We begin our examination of radiation physics with a test problem emphasizing
the coupling between matter and radiation.  The Marshak wave problem we compute
was formulated by~\citet{su96} after a description in~\citet{pomraning79}.  The
problem considers the heating of a uniform, semi-infinite slab initially at
$T = 0$ everywhere.  The material is characterized by a $T$-independent (and
therefore constant) opacity ($\kappa$) and a specific heat ($\alpha$) proportional
to $T^3$, in which case the gas and radiation energy equations become linear in
the quantities $\erad$ and $T^4$. Pomraning defined
dimensionless space and time coordinates as
\begin{equation}\label{xvar}
x \ \equiv \ \sqrt{3}\kappa z,
\end{equation}
and
\begin{equation}\label{tvar}
\tau \ \equiv \ \left({4ac\kappa \over \alpha}\right) t,
\end{equation}
and introduced dimensionless dependent variables, defined as
\begin{equation}\label{uvar}
u(x,\tau) \ \equiv \ \left(c \over 4\right) \left[E(z,t) \over
F_{\rm inc}\right],
\end{equation}
and
\begin{equation}\label{vvar}
v(x,\tau) \ \equiv \ \left(c \over 4\right) \left[aT^{4}(z,t) \over
F_{\rm inc}\right].
\end{equation}

In (\ref{uvar}) and (\ref{vvar}), $F_{\rm inc}$ is the incident
boundary flux.  With the definitions given by (\ref{xvar})
through (\ref{vvar}), Pomraning showed that the radiation and gas
energy equations could be rewritten, respectively, as
\begin{equation}\label{ueq}
\epsilon{\partial u(x,\tau) \over \partial \tau} -
{\partial^{2} u(x,\tau) \over \partial x^{2}} \ = \
v(x,\tau) - u(x,\tau),
\end{equation}
and
\begin{equation}\label{veq}
{\partial v(x,\tau) \over \partial \tau} \ = \
u(x,\tau) - v(x,\tau),
\end{equation}
subject to the following boundary conditions:
\begin{equation}
u(0,\tau) - {2 \over \sqrt{3}}{\partial u(0,\tau)\over \partial x}
\ = \ 1,
\end{equation}
and
\begin{equation}
u(\infty,\tau) \ = \ u(x,0) \ = \ v(x,0) \ = \ 0.
\end{equation}

The user-specified parameter $\epsilon$ is related to the radiation constant
and specific heat through
\begin{equation}
\epsilon \ = \ {4a \over \alpha}.
\end{equation}

\begin{figure}
\leavevmode
\includegraphics[width=\columnwidth]{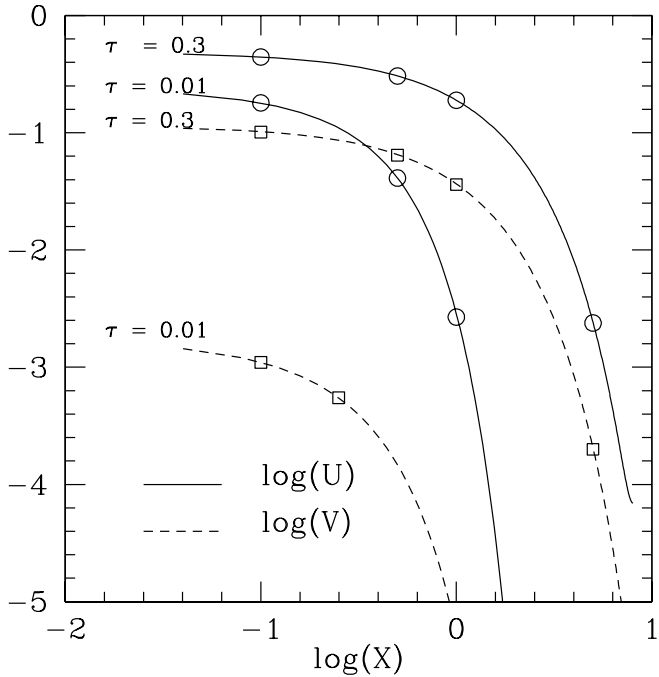}
\caption{Curves of log(U) and log(V) vs. log(X) for 2 values of $\tau$.
Curves show analytic solutions; circles indicate numerical data.}
\label{marshak}
\end{figure}

With a choice of $\epsilon$, the problem is completely specified and may be
solved both numerically and analytically.  For the~\zmp~test, we chose
a 1-D Cartesian grid with 200 zones and a uniform density of 1 g cm$^{-3}$.
The domain length is set to 8 cm, and the photon mean-free path ($\kappa^{-1}$)
is chosen to be 1.73025 cm.  Because this problem was designed for a pure diffusion
equation, no flux limiters were used in the FLD module.  $\epsilon$ was chosen
to be 0.1, allowing direct comparison between our results and those given by
\citet{su96}.  Our results are shown in Figure~\ref{marshak}, in which the
dimensionless energy variables $u$ and $v$ are plotted against the dimensionless
space coordinate $x$ at two different values of the dimensionless time, $\tau$.
The open circles indicate benchmark data taken from the tabulated solutions
of~\citet{su96}; solid curves indicate ZEUS-MP results.  The agreement is 
excellent.

      \subsubsection{Radiating Shock Waves}

\begin{figure}
\leavevmode
\includegraphics[width=\columnwidth]{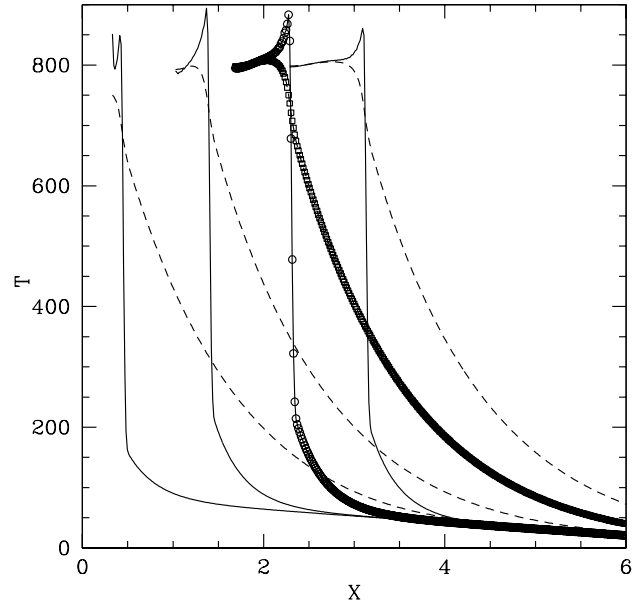}
\caption{Subcritical radiating shock; matter and
radiation temperatures vs. comoving X coordinate.
Plot times
are 5400, 1.7$\times 10^4$, 2.8$\times 10^4$, and 3.8$\times 10^4$ seconds.}
\label{subrshk}
\end{figure}

The classic text on the theory of shock waves and associated radiative phenomena
is due to~\citet{zeldovich67} (see also~\citet{zeldovich69} for a short review
article on shock waves and radiation).  A more recent summary of basic concepts
is available in~\citet{mihalas84}.  Radiating shock waves differ qualitatively
from their purely hydrodynamic counterparts due the presence of a radiative
precursor created by radiative preheating of material upstream from the shock
front.  The existence of this precursor gives rise to the identification of
so-called {\it subcritical} and {\it supercritical} radiating shocks,
which are distinguished by a comparison of the gas temperature behind the shock
front to that in the material immediately upstream from the shock.  In the
case of subcritical shocks, the post-shock gas temperature exceeds the upstream
value, and the radiative precursor is relatively weak.  As the shock velocity
is increased beyond a critical value, however, the upstream gas temperature
becomes equal to (but never exceeds) the post-shock temperature; such shocks
show very strong radiative preheating of the unshocked gas and are identified
as supercritical shocks.

\begin{figure}
\leavevmode
\includegraphics[width=\columnwidth]{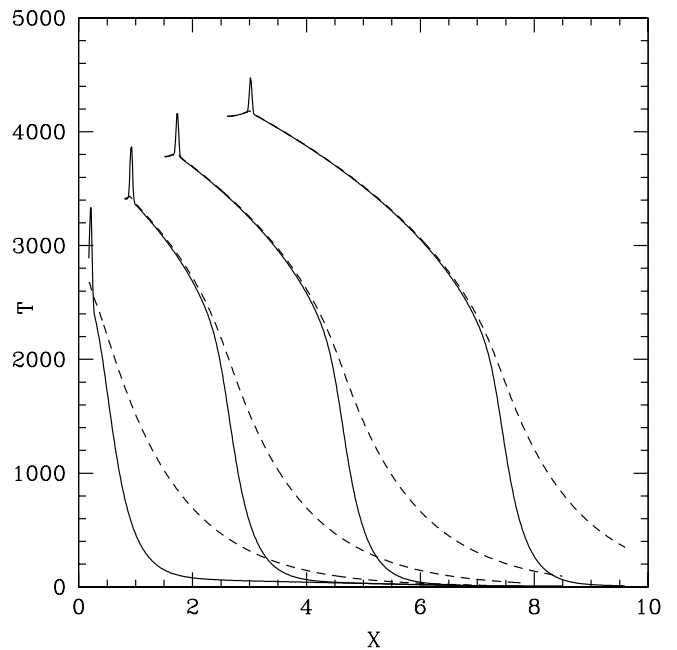}
\caption{Supercritical radiating shock; matter and
radiation temperatures vs. comoving X coordinate.
Plot times are 860, 4000, 7500, and 1.3$\times 10^4$
seconds.}
\label{suprshk}
\end{figure}

A numerical prescription for radiating shock test problems appropriate for
astrophysical simulation codes was published by~\citet{ensman94}; this
configuration was revisited by~\citet{gehmeyr94} and again by~\citet{sincell99a,
sincell99b} and~\citet{hayes03}.  In this model, a domain of length or radius
$7 \times 10^{10}$ cm and an initially uniform density of $7.78 \times 10^{-8}$
g cm$^{-3}$ is given an initial temperature profile such that $T$ falls
smoothly from a value of 85 K at the inner boundary to 10 K at the outer boundary.
The non-zero gradient was necessary to avoid numerical difficulties in Ensman's
VISPHOT code.  A constant opacity of $3.1 \times 10^{-10}$ is chosen, which
yields a photon mean-free path roughly 5\% of the domain length.  Because the VISPHOT
code uses a Lagrangean mesh, the shock is created by a ``piston'' affected by choosing
an inner boundary condition on the fluid velocity.  \zmp~recreates this condition
on an Eulerian grid by initializing the fluid velocity throughout the domain and
outer boundary to the (negative of) the required piston velocity.  The subcritical
shock and supercritical shock tests share all problem parameters save for the piston
velocity, chosen to be 6 km/s in the former case and 20 km/s for the latter.
512 zones were used to execute the problem on a 1-D mesh.

Figures~\ref{subrshk} and~\ref{suprshk} present temperature profiles for the subcritical
and supercritical cases, respectively.  To aid comparison of our Eulerian results
to the Lagrangean results of~\citet{ensman94}, we transform the coordinate axis
into the rest frame of the unshocked matter.  Solid lines indicate gas temperature;
dashed lines indicate a radiation ``temperature'' defined by $T_r \equiv 
\left(\erad/a_r\right)^{1/4}$, where $a_r$ is the radiation constant.  Note the
strongly preheated material ahead of the shock front in the supercritical case.
Our results were computed in Cartesian geometry; those from~\citet{ensman94} were
computed in a thin spherical shell with a large radius of curvature.  This problem
was also treated by~\citet{hayes03} using ZEUS-MP coupled to a parallel VTEF algorithm.
Because that code was designed specifically for 2-D cylindrically-symmetric problems, 
the problem geometry was different in that the radiating surface was planar yet of finite 
transverse extent, whereas these results consider a formally infinite plane.  This
difference results in somewhat different peak values for temperature, but otherwise
the results are qualitatively consistent.

\section{Performance Tests}\label{perf}

Section~\ref{verify} considered test problems which gauge the accuracy of the
code.  This section considers issues of numerical performance, with a particular
emphasis on problems distributed among large numbers of parallel processors.

   \subsection{Aspects of Scalability}

The topic of parallel performance is most often encapsulated in the notion of
{\it scalability}, which in this context is typically assessed by measuring the
reduction in CPU time for a given quantity of numerical work as this work is
distributed among an increasing number of processors.  Relative to the cost
on one CPU, perfect scalability would be represented by a cost reduction factor
of $1/N$ when the same job is distributed across N processors.  For tasks in which
each processor can operate upon its portion of data independently of all other
processors (a so-called {\it embarrassingly parallel} operation), perfect scalability
is trivially achieved.  Algorithms which compute solutions to spatially-discretized
PDE's are by construction not embarrassingly parallel because the discrete spatial
derivative operators employ stencils that overlap processor tile boundaries
along the tile edges.  On distributed-memory computers, data communication will 
therefore be required.  Efficient management of this communication is thus a key
ingredient to an efficient parallelization strategy.

More generally, scalability
describes the sensitivity of an algorithm's CPU cost to a number of factors,
of which parallelism is a leading but by no means unique member.  Section~\ref{mgmeth}
compared the cost of MG linear solvers to traditional stationary methods
for a given problem size; the costs of the two methods exhibit very different
dependencies upon the number of unknowns in the linear system.  For iterative
methods such as CG and MG, the required number of iterations for a converged
solution is the primary factor in algorithm cost.  Solvers whose iteration
counts vary more weakly with problem size are to be favored for very large problems.
In ideal cases, the required number of iterations for convergence of an MG
solver can be virtually independent of the problem size, thus MG is often
said to scale well to large problems.  Because this behavior is orthogonal to the
issue of parallel decomposition, we identify this as an independent definition of
scalability.  

\begin{deluxetable}{lccc}
\tablecaption{Radiation Diffusion Test Parameters.\label{icfpar}}
\tablehead{
\colhead{Medium} & \colhead{Outer Radius (cm)} & \colhead{$\rho$ (g/cm$^3$)} &
\colhead{Temperature (eV)}  }
\startdata
D-T gas  & 0.087 & 0.025     & \twospc 0.025 \\
D-T ice  & 0.095 & 0.25\phn  & \twospc 0.025 \\
C-H foam & 0.111 & 1.20\phn  & \twospc 0.025 \\
He gas   & 0.490 & 0.01\phn  &       300\phd\thrspc   \\
\enddata
\end{deluxetable}

An additional factor bearing on the cost of an iterative linear system solution
is diagonal dominance, a condition in which matrix elements along the main diagonal
are much larger in magnitude than off-diagonal elements along a given row.  Matrices
resulting from the discretization of the time-dependent diffusion equation
exhibit diagonal dominance that varies directly with the size of the time step.
To see this, consider the static diffusion equation discretized on a 1-D mesh
with uniform spacing, $\Delta x$.  For a static medium, the radiation
energy equation becomes
\begin{equation}
{\partial\erad \over \partial t}  - \nabla\cdot D\nabla\erad\ = \ S,
\end{equation}
where $S$ contains local source terms.  We assume a spatially uniform medium
(yielding a spatially constant D) and write this equation in discrete form as
\begin{equation}
{{\erad\supnp\subi - \erad\supn\subi}\over \dt} - \left[D\over(\Delta x)^2\right]
\left(\erad\supnp\subip - 2\erad\supnp\subi + \erad\supnp\subim\right) \ = \ S\subi,
\end{equation}
which may be rearranged to make the linear system structure obvious:
\begin{eqnarray}\label{toyprob}
& - & \left[D\dt\over(\Delta x)^2\right]\erad\supnp\subim  +
\left[1 + {2D\dt\over(\Delta x)^2}\right]\erad\supnp\subi \nonumber \\ & - &
\left[D\dt\over(\Delta x)^2\right]\erad\supnp\subip \ = \
\erad\supn\subi + S\subi\dt.
\end{eqnarray}
The relationship between time step and diagonal dominance is manifest:
in the limit that $\dt \rightarrow 0$, the linear system represented by (\ref{toyprob})
reduces to the identity matrix!  In the opposite limit, the off-diagonal elements
are comparable in magnitude to the main diagonal, a situation which results in
greatly increased numbers of iterations required for convergence in a CG linear
solver such as that implemented in our FLD module.

\begin{figure}
\leavevmode
\includegraphics[width=\columnwidth]{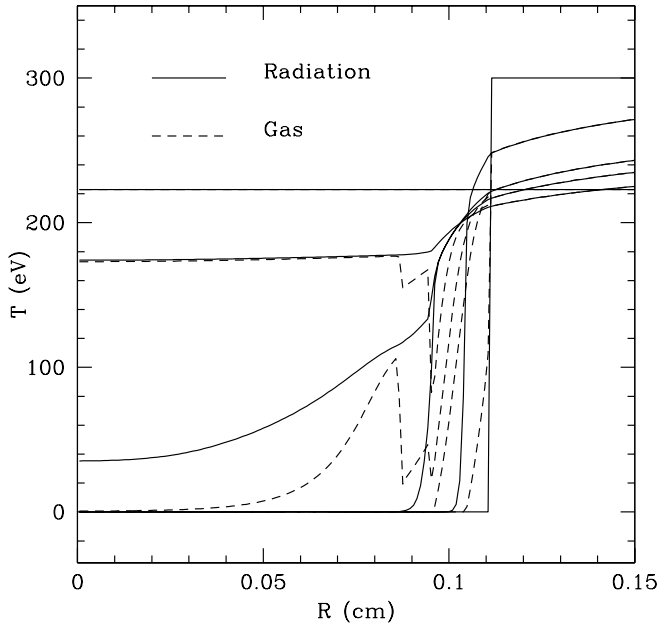}
\caption{Radiation diffusion (1D): profiles of radiation temperature (solid lines)
and gas temperature (dashed lines) at times of 0.0, 0.1, 0.5, 0.7, 1.0, and 10 ns.
The initial profiles are step functions with the discontinuities located at R = 0.11 cm.
The radiation wave propagates leftward, with the gas temperature lagging the radiation
temperature, particularly in the optically thick regions.}
\label{icftemp}
\end{figure}

\begin{figure}
\leavevmode
\includegraphics[width=\columnwidth]{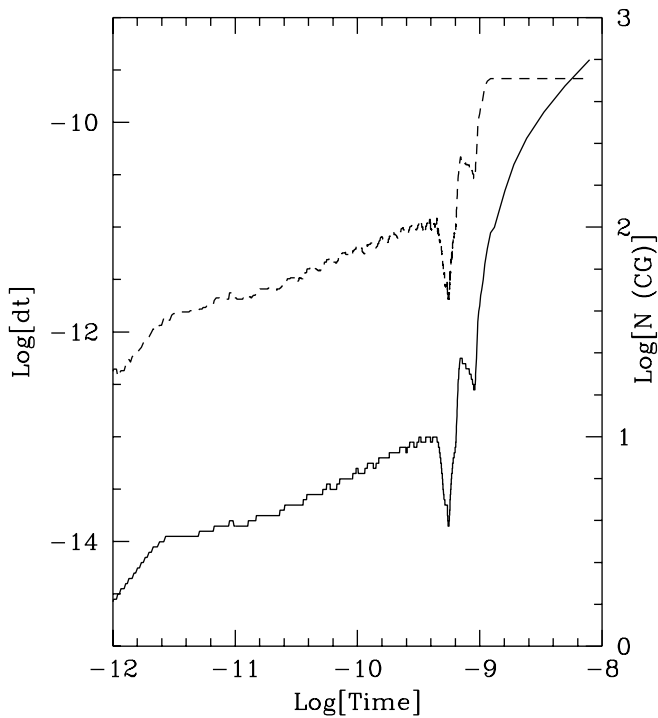}
\caption{Radiation diffusion (1D): logarithms of the time step (solid line)
and mean number of CG iterations per N-R iteration (dashed line) as functions
of the evolution time.  Times are measured in seconds.}
\label{icfhist}
\end{figure}

We demonstrate this behavior with a radiation diffusion problem in which a cold
sphere is immersed in a high-temperature radiation field.  Problem parameters
are given in Table~\ref{icfpar}.  The initial values of density and temperature
are taken from a test problem given in~\citet{hayes99} designed to
qualitatively mimic features of an Inertial Confinement Fusion (ICF) simulation
such as that used by~\citet{baldwin99} to compare numerical performance of CG-based
and MG-based linear system solvers. 
The irradiated sphere is constructed of layers
with strongly disparate densities and photon mean-free paths.  The actual physical
system this problem imitates (albeit crudely) is a sphere of D-T gas surrounded
by a solid D-T ``ice'' of higher density, itself surrounded by a carbon-hydrogen
foam of yet higher density.  This assembly is immersed in a low density He gas
subjected at t = 0 to an intense radiation field with a characteristic temperature
of 300 eV. Opacities for real ICF materials have complex dependencies
on energy and composition; our toy problem captures the gross features of the
mean-free path variation via the following expression:
\begin{equation}
\Lambda(\rho,T) \ = \ \Lambda_0 \left(\rho\over\rho_0\right)^\nu 
                                \left(T\over T_0\right)^\mu,
\end{equation}
with $\Lambda_0$, $\rho_0$, $T_0$, $\nu$, and $\mu$ given by 10$^{-6}$, 1.2 g/cm$^{3}$,
0.025 eV, 2.0, and 1.2, respectively.  A further restriction is placed on the resulting
opacities such that the minimum and maximum allowed values of $\chi$ are 10 and
10$^6$ cm$^{-1}$.  This restriction filters out unphysically high and low values
of the absorption coefficients.  The important feature of this problem is that
low-density gas is surrounded by two solids of much higher (and differing)
densities.  This construction results in an inward-propagating radiation diffusion
wave with a highly variable rate of progress.
Snapshots of the gas and radiation temperatures at times
of 0, 0.1, 0.5, 0.7, 1.0, and 10 nanoseconds are given in Figure~\ref{icftemp}.
Histories of the time step and average number of CG iterations required
for NR iteration in the FLD module are plotted against evolution time in
Figure~\ref{icfhist}.  The choppy appearance of the plots is an artifact of sampling
(every tenth cycle was archived for plotting).
For static diffusion problems or RHD problems characterized by rapidly time-varying
radiation fields, evolution of the time step will be strongly constrained by the
maximum allowed fractional change in the radiation energy (= 0.01 in this problem).
The initial time step progression is upward as the exterior radiation field slowly
diffuses through the opaque foam layer.
By 0.5 ns, the radiation has diffused through the foam layer
and begun to penetrate the less opaque D-T ice layer, at which time the time step
drops sharply owing to the more rapid evolution of the radiation energy.  The time step
trends upward again until the radiation wave breaks through the D-T gas/ice
boundary.  The time step then drops again as the radiation streams into the central
region.  The final evolution of the time step is upward as the problem domain reaches
its final equilibrium.  The equilibrium temperature is lower than the initial exterior
value because a reflecting outer boundary was chosen rather than an imposed boundary flux.

The dashed line in Figure~\ref{icfhist} shows that the number of CG linear system
iterations required to solve the FLD matrix during an outer N-R iteration closely
parrots the time step behavior.  This is understood as a natural consequence of
time-dependent diagonal dominance of the matrix as illustrated by equation~\ref{toyprob}.
This exercise demonstrates the existence of a third dimension of scalability of
particular relevance to time-dependent simulations: the dependence of CPU cost upon
time step size.  When a linear system is very nearly diagonal, both CG and MG will
converge rapidly, but because a single iteration of full MG is more expensive than
a single CG iteration, one may expect situations in which CG presents a more economical
solution strategy for a given problem size.  That increasing time step size could
provide a ``cross over'' point with regard to optimal method is a logical consequence.
This issue was investigated extensively by~\citet{baldwin99} in
the context of 2-D RHD simulations in spherical geometry.  They considered RHD
problems in which the time step varied naturally by orders of magnitude during the
course of the simulation, and noted indeed that no one method provided the best economy
over the entire calculation.  While adaptive selection of linear solvers in a particular
physics module has not been implemented in~\zmp, we note that experimention along
such lines in the context of astrophysical problems is an enticing candidate for
future research.

   \subsection{Parallel Performance Results}

\begin{figure}
\leavevmode
\includegraphics[width=\columnwidth]{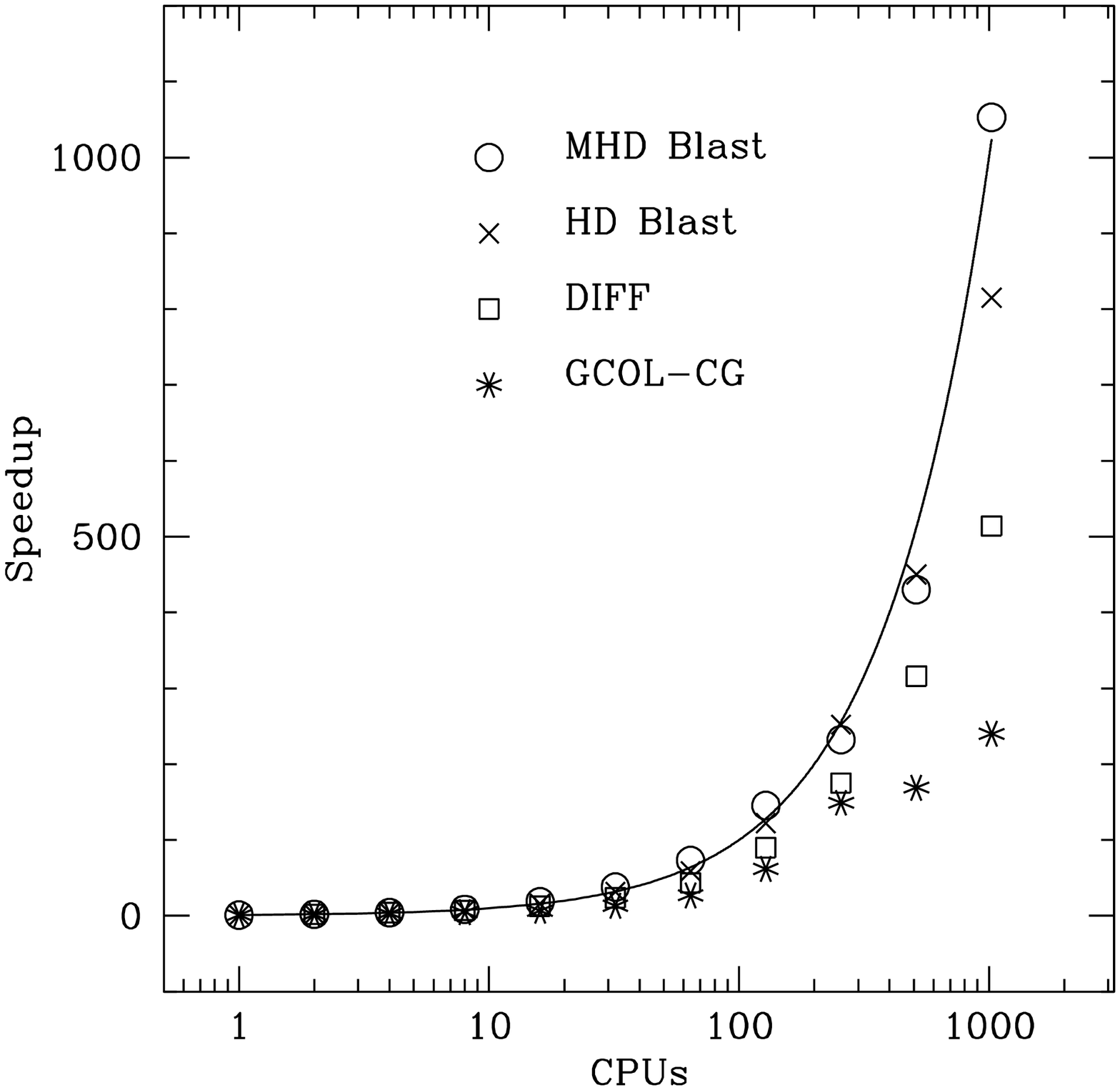}
\caption{Strong-scaling comparison: relative speedup vs. number of CPUs for problems
of fixed size.  The solid curves indicates perfect N-fold scaling with CPU count.
256$^3$ zones were used for each problem; the maximum CPU number was 1024 for each
case.}
\label{strscal}
\end{figure}

We explore additional aspects of algorithm performance with a quartet of test problems
computed on 3-D grids with 256$^3$ zones.  The first two problems used are
non-magnetic and magnetic variants of a simple blast wave test in which a sphere
with initial overdensity and overpressure ratios of 100 and 10$^6$ is defined with
respect to a uniform background medium.
The problem is defined on a Cartesian grid.  The magnetic version
augments the problem setup with a uniform magnetic field aligned with the Z axis.
The third problem is a 3-D calculation of radiation diffusion into an ICF capsule,
with problem parameters as given previously.
The fourth problem is the gravitational collapse of a pressureless cloud, using
problem parameters given in~\citet{stone92a}.

\begin{deluxetable}{cccc}
\tablecaption{Fixed-Work Scaling: 256$^3$ MHD (30 Time Steps)\label{mhd256}}
\tablehead{
\colhead{Processors} & \colhead{CPU Time (sec)} & \colhead{Speedup} &
\colhead{Parallel Efficiency (\%)} }
\startdata
\thrspc  1   &   9391\phd\phn\phn   &    ---                & --- \\
\thrspc  2   &   4624\phd\phn\phn   & \thrspc 2.031    & 102 \\
\thrspc  4   &   2236\phd\phn\phn   & \thrspc 4.200    & 105 \\
\thrspc  8   &   1097\phd\phn\phn   & \thrspc 8.561    & 107 \\
\twospc 16   &   \phn 504.9\phn     & \twospc 18.60\phn    & 116 \\
\twospc 32   &   \phn 248.1\phn     & \twospc 37.85\phn    & 118 \\
\twospc 64   &   \phn 128.4\phn     & \twospc 73.14\phn    & 114 \\
\phn   128   &   \twospc 64.60     & \phn 145.4\twospc    & 114 \\
\phn   256   &   \twospc 40.45     & \phn 232.2\twospc    & \phn 90 \\
\phn   512   &   \twospc 21.84     & \phn 430.0\twospc    & \phn 84 \\
      1024   &   \thrspc  8.91  & 1053\phd\thrspc          & 103 \\
\enddata
\end{deluxetable}

\begin{deluxetable}{cccc}
\tablecaption{Fixed-Work Scaling: 256$^3$ HD (50 Time Steps)\label{hd256}}
\tablehead{
\colhead{Processors} & \colhead{CPU Time (sec)} & \colhead{Speedup} &
\colhead{Parallel Efficiency (\%)} }
\startdata
\thrspc   1   &  5256\phd\thrspc   &     ---          & --- \\
\thrspc   2   &  2746\phd\thrspc   & \thrspc   1.91         & 96 \\
\thrspc   4   &  1323\phd\thrspc   & \thrspc   3.97         & 99 \\
\thrspc   8   &   669.8\twospc     & \thrspc   7.85         & 98 \\
\twospc  16   &   335.4\twospc     & \twospc  15.7\phn      & 98 \\
\twospc  32   &   165.4\twospc     & \twospc  31.8\phn      & 99 \\
\twospc  64   &    89.25\phn       & \twospc  58.9\phn      & 92 \\
\phn    128   &    43.14\phn       & \phn    122\phd\twospc & 95 \\
\phn    256   &    20.85\phn       & \phn    252\phd\twospc & 98 \\
\phn    512   &    11.69\phn       & \phn    450\phd\twospc & 88 \\
       1024   &     6.450          & \phn    815\phd\twospc & 80 \\
\enddata
\end{deluxetable}

For each problem, parallel performance is measured by a so-called {\it
strong-scaling} test in which the total number of zones (and therefore the total
amount of computational work) is held constant as the problem is repeated with
increasing numbers of CPU's.  Each problem is run for a small number of cycles
(typically 30 to 50) which is held fixed for each trial.  Figure~\ref{strscal}
and Tables~\ref{mhd256},~\ref{hd256},~\ref{fld256}, and~\ref{gcol256}
summarize the results for the MHD blast wave, HD blast wave, radiation diffusion,
and gravitational collapse tests.  The number of timesteps for which each test
was run is indicated in the title of each table, from which single time step
costs may be derived.  In this example, the gravitational collapse
problem solved Poisson's equation with the CG linear solver, which is also used
in the diffusion test.  It is important to note that in this type of scaling
study where the total problem size is held fixed, parallel scalability will
inevitably break down for a sufficiently large number of processors, due in large
part to surface-to-volume effects: when the local processor data block is too small,
the communication cost of shipping data along the block's surfaces will compete with
the computational cost of processing the full block volume.  The number processors
necessary to induce a turnover in a code's parallel scalability behavior will
depend strongly on the level of communication required by the algorithm, a point
we demonstrate in the experiments that follow.  A competing technique
for measuring scalability, known as a {\it weak-scaling} test, holds the processor
block size constant and thus scales the total problem size with the number of processors.
This alternative has some utility: if, for example, one determines that twice the
grid resolution is required to satisfy a given accuracy metric, one may investigate
if doubling the number of processors along the axis preserves the cost of computing
a time step without degrading parallel performance.  While this is a relevant
consideration, we eschew weak-scaling studies in this paper because (1) with a
sufficiently large block of data on each processor, even poor message-passing implementations 
of parallelism can perform reasonably well, and (2) the characteristics of the
problem under study change as the zone number increases.  For Courant-limited
calculations, doubling the resolution will double the number of time steps needed
to complete a calculation, which rather offsets the virtues of maintaining a
constant cost per time step.  For problems using implicit linear solvers, increasing
the total number of zones will, to a degree depending on the solver method, increase
the number of iterations required for convergence at each cycle. Strong-scaling
studies, while providing a harsher test of a parallel implementation, speak directly
to the question: how rapidly may a research problem of a given size be solved?

\begin{deluxetable}{cccc}
\tablecaption{Fixed-Work Scaling: 256$^3$ FLD (20 Time Steps) \label{fld256}}
\tablehead{
\colhead{Processors} & \colhead{CPU Time (sec)} & \colhead{Speedup} &
\colhead{Parallel Efficiency (\%)} }
\startdata
\thrspc    2   & 3184\phd\thrspc   & ---             & --- \\
\thrspc    4   & 1569\phd\thrspc   &  \twospc 2.03   & 102 \\
\thrspc    8   & \phn 887.1\twospc &  \twospc 3.59   & \phn 90 \\
\twospc   16   & \phn 509.8\twospc &  \twospc 6.25   & \phn 78 \\
\twospc   32   & \phn 269.9\twospc &  \phn 11.8\phn  & \phn 74 \\
\twospc   64   & \phn 146.9\twospc &  \phn 21.7\phn  & \phn 68 \\
\phn     128   & \twospc 71.12\phn &  \phn 44.8\phn  & \phn 70 \\
\phn     256   & \twospc 36.38\phn &  \phn 87.5\phn  & \phn 68 \\
\phn     512   & \twospc 20.15\phn &  158\phd\twospc & \phn 62 \\
        1024   & \twospc 12.38\phn &  257\phd\twospc & \phn 50 \\
\enddata
\end{deluxetable}

\begin{deluxetable}{cccc}
\tablecaption{Fixed-Work Scaling: 256$^3$ Grav-CG (5 Time Steps) \label{gcol256}}
\tablehead{
\colhead{Processors} & \colhead{CPU Time (sec)} & \colhead{Speedup} &
\colhead{Parallel Efficiency (\%)} }
\startdata
\thrspc   1   &       11430\phd\twospc  &   ---                 &  ---  \\
\thrspc   2   & \phn   7013\phd\twospc  & \twospc 1.63          &  82   \\
\thrspc   4   & \phn   3696\phd\twospc  & \twospc 3.09          &  77   \\
\thrspc   8   & \phn   2300\phd\twospc  & \twospc 4.97          &  62   \\
\twospc  16   & \phn   1666\phd\twospc  & \twospc 6.86          &  83   \\
\twospc  32   & \twospc 866.5\phn       & \phn   13.2\phn       &  41   \\
\twospc  64   & \twospc 422.8\phn       & \phn   27.0\phn       &  42   \\
\phn    128   & \twospc 184.8\phn       & \phn   61.9\phn       &  48   \\
\phn    256   & \thrspc  76.97          &       149\phd\twospc  &  58   \\
\phn    512   & \thrspc  67.47          &       169\phd\twospc  &  33   \\
       1024   & \thrspc  47.64          &       240\phd\twospc  &  23   \\
\enddata
\end{deluxetable}

The behaviors in Figure~\ref{strscal} reflect the relative impact of MPI communication
operations on each module.  The superlative scaling of the MHD tests derives from 
the highly computation-intensive nature of the algorithm.  The HD test is actually
a subset of the MHD problem, as both the MHD-specific routines and the HD advection
algorithms must be used in any MHD problem.  The radiation and gravity problems
are both dominated by the cost of CG linear solver.  The diffusion problem was run
for a sufficiently limited number of time steps such that an average of eight CG
iterations were required at each time step.  In contrast, when used for the Poisson
equation, of order 10$^2$ iterations are required for a mesh size of 256$^3$.
Because each CG iteration requires both MPI data exchanges at tile boundaries and
global searches for error minima, high iteration counts result in very 
communication-intensive operations.  Parallel efficiency, which is computed by dividing
the speedup relative to 1 processor by the processor number, is displayed in the
fourth column of Tables~\ref{mhd256}-\ref{gcol256}.  Superlinear speedup is observed
most dramatically for the MHD test; this behavior is a by-product of strong-scaling
studies and arises because single-CPU performance is degraded when the local data
chunk is too large to fit in a processor's cache memory.  This effect decreases
as the per-CPU data size shrinks; the deleterious effects of communication then
begin to appear as the processor counts run into the hundreds.  Some of the peculiar
variations in parallel efficiency in the MHD example are likely consequences of
system and network effects associated with the particular machine used. Memory,
bandwidth, and latency characteristic vary tremendously among different architectures;
but the major trends shown in Figure~\ref{strscal} and the associated tables are
representative and instructive, and are internally consistent with the
relative reliance of each module upon data exchange among processors.

\begin{figure}
\leavevmode
\includegraphics[width=\columnwidth]{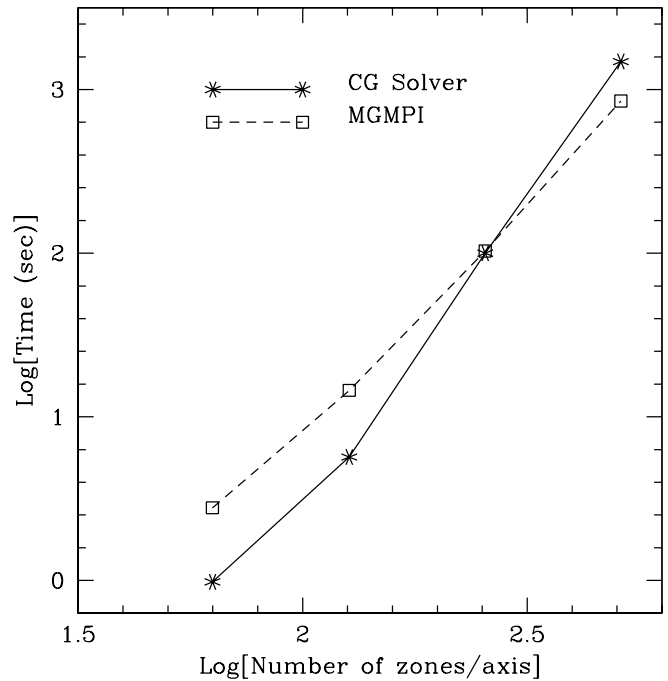}
\caption{A comparison of timings for the gravitational collapse problem in 3-D
using the general CG solver and MGMPI's multigrid solver for the Poisson equation.}
\label{sizescal}
\end{figure}

The fact that the CG solver requires -- irrespective of parallelism -- more
iterations for larger problems brings with it two liabilities: increased iteration
counts boost both the total operation cost {\it and} the number of MPI messages sent
and received. The fact that multi-grid methods exhibit convergence behavior with little
sensitivity to problem size motivated us to implement the independently-developed MGMPI
package for use as an alternative
Poisson solver in ZEUS-MP.  In its current form, MGMPI is restricted to 3-D Cartesian
grids (which are nevertheless a common choice in large astrophysical calculations)
with Dirichlet or Neumann boundary conditions (recall that the FFTw solver is
offered for triply-periodic Cartesian meshes).  Figure~\ref{sizescal} shows the
behavior of solution time against problem size for the gravitational
collapse problem computed on a 3-D Cartesian mesh.  Grid sizes of 63$^3$, 127$^3$,
255$^3$, and 511$^3$ zones were run.  (Odd numbers of zones are required by the
multigrid V-cycle in MGMPI.)  Each trial was distributed across 64 processors to
ensure that the larger problems would not exceed single-CPU memory limits.
At small grid sizes, the CG solver is less expensive
than the MGMPI solution, but at a mesh size of 511$^3$ the CG solution has clearly
diverged with respect to the MGMPI solver. The fundamental difference lies in the
average number of solver iterations required per time step.  For the CG solver, this
number was 32, 56, 99, and 190, respectively, for the four problem sizes tested.
For MGMPI, this number is 2.2 for the smallest problem and grows only to 3 for
the 511$^3$ run.  Despite MGMPI's fairly high operation cost per iteration, the
insensitivity of its convergence behavior to problem size guarantees, for a given parallel
distribution, a performance advantage over the CG solver for a sufficiently large
problem.  In its current form, MGMPI does not employ asynchronous MPI calls for
its message passing, as does the CG solver.  The problem size for which MGMPI enjoys
a clear advantage over the CG solver may therefore depend in part on the number
of processors chosen.  Nonetheless,
for very large problems involving self-gravity on a Cartesian
mesh, MGMPI is likely to be the preferred option in a ZEUS-MP calculation.

\section{Summary}\label{summary}

In the introduction, we advertised the theme of this paper as ``physics, flexibility,
and parallelism.''  That these features are defining traits of~\zmp~is manifest:
hydrodynamics, MHD, and radiation diffusion may be deployed, singly or in concert,
on Cartesian, cylindrical, or spherical meshes in one to three dimensions.  ZEUS-MP
demonstrates parallel scalability suitable for computing platforms ranging from
small clusters to the largest platforms currently available for unclassified
research.  Features of a code designed for community use must also include
accuracy and computational expediency.  The accuracy of ZEUS-MP has been verified
both by traditional test problems and a multidimensional MHD problem 
frequently touted by developers of Godunov-based MHD codes.  Even when additional
resolution is required to ensure accuracy of a calculation, ZEUS-MP's parallel
performance provides a powerful mechanism for keeping the required solution times
manageable.

Virtues notwithstanding, we note that there are several ways in which ZEUS-MP
may be modified and improved within its solution paradigm.  Non-ideal MHD
effects such as Ohmic dissipation and ambipolar diffusion are requisite in a
variety of topics in interstellar physics such as star formation and interstellar shocks;
methodologies for including these effects in the ZEUS framework have been
documented by~\citet{stone99} in the case of Ohmic dissipation and likewise by~\citet{stone97}
for ambipolar diffusion.  A 3-D version of the VTEF algorithm described by~\citet{hayes03}
would be a major undertaking, but we note an approximation suggested by~\citet{ruszkowski03}
as an improvement to FLD suitable for the ZEUS codes.  Because ZEUS-MP is intended
for public distribution, one mission of this paper is to provide reference documentation
at a sufficiently high level of detail so that ambitious code developers may modify
it for their particular needs.

An additional area of improvement in which we are currently engaged concerns the
iterative solvers offered with the code.  The much higher computational cost of
simulations with FLD and the CG-based self-gravity module derives from the very
high numbers of iterations required for the CG linear solver to converge when
the matrix loses its diagonal dominance.  Because the matrix generated by the
discrete Poisson equation is {\it never} strongly diagonally dominant, CG methods
lose favor as the tool of choice for the Poisson problem on large grids.
The suitability of CG to radiation problems is very dependent on the physical and
temporal character of the problem at hand.  As shown by~\citet{baldwin99},
suitably optimized MG methods may be preferable to CG for some classes of radiation
problems.  Our current MGMPI solver is not yet flexible enough for use in radiation
applications, but enlarging its scope of applicability is a high priority item
for future research.

We also note that the convergence requirements of our current CG solver may be
dramatically improved with the use of a more effective preconditioner (recall
the discussion in~\S\ref{cgmeth}).  Our solver uses diagonal preconditioning,
which simultaneously boasts maximal ease of implementation and minimal range
of effectiveness with respect to the condition numbers for which convergence is
notably improved.  Despite the importance of preconditioning to the performance
of linear solvers upon which many astrophysical simulations must depend,
this topic has received relatively little attention in the numerical astrophysics
literature.  One study which has focused on astrophysical applications was
performed by~\citet{swesty04}, who considered a class of
preconditioners known as ``sparse approximate inverse'' (SPAI) preconditioners.
As the name implies, SPAI preconditioners
attempt to construct an approximation to the inverse of a matrix which is more
sophisticated than the purely diagonal approximation, but far less expensive
to compute than the full inverse. \citet{swesty04} have constructed SPAI preconditioners
designed for the linearized, discrete, energy-dependent version of the
FLD equation, the so-called {\it multigroup} flux-limited diffusion (MGFLD)
equation. The scientific focus in the~\citet{swesty04} paper is on 2-D and 3-D MGFLD
linear systems written to compute multidimensional neutrino diffusion
coupled to hydrodynamic flows in core-collapse supernova simulations.
While their analysis is designed for the MGFLD equations, they consider special
cases of isoenergetic diffusion directly analogous to the reduced system of
energy-averaged (or ``grey'') FLD equations adopted in ZEUS-MP (and ZEUS-2D).
The results reported in~\citet{swesty04}
suggest that SPAI preconditioners may offer a very profitable line of research for future
FLD implementations in~\zmp~or other application codes.

\acknowledgments

The release of a vastly redesigned and augmented ``Version 2'' of ZEUS-MP was
occassioned by ZEUS-MP's adoption by the Terascale Supernova Initiative
as a computational platform for simulating core-collapse supernova explosions
in multidimensions.  The TSI project, led by Dr. Anthony Mezzacappa, provided both
the demand and much of the financial support for the effort which created this code.  We are indebted
to Tony Mezzacappa for his continued support, enthusiasm, and remarkable patience during
the development phase of ZEUS-MP 2.0.  Additionally, we gratefully acknowledge the
continued support and wise counsel from Dr. Frank Graziani at the Lawrence Livermore
National Laboratory.  This work was supported by SciDAC grants from the
DOE Office of Science High-Energy, Nuclear, and Advanced Scientific Computing Research
Programs, and by DOE contract W-7405-ENG-48.

\begin{appendix}

\section{A Map of ZEUS-MP}\label{mapapp}

\LongTables
\begin{deluxetable}{lclccc}
\tablecaption{The Mapping Between ZEUS-MP Subroutines and Equations Solved.
\label{codemap}}
\tablehead{\colhead{Action} &
\colhead{Equation} &  \colhead{Routine} & \colhead{Hydro } & \colhead{MHD }
                   & \colhead{RHD }  }
\startdata
Compute $\Phi$       & \ref{sympoi}   & {\bf GRAVITY}  &   $\surd$   &   $\surd$   &   $\surd$ \\
\sidehead{{\bf SRCSTEP} Updates}
$\vone$ force update & \ref{v1force}  & {\bf FORCES}   &   $\surd$   &   $\surd$   &   $\surd$ \\
$\vtwo$ force update & \ref{v2force}  & {\bf FORCES}   &   $\surd$   &   $\surd$   &   $\surd$ \\
$\vthr$ force update & \ref{v3force}  & {\bf FORCES}   &   $\surd$   &   $\surd$   &   $\surd$ \\
$\vone$ viscosity update & \ref{qav1}     & {\bf AVISC}    &   $\surd$   &   $\surd$   &   $\surd$ \\
$\vtwo$ viscosity update &\ref{qav2}     & {\bf AVISC}    &   $\surd$   &   $\surd$   &   $\surd$ \\
$\vthr$ viscosity update & \ref{qav3}     & {\bf AVISC}    &   $\surd$   &   $\surd$   &   $\surd$ \\
$\egas\subijk$ viscosity update & \ref{eav} & {\bf AVISC}    &   $\surd$   &   $\surd$   &   $\surd$ \\
FLD solution & \ref{redjac}   & {\bf GREY\_FLD} &             &             &   $\surd$ \\
$\egas\subijk~pdV$ update & \ref{egpdv}    & {\bf PDV}    &  $\surd$    &   $\surd$   &  in {\bf GREY\_FLD}  \\
\sidehead{{\bf TRANSPRT} Updates}
$\vone$ Lorentz acc. & \ref{vonelr} & {\bf LORENTZ} &            &   $\surd$   &           \\
$\vtwo$ Lorentz acc. & \ref{vtwolr} & {\bf LORENTZ} &            &   $\surd$   &           \\
$\vthr$ Lorentz acc. & \ref{vthrlr} & {\bf LORENTZ} &            &   $\surd$   &           \\
$\emfone$ update & \ref{emf1a}    & {\bf HSMOC}    &             &   $\surd$   &           \\
$\emftwo$ update & \ref{emf2a}    & {\bf HSMOC}    &             &   $\surd$   &           \\
$\emfthr$ update & \ref{emf3a}    & {\bf HSMOC}    &             &   $\surd$   &           \\
$\bone$ update & \ref{b1evol}   & {\bf CT}       &             &   $\surd$   &           \\
$\btwo$ update & \ref{b2evol}   & {\bf CT}       &             &   $\surd$   &           \\
$\bthr$ update & \ref{b3evol}   & {\bf CT}       &             &   $\surd$   &           \\
1-Advect $\rho\subijk$  & \ref{rhoadv1}  & {\bf TRANX1}   &   $\surd$   &   $\surd$   &   $\surd$ \\
1-Advect $\egas\subijk$ & \ref{egadv1}   & {\bf TRANX1}   &   $\surd$   &   $\surd$   &   $\surd$ \\
1-Advect $\erad\subijk$ & \ref{eradv1}   & {\bf TRANX1}   &             &             &   $\surd$ \\
1-Advect X$\subijk$   & \ref{xadv1}    & {\bf TRANX1}   &   $\surd$   &   $\surd$   &   $\surd$ \\
1-Advect $\vone$ & \ref{v1adv1}   & {\bf MOMX1}    &   $\surd$   &   $\surd$   &   $\surd$ \\
1-Advect $\vtwo$ & \ref{v2adv1}   & {\bf MOMX1}    &   $\surd$   &   $\surd$   &   $\surd$ \\
1-Advect $\vthr$ & \ref{v3adv1}   & {\bf MOMX1}    &   $\surd$   &   $\surd$   &   $\surd$ \\
2-Advect $\rho\subijk$  & \ref{rhoadv2}  & {\bf TRANX2}   &   $\surd$   &   $\surd$   &   $\surd$ \\
2-Advect $\egas\subijk$ & \ref{egadv2}   & {\bf TRANX2}   &   $\surd$   &   $\surd$   &   $\surd$ \\
2-Advect $\erad\subijk$ & \ref{eradv2}   & {\bf TRANX2}   &             &             &   $\surd$ \\
2-Advect X$\subijk$ & \ref{xadv2}    & {\bf TRANX2}   &   $\surd$   &   $\surd$   &   $\surd$ \\
2-Advect $\vone$ & \ref{v1adv2}   & {\bf MOMX2}    &   $\surd$   &   $\surd$   &   $\surd$ \\
2-Advect $\vtwo$ & \ref{v2adv2}   & {\bf MOMX2}    &   $\surd$   &   $\surd$   &   $\surd$ \\
2-Advect $\vthr$ & \ref{v3adv2}   & {\bf MOMX2}    &   $\surd$   &   $\surd$   &   $\surd$ \\
3-Advect $\rho\subijk$  & \ref{rhoadv3}  & {\bf TRANX3}   &   $\surd$   &   $\surd$   &   $\surd$ \\
3-Advect $\egas\subijk$ & \ref{egadv3}   & {\bf TRANX3}   &   $\surd$   &   $\surd$   &   $\surd$ \\
3-Advect $\erad\subijk$ & \ref{eradv3}   & {\bf TRANX3}   &             &             &   $\surd$ \\
3-Advect X$\subijk$      & \ref{xadv3}    & {\bf TRANX3}   &   $\surd$   &   $\surd$   &   $\surd$ \\
3-Advect $\vone$ & \ref{v1adv3}   & {\bf MOMX3}    &   $\surd$   &   $\surd$   &   $\surd$ \\
3-Advect $\vtwo$ & \ref{v2adv3}   & {\bf MOMX3}    &   $\surd$   &   $\surd$   &   $\surd$ \\
3-Advect $\vthr$ & \ref{v3adv3}   & {\bf MOMX3}    &   $\surd$   &   $\surd$   &   $\surd$ \\
\sidehead{{\bf Time Step} Control}
New $\dt$        & \ref{dtcon}    & {\bf NUDT}    &   $\surd$   &   $\surd$   &   $\surd$ \\
\enddata
\end{deluxetable}

Table~\ref{codemap} provides a reference listing of the major equations derived in
the following appendices and the ZEUS-MP subroutines which compute them.  The
first three columns of the table indicate the solution substep, pertinent equation,
and associated subroutine, respectively.  The latter three columns are headed by
labels defining three classes of simulation: purely hydrodynamic, MHD, and RHD.
In each column a ``$\surd$'' mark indicates that the equation on that line is
include in the solution update.  Minor headings reading ``SRCSTEP'' and ``TRANSPRT''
(which reference subroutines with those names), respectively, indicate the two major
groups of solution substeps introduced in section~\ref{methods}

Entries in the table are ordered corresponding to the
sequence in which these operations occur during exection, save that advection
operations along each coordinate axis in the ``TRANSPRT'' section are cyclically permuted
from one time step to the next.

\section{The 3-D Discrete Gas Hydrodynamic Equations}\label{hdapp}

   \subsection{Metric Factors}

\zmp~expresses the discrete fluid equations in the coordinate-independent
fashion documented in~\citet{stone92a}.  For convenience, we reproduce the
basic metric definitions here.  The {\it metric tensor}, $g\subij$, relates
the length, $ds$, of a line element in one coordinate space, $y^k$, to the
equivalent expression in a second coordinate space, $x^i$, where we assume
that the $y^k$ can be expressed as functions of the $x^i$.  Thus:
\begin{equation}
ds^2 \ = \ \left(\partial y^k\over\partial x^i\right)
           \left(\partial y^k\over\partial x^j\right)dx^i dx^j \ \equiv \
g\subij dx^i dx^j,
\end{equation}
where k is summed from 1 to n, where n is the number of dimensions of $x$.
For orthogonal coordinate bases, $g\subij$ is diagonal; following the 
convention in~\citet{stone92a} we write:
\begin{equation}\label{metric}
g\subij\ = \ \left(
             \begin{array}{ccc}
             h^{2}_{1} &     0     & 0 \\
                 0     & h^{2}_{2} & 0 \\
                 0     &     0     & h^{2}_{3}
             \end{array}
             \right).
\end{equation}
In Cartesian coordinates, we then have
\begin{equation}\label{hxyz}
\left(x_{1},x_{2},x_{3}\right) \ = \ (x, y, z); \ \ \ \left(h_{1},h_{2},h_{3}\right)
                               \ = \ (1, 1, 1),
\end{equation}
while in cylindrical coordinates, we have
\begin{equation}\label{hzrp}
\left(x_{1},x_{2},x_{3}\right) \ = \ (z, r, \phi); \ \ \ \left(h_{1},h_{2},h_{3}\right)
                               \ = \ (1, 1, r),
\end{equation}
and in spherical coordinates, we have
\begin{equation}\label{hrtp}
\left(x_{1},x_{2},x_{3}\right) \ = \ (r, \theta, \phi); \ \ \ \left(h_{1},h_{2},h_{3}\right)
                               \ = \ (1, r, r\sin\theta).
\end{equation}
Following the convention introduced in the~\ztwd~papers, the $h$ factors are
re-expressed as separable functions of $g$ factors which are not to be confused
with $g\subij$ defined above:
\begin{equation}
h_{1} \ = \ 1 \ \equiv \ g_{1}, \label{g1}
\end{equation}
\begin{equation}
h_{2} \ = \ f\left(x_{1}\right) \ \equiv \ g_{2}, \label{g2}
\end{equation}
\begin{equation}
h_{3} \ = \ f\left(x_{1}\right)f\left(x_{3}\right) \ \equiv \ g_{31}g_{32}.
\label{g31g32}
\end{equation}
The explicit expressions for $g_2$, $g_{31}$, and $g_{32}$ are apparent by
comparing expressions (\ref{g1}) - (\ref{g31g32}) with (\ref{hxyz}) - (\ref{hrtp}).

   \subsection{Coordinate Meshes}

The staggered-mesh formalism relies upon an ``A'' mesh, whose points are centered
on zone faces, and a ``B'' mesh, whose points are located at zone centers.  
The coordinates of the A mesh along each axis are given by $x1a\subi$, $x2a\subj$,
and $x3a\subk$, with corresponding arrays for the B mesh.  Associated values
for the metric coefficients $g2$, $g31$, $g32$, and the derivatives of these
coefficients with respect to $x1$ and $x2$ are likewise evaluated on both
meshes and stored in 1-D arrays.

In many (but not all) instances, spatial derivatives are written as functions
of volume differences rather than coordinate differences.  Along the three
axes, transformation from coordinate to volume derivatives are written as
\begin{equation}
\begin{array}{lcrl}
{\partial / \partial x_1} & \rightarrow & g_2g_{31} &
{\partial / \partial V_1}, \\
{\partial / \partial x_2} & \rightarrow & g_{32}    &
{\partial / \partial V_2}, \\
{\partial / \partial x_3} & \rightarrow &           &
{\partial / \partial V_3}. 
\end{array}
\end{equation}

Scalar
field variables ($\rho$, e, E, and $\abun$) are centered on the B mesh.  Velocity
and magnetic field vector arrays $(\vone,\vtwo,\vthr); (\bone,\btwo,\bthr)$
are centered on the appropriate zone faces.  Magnetic EMF's $(\emfone,\emftwo,
\emfthr)$ are defined at midpoints of zone edges.

   \subsection{The ``Source Step'' Equations}

      \subsubsection{Body Forces}

In this subsection we document the updates to velocity due to body forces and
artificial viscosity, and the updates to internal energy due to artificial viscosity
and compressional heating.
The three components of fluid velocity are updated from body forces due to
pressure gradients, self-gravity, rotational pseudo-forces, magnetic pressure,
and radiation stress according to the following expressions:
%%
%% V1 forces update
%%
\begin{eqnarray}
{v1\supna\subijk - v1\supn\subijk \over \dt} & = &
-\left(2\over{\rho\subijk + \rho\subimjk}\right)
{{\pg\subijk - \pg\subimjk}\over{\Delta x1b\subi}}
\ + \ {{\Phi\subijk - \Phi\subimjk}\over{\Delta x1b\subi}} \nonumber \\
& + &
{\partial g2a\subi\over\partial x1}
\left[{\cal S}2^{oo}\subi + {\cal S}2^{po}\subi + 
      {\cal S}2^{oo}\subim + {\cal S}2^{po}\subim\right] \ \times \nonumber \\
&   & 
\left(1\over 8\right){\left[v2\supn\subijk + v2\supn\subijpk +
                            v2\supn\subimjk + v2\supn\subimjpk\right]\over
g2a^2\subi\left(\rho\subijk +\rho\subimjk\right)}
\nonumber \\
& + &
{\partial g31a\subi\over\partial x1}
\left[{\cal S}3^{oo}\subi + {\cal S}3^{op}\subi + 
      {\cal S}3^{oo}\subim + {\cal S}3^{op}\subim\right] \ \times \nonumber \\
&   & 
\left(1\over 8\right){\left[v3\supn\subijk  + v3\supn\subijk +
                            v3\supn\subimjk + v3\supn\subimjkp\right]\over
g31a^2\subi g32b\subj\left(\rho\subijk +\rho\subimjk\right)}
\nonumber \\
& - & {\left(d1b2^{oo}\subi + d1b2^{po}\subi + d1b3^{oo}\subi + d1b3^{op}\subi\right)
       \over 2\left(\rho\subijk + \rho\subimjk\right)\Delta x1b\subi}
\nonumber \\
& - & 
2\flim^{(1)} \ {{\erad\subijk - \erad\subimjk}\over
                {\left(\rho\subijk  + \rho\subimjk\right)\Delta x1b\subi}};
\label{v1force}
\end{eqnarray}
%%
%% V2 forces update
%%
\begin{eqnarray}
{v2\supna\subijk - v2\supn\subijk \over \dt} & = &
-\left(2\over{\rho\subijk + \rho\subijmk}\right)
{{\pg\subijk - \pg\subijmk}
\over{g2b\subi\Delta x2b\subj}}
\ + \ {{\Phi\subijk - \Phi\subijmk}\over{g2b\subi \Delta x2b\subj}} \nonumber \\
& + & 
{\partial g32a\subj\over\partial x2}
\left[{\cal S}3^{oo}\subi + {\cal S}3^{op}\subi + 
      {\cal S}3^{mo}\subi + {\cal S}3^{mp}\subi\right] \ \times
\nonumber \\
&  &
\left(1\over 8\right){\left[v3\supn\subijk + v3\supn\subijkp +
                       v3\supn\subijmk + v3\supn\subijmkp\right]
\over g31b\subi g32a^2\subj\left(\rho\subijk +\rho\subijmk\right)}
\nonumber \\
& - & {\left(d2b3^{oo}\subi + d2b3^{op}\subi + d2b1^{oo}\subi + d2b1^{po}\subi\right)
       \over 2\left(\rho\subijk + \rho\subijmk\right)g2b\subi\Delta x2b\subj}
\nonumber \\
& - & 
2\flim^{(2)} \ {{\erad\subijk - \erad\subijmk}\over
                {\left(\rho\subijk  + \rho\subijmk\right)g2b\subi\Delta x2b\subj}};
\label{v2force}
\end{eqnarray}
%%
%% V3 forces update
%%
\begin{eqnarray}
{v3\supna\subijk - v3\supn\subijk \over \dt} & = &
-\left(2\over{\rho\subijk + \rho\subijkm}\right)
{{\pg\subijk - \pg\subijkm}
\over{g31b\subi g32b\subj\Delta x3b\subk}}
\ + \ {{\Phi\subijk - \Phi\subijkm}\over
       {g31b\subi g32b\subj \Delta x3b\subk}} \nonumber \\
& - & {\left(d3b1^{oo}\subi + d3b1^{po}\subi + d3b2^{oo}\subi + d3b2^{po}\subi\right)
       \over 2\left(\rho\subijk + \rho\subijkm\right)g31b\subi g32b\subj\Delta x3b\subk}
\nonumber \\
& - & 
2\flim^{(3)} \ {{\erad\subijk - \erad\subijkm}\over
         {\left(\rho\subijk  + \rho\subijkm\right)g31b\subi g32b\subj\Delta x3b\subk}}.
\label{v3force}
\end{eqnarray}

Equations (\ref{v1force}) through (\ref{v3force}) make use of the following functions 
in the rotational pseudo-force terms:
\begin{eqnarray}
{\cal S}2^{oo}\subi & = & \left(1\over 2\right)v2\supn\subijk g2b\subi
                          \left(\rho\subijk + \rho\subijmk\right), \\
{\cal S}2^{po}\subi & = & \left(1\over 2\right)v2\supn\subijpk g2b\subi
                          \left(\rho\subijpk + \rho\subijk\right), \\
{\cal S}3^{oo}\subi & = & \left(1\over 2\right)v3\supn\subijk g31b\subi g32b\subj
                          \left(\rho\subijk + \rho\subijkm\right), \\
{\cal S}3^{op}\subi & = & \left(1\over 2\right)v3\supn\subijkp g31b\subi g32b\subj
                          \left(\rho\subijkp + \rho\subijk\right), \\
{\cal S}3^{mo}\subi & = & \left(1\over 2\right)v3\supn\subijmk g31b\subi g32b\subjm
                          \left(\rho\subijk + \rho\subijkm\right), \\
{\cal S}3^{mp}\subi & = & \left(1\over 2\right)v3\supn\subijmkp g31b\subi g32b\subjm
                          \left(\rho\subijmkp + \rho\subijmk\right).
\end{eqnarray}
Similarly, the magnetic pressure terms employ the following:
\begin{eqnarray}
d1b2^{oo} & = & {\left[\left(g2b\subi  b2\subijk \right)^2 - 
                       \left(g2b\subim b2\subimjk\right)^2
                 \right]\over g2a^2\subi}, \label{mpterm1} \\
d1b2^{po} & = & {\left[\left(g2b\subi  b2\subijpk \right)^2 - 
                       \left(g2b\subim b2\subimjpk\right)^2
                 \right]\over g2a^2\subi}, \label{mpterm2} \\
d1b3^{oo} & = & {\left[\left(g31b\subi  b3\subijk \right)^2 - 
                       \left(g31b\subim b3\subimjk\right)^2
                 \right]\over g31a^2\subi}, \label{mpterm3} \\
d1b3^{op} & = & {\left[\left(g31b\subi  b3\subijkp \right)^2 - 
                       \left(g31b\subim b3\subimjkp\right)^2
                 \right]\over g31a^2\subi}, \label{mpterm4} \\
d2b3^{oo} & = & {\left[\left(g32b\subj  b3\subijk \right)^2 - 
                       \left(g32b\subjm b3\subijmk\right)^2
                 \right]\over g32a^2\subj}, \label{mpterm5} \\
d2b3^{op} & = & {\left[\left(g32b\subj  b3\subijkp \right)^2 - 
                       \left(g32b\subjm b3\subijmkp\right)^2
                 \right]\over g32a^2\subj}, \label{mpterm6} \\
d2b1^{oo} & = & (b1\subijk )^2 - (b1\subijmk )^2, \label{mpterm7} \\
d2b1^{po} & = & (b1\subipjk)^2 - (b1\subipjmk)^2, \label{mpterm8} \\
d3b1^{oo} & = & (b1\subijk )^2 - (b1\subijkm )^2, \label{mpterm9} \\
d3b1^{po} & = & (b1\subipjk)^2 - (b1\subipjkm)^2, \label{mpterm10} \\
d3b2^{oo} & = & (b2\subijk )^2 - (b2\subijkm )^2, \label{mpterm11} \\
d3b2^{op} & = & (b2\subijpk)^2 - (b2\subijpkm)^2. \label{mpterm12}
\end{eqnarray}

      \subsubsection{Artificial Viscosity}

Once the velocity update from forces is complete, velocities which were known
at time level ``$n$'' are now known at an intermediate time level which we
designate as level ``$n+a$''.  These intermediate velocity components are then
updated due to the Von Neumann and Richtmyer prescription as follows: define
\begin{eqnarray}
\Delta v1\subijk & = & \left\{\begin{array}{lr}
                       v1\supna\subijk - v1\supna\subimjk, &
                       \mbox{$v1\supna\subijk < v1\supna\subimjk$;} \\
                       0, &
                       \mbox{$v1\supna\subijk \geq v1\supna\subimjk$;}
                       \end{array}
                       \right. \\
\Delta v2\subijk & = & \left\{\begin{array}{lr}
                       v2\supna\subijk - v2\supna\subimjk, &
                       \mbox{$v2\supna\subijk < v2\supna\subijmk$;} \\
                       0, &
                       \mbox{$v2\supna\subijk \geq v2\supna\subijmk$;}
                       \end{array}
                       \right. \\
\Delta v3\subijk & = & \left\{\begin{array}{lr}
                       v3\supna\subijk - v3\supna\subijkm, &
                       \mbox{$v3\supna\subijk < v3\supna\subijkm$;} \\
                       0, &
                       \mbox{$v3\supna\subijk \geq v3\supna\subijkm$;}
                       \end{array}
                       \right.
\end{eqnarray}
and
\begin{equation}
q1\subijk \ = \ C_{av}\rho\left(\Delta v1\subijk\right)^2; \ \ \
q2\subijk \ = \ C_{av}\rho\left(\Delta v2\subijk\right)^2; \ \ \
q3\subijk \ = \ C_{av}\rho\left(\Delta v3\subijk\right)^2.
\end{equation}
The velocity updates are then computed as
\begin{eqnarray}
v1\supnb\subijk & = & v1\supna\subijk \ - \ {{q1\subijk - q1\subimjk}\over
                {\Delta x1b\subi\left(\rho\subijk+\rho\subimjk\right)/2}}, \label{qav1}\\
v2\supnb\subijk & = & v2\supna\subijk \ - \ {{q2\subijk - q2\subijmk}\over
             {g2b\subi\Delta x2b\subj\left(\rho\subijk+\rho\subijmk\right)/2}}, 
             \label{qav2} \\
v3\supnb\subijk & = & v3\supna\subijk \ - \ {{q3\subijk - q3\subijkm}\over
          {g31b\subi g32b\subj\Delta x3b\subk\left(\rho\subijk+\rho\subijkm\right)/2}}.
          \label{qav3}
\end{eqnarray}
The gas internal energy is simultaneously updated via
\begin{equation}\label{eav}
\egas\supnb\subijk \ = \ \egas\supn\subijk \ - \ 
{q1\subijk\Delta v1\subijk\over dx1a\subi} \ - \
{q2\subijk\Delta v2\subijk\over g2b\subi dx2a\subj} \ - \
{q3\subijk\Delta v3\subijk\over g31b\subi g32b\subj dx3a\subk}.
\end{equation}

For problems with strong shocks, an additional linear viscosity may used.  ZEUS-MP
includes a linear viscosity of the form described in~\citet{stone92a}, in which
the linear viscosity depends upon the local sound speed:
\begin{equation}\label{qlin}
{\rm qlin}\subijk \ = \ C_l \left(\gamma P \over \rho\right)^{1/2}~\rho~\Delta v,
\end{equation}
where $C_l$ is a constant (typically of order 0.1) and $\Delta v$ is the difference
in neighboring velocities along the coordinate under consideration.
As with the quadratic viscosity, qlin$\subijk$ is evaluated independently along
each axis.  The updates to velocity and gas energy are identical to those for
the quadratic viscosity save for the replacement of ``$q$'' with ``qlin'' in
equations (\ref{qav1}) through (\ref{eav}).

      \subsubsection{Compressional Heating}

For an ideal EOS the $pdV$ compressional heating term is evaluated exactly
as outline in~\citet{stone92a}: to improve energy conservation, the updated
gas energy can be written as an implicit function of a time-centered 
pressure, whence
\begin{equation}\label{eimp}
\left(\egas\supnp - \egas\supn\right)/\dt \ = \ -\pg\supnh\divv,
\end{equation}
where $\pg\supnh \equiv 0.5\left(\pg\supn + \pg\supnp\right)$.  Using the
equation of state, $\pg = \left(\gamma - 1\right)\egas$, (\ref{eimp}) may
be rewritten to yield
\begin{equation}\label{egpdv}
\egas\subijk\supnc \ = \ 
\left[{1-\left(\dt/2\right)\left(\gamma - 1\right)\divv\subijk}\over
      {1+\left(\dt/2\right)\left(\gamma - 1\right)\divv\subijk}\right]
       \egas\subijk\supnb,
\end{equation}
where $\egas\supnb$ and $\egas\supnc$ are the gas energies immediately
prior to and after the $pdV$ update.  For non-ideal equations of state,
predictor-corrector techniques or Newton-Raphson iterations over temperature
may be employed.

   \subsection{The ``Transport Step'' Equations}

In the transport step,
ZEUS field variables are advected through the computational mesh using the
technique of {\it consistent transport}, introduced by~\citet{norman80b}.
Consistent transport attempts to minimize local conservation errors due to
numerical diffusion by defining face-centered fluxes of each field variable
consistent with the mass flux used to advect the matter density.  In this
procedure, the quantities advected are the mass density ($\rho$), the
specific internal energy ($\egas / \rho$), the specific radiation energy
($\erad / \rho$), and the specific momenta $S1~=~\rho~v1$, $S2
~=~\rho~g2~v2$, and $S3~=~\rho~g31~g32~v3$.  
The metric factors introduced
into the definitions of $S2$ and $S3$ transform these quantities into
angular momenta in curvilinear coordinates.

      \subsubsection{Scalar Variables}

We first consider the advection of mass density along the $i$ coordinate.  The amount
of mass crossing a cell face perpendicular to the $i$ axis in a time step,
$\dt$, is given by
\begin{equation}\label{mflux1}
\mdotone\subijk \dt \ = \ \dtwid\subijk\atwid\subi \left(\vone - \vgone\right)\dt,
\end{equation}
where $\atwid\subi$ is the time-centered area factor for cell i-face $i$, 
and $\dtwid$ is the matter density average to cell face $i$.  \zmp~uses
second-order Van Leer~\citep{vanleer77} averaging to construct monotonic,
upwinded averages of all advected quantities.  For advection across the
$i$ faces, the time-centered area factor (which accounts for grid motion)
is
\begin{equation}
\atwid\subi \ = \ g2a\subi\supnh g31a\subi\supnh.
\end{equation}
The computed mass flux, $\mdotone\subijk$, is then used to advect $\rho\subijk$
according to
\begin{equation}\label{rhoadv1}
\rho\supnp\subijk \ = \ \left[ \rho\supn\subijk dvl1a\supn\subi + 
          \left(\mdotone\subijk - \mdotone\subipjk\right)\dt\right] / dvl1a\subi\supnp.
\end{equation}

Consistent transport of the gas and radiation energy densities proceeds by defining
specific energies (erg/gm) for each of these quantities, averaging the specific
energy to cell faces via Van Leer interpolation, and computing fluxes across each
face with the mass fluxes computed in (\ref{mflux1}).  We thus have:
\begin{eqnarray}
\egas\supnp\subijk & = & \left\{\egas\subijk\supn dvl1a\supn\subi
 \ + \ \left[\etwid\subi\mdotone\subijk - 
             \etwid\subip\mdotone\subipjk\right]\dt
                         \right\} / dvl1a\subi\supnp; \label{egadv1}\\
\erad\supnp\subijk & = & \left\{\erad\subijk\supn dvl1a\supn\subi
 \ + \ \left[\ertwid\subi\mdotone\subijk - 
             \ertwid\subip\mdotone\subipjk\right]\dt
                         \right\} / dvl1a\subi\supnp. \label{eradv1}
\end{eqnarray}
The multi-species composition advection uses the $\abun$ variables to define
partial densities, which are then advected and converted back to dimensionless
mass fractions.  Thus:
\begin{equation}\label{xadv1}
\abun\supnp\subijk \ = \ \left[ \abun\supn\subijk\rho\supn\subijk dvl1a\supn\subi + 
        \left(\xtwid\subi\mdotone\subijk - 
              \xtwid\subip\mdotone\subipjk\right)\dt\right] / 
                \left(\rho\supnp\subijk dvl1a\subi\supnp\right).
\end{equation}

For the advection of scalar variables across cell faces perpendicular to the
$j$ axis, we write mass fluxes and time-centered face areas as
\begin{equation}\label{mflux2}
\mdottwo\subijk \ = \ \dtwid\subijk\atwid\subj \left(\vtwo - \vgtwo\right),
\end{equation}
and
\begin{equation}
\atwid\subj \ = \ g31b\subi\supn~g32a\subj\supnh~dx1a\supn\subi / dvl1a\subi\supn.
\end{equation}
Advection of $\rho$, $\egas$, $\erad$, and $\abun$ along the $j$ coordinate
then proceeds as
\begin{equation}\label{rhoadv2}
\rho\supnp\subijk \ = \ \left[ \rho\supn\subijk dvl2a\supn\subj +
          \left(\mdottwo\subijk - \mdottwo\subijpk\right)\dt\right] / dvl2a\subj\supnp,
\end{equation}
\begin{eqnarray}
\egas\supnp\subijk & = & \left\{\egas\subijk\supn dvl2a\supn\subj
 \ + \ \left[\etwid\subj\mdottwo\subijk -
             \etwid\subjp\mdottwo\subijpk\right]\dt
                         \right\} / dvl2a\subj\supnp, \label{egadv2}\\
\erad\supnp\subijk & = & \left\{\erad\subijk\supn dvl2a\supn\subj
 \ + \ \left[\ertwid\subj\mdottwo\subijk -
             \ertwid\subjp\mdottwo\subijpk\right]\dt
                         \right\} / dvl2a\subj\supnp, \label{eradv2}
\end{eqnarray}
and
\begin{equation}\label{xadv2}
\abun\supnp\subijk \ = \ \left[ \abun\supn\subijk\rho\supn\subijk dvl2a\supn\subj +
        \left(\xtwid\subj\mdottwo\subijk - 
              \xtwid\subjp\mdottwo\subijpk\right)\dt\right] /
                \left(\rho\supnp\subijk dvl2a\subj\supnp\right). 
\end{equation}

Similarly, the advection of scalar quantities along the $k$ axis is done as follows:
define
\begin{equation}\label{mflux3} 
\mdotthr\subijk \ = \ \dtwid\subijk\atwid \left(\vthr - \vgthr\right),
\end{equation}
and
\begin{equation}
\atwid \ = \ g2b\subi\supn~dx1a\supn\subi~dx2a\subj\supn / 
             \left(dvl1a\subi\supn~dvl2a\subj\supn\right).
\end{equation}
Thus
\begin{equation}\label{rhoadv3}
\rho\supnp\subijk \ = \ \left[ \rho\supn\subijk dvl3a\supn\subk +
          \left(\mdotthr\subijk - \mdotthr\subijkp\right)\dt\right] / dvl3a\subk\supnp,
\end{equation}
\begin{eqnarray}
\egas\supnp\subijk & = & \left\{\egas\subijk\supn dvl3a\supn\subk
 \ + \ \left[\etwid\subk\mdotthr\subijk -
             \etwid\subkp\mdotthr\subijkp\right]\dt
                         \right\} / dvl3a\subk\supnp, \label{egadv3}\\
\erad\supnp\subijk & = & \left\{\erad\subijk\supn dvl3a\supn\subk
 \ + \ \left[\ertwid\subk\mdotthr\subijk -
             \ertwid\subkp\mdotthr\subijkp\right]\dt
                         \right\} / dvl3a\subk\supnp,\label{eradv3}
\end{eqnarray}
and
\begin{equation}\label{xadv3}
\abun\supnp\subijk \ = \ \left[ \abun\supn\subijk\rho\supn\subijk dvl3a\supn\subk +
        \left(\xtwid\subk\mdotthr\subijk -
              \xtwid\subkp\mdotthr\subijkp\right)\dt\right] /
                \left(\rho\supnp\subijk dvl3a\subk\supnp\right).
\end{equation}

      \subsubsection{Momentum Variables}

Each component of the specific momentum is computed (modulo metric factors) by
dividing the appropriate velocity component by an arithmetic average of the
density at the corresponding cell face.  Thus
\begin{eqnarray}
\sone & = & 0.5~\vone\left(\rho\subijk\supn + \rho\subimjk\supn\right), \\
\stwo & = & 0.5~\vtwo\left(\rho\subijk\supn + \rho\subijmk\supn\right)
            ~g2b\subi, \\
\sthr & = & 0.5~\vthr\left(\rho\subijk\supn + \rho\subijkm\supn\right)
            ~g31b\subi~g32b\subj.
\end{eqnarray}

Along the $i$ coordinate, the specific momenta are transported according to the
following:
\begin{eqnarray}
\sone\supnp & = & \left[\sone\supn~dvl1b\subi\supn
                    + \sdotone\subi - \sdotone\subim
                  \right] / dvl1b\subi\supnp; \label{v1adv1}\\
\stwo\supnp & = & \left[\stwo\supn~dvl1a\subi\supn
                    + \sdottwo\subi - \sdottwo\subim
                  \right] / dvl1a\subi\supnp; \label{v2adv1}\\
\sthr\supnp & = & \left[\sthr\supn~dvl1a\subi\supn
                    + \sdotthr\subi - \sdotthr\subim
                  \right] / dvl1a\subi\supnp.\label{v3adv1}
\end{eqnarray}
Note that the volume factors used to transport $\stwo$ and $\sthr$ differ from
those used to transport $\sone$ owing to the different centering of $\stwo$ and
$\sthr$ with respect to the staggered $i$ mesh.  The momentum fluxes are
constructed from the previously computed $i$ components of the mass flux as:
\begin{eqnarray}
\sdotone\subi & = & \left(\mdotone\subijk + \mdotone\subipjk\right)\tilde{v}_1
\left(0.5~g2b\subi\supnh~g31b\subi\supnh\right), \\
\sdottwo\subi & = & \left(\mdotone\subijmk + \mdottwo\subijk\right)\tilde{v}_2
\left(0.5~g2a\subi\supn ~g2a\subi\supnh~g31a\subi\supnh\right), \\
\sdotthr\subi & = & \left(\mdotone\subijkm + \mdotthr\subijk\right)\tilde{v}_3
\left(0.5~g31a\subi\supn ~g2a\subi\supnh~g31a\subi\supnh\right).
\end{eqnarray}
In the definition of the momentum fluxes, $\tilde{v}_1$, $\tilde{v}_2$, and
$\tilde{v}_3$ denote cell-centered Van Leer averages of the three relative velocity
components: $\vone - \vgone$, $\vtwo - \vgtwo$, and $\vthr - \vgthr$.

Along the $j$ coordinate, momentum advection is computed via
\begin{eqnarray}
\sone\supnp & = & \left[\sone\supn~dvl2a\subj\supn
                    + \sdotone\subj - \sdotone\subjp
                  \right] / dvl2a\subj\supnp; \label{v1adv2}\\
\stwo\supnp & = & \left[\stwo\supn~dvl2b\subj\supn
                    + \sdottwo\subjm - \sdottwo\subj
                  \right] / dvl2b\subj\supnp; \label{v2adv2}\\
\sthr\supnp & = & \left[\sthr\supn~dvl2a\subj\supn
                    + \sdotthr\subj - \sdotthr\subjp
                  \right] / dvl2a\subj\supnp,\label{v3adv2}
\end{eqnarray}
with
\begin{eqnarray}
\sdotone\subj & = & \left(\mdottwo\subimjk + \mdottwo\subijk\right)\tilde{v}_1
\left(0.5~g31a\subi\supn~g32a\subj\supnh~dx1b\subi\supn~dvl1b\subi\supn\right), \\
\sdottwo\subj & = & \left(\mdottwo\subijk + \mdottwo\subijpk\right)\tilde{v}_2
\left(0.5~g2b\subi\supn~g31b\subi\supn~g32b\subi\supnh~dx1a\subi\supn~
dvl1a\subi\supn\right), \\
\sdotthr\subj & = & \left(\mdottwo\subijkm + \mdottwo\subijk\right)\tilde{v}_3
\left(0.5~\left(g31b\subi\supn\right)^{2}~g32a\subi\supn~g32a\subi\supnh~dx1a\subi\supn~
dvl1a\subi\supn\right). \\
\end{eqnarray}
As with the definitions of momentum fluxes along the $i$ axis, the $\tilde{v}$ terms
in the j-flux expressions represent Van Leer averages of the relative velocity components,
but the numerical values differ owing to the change of axis.

Finally, the $k$-axis equations for momentum advection are written as
\begin{eqnarray}
\sone\supnp & = & \left[\sone\supn~dvl3a\subk\supn
                    + \sdotone\subk - \sdotone\subkp
                  \right] / dvl3a\subk\supnp; \label{v1adv3}\\
\stwo\supnp & = & \left[\stwo\supn~dvl3a\subk\supn
                    + \sdottwo\subk - \sdottwo\subkp
                  \right] / dvl3a\subk\supnp; \label{v2adv3}\\
\sthr\supnp & = & \left[\sthr\supn~dvl3b\subk\supn
                    + \sdotthr\subkm - \sdotthr\subk
                  \right] / dvl3b\subk\supnp,\label{v3adv3}
\end{eqnarray}
with
\begin{eqnarray}
\sdotone\subk & = & \left(\mdotthr\subim + \mdotthr\subi\right)\tilde{v}_1
\left(0.5~g2a\subi\supn~dx1b\subi\supn~dx2a\subj\supn / dvl1b\subi\supn
~dvl2a\subj\right); \\
\sdottwo\subk & = & \left(\mdotthr\subjm + \mdotthr\subj\right)\tilde{v}_2
\left(0.5~\left(g2b\subi\supn\right)^2~dx1a\subi\supn~dx2b\subj\supn / 
dvl1a\subi\supn~dvl2b\subj\supn\right); \\
\sdotthr\subk & = & \left(\mdotthr\subk + \mdotthr\subkp\right)\tilde{v}_3 
\left(0.5~g31b\subi\supn~g2b\subi\supn~dx1a\subi\supn~g32b\subj\supn~dx2a\subj\supn / 
dvl1a\subi\supn~dvl2a\subj\supn\right).
\end{eqnarray}

\section{The 3-D Discrete MHD Equations}\label{mhdapp}

   \subsection{Construction of the EMF's}

In a 3-D geometry expressed upon covariant mesh variables, the characteristic
equations for Alfv\'{e}n wave propagation (\ref{chareqn}) along the 1-axis become
\begin{equation}\label{alfchar1}
{1\over\srd}{\mathcal{D}\over\mathcal{D}t}\left(\bone\right) \ \pm \ 
{\mathcal{D}^{\pm}\over\mathcal{D}t}\left(\vone\right) \ = \ \pm\mathcal{S},
\end{equation}
in which the Lagrangean derivative is expanded as
\begin{equation}\label{clagder}
{\mathcal{D}^{\pm}\over\mathcal{D}t} \ = \ {\partial\over\partial t} \ + \ 
\left(v2 \mp \vatwo\right){\partial\over\partial x2} \ + \ 
\left(v3 \mp \vathr\right){\partial\over\partial x3},
\end{equation}
the Alfv\'{e}n velocities are given by
\begin{eqnarray}
\vatwo & = & b2 / \srd, \label{va2} \\
\vathr & = & b3 / \srd, \label{va3}
\end{eqnarray}
and $\mathcal{S}$ is a source term arising from derivatives of the coordinate
metric factors ($\equiv 0$ in Cartesian geometry).  We difference the temporal
derivatives along each characteristic as
\begin{eqnarray}
\mathcal{D}^+[b1] & = & \bonestr - b1^+, \\
\mathcal{D}^-[b1] & = & \bonestr - b1^-, \\
\mathcal{D}^+[v1] & = & \vonestr - v1^+, \\
\mathcal{D}^-[v1] & = & \vonestr - v1^-, \\
\end{eqnarray}
and solve the characteristic equations for $\vonestr$ and $\bonestr$:
\begin{equation}\label{bchar}
\bonestr \ = \ {\sqrt{\rho^+\rho^-}\over \srdp + \srdm}\left[
                {b1^+\over\srdp} + {b1^-\over\srdm} + v1^+ - v1^-
               \right],
\end{equation}
and
\begin{equation}\label{vchar}
\vonestr \ = \ {1\over \srdp + \srdm}\left[
                {v1^+\srdp} + {v1^- \srdm} + b1^+ - b1^-
               \right] + \mathcal{S}\Delta t.
\end{equation}
As discussed in the main text, $\rho^+$ and $\rho^-$ are the densities at
the footpoints of the respective characteristics.  Equation~\ref{alfchar1},
in view of equations (\ref{clagder})-(\ref{va3}), suggests that when
evaluating $\emfthr$ as outlined in section~\ref{mhdmeth}, $v1$ and $b1$
should be upwinded along both the 2- and 3-components of the characteristic
velocity.  The numerical impracticality of this approach leads us to adopt
the approach of~\citet{hawley95}, in which only partial characteristics are
used to upwind velocity and magnetic field components.  Quantities upwinded
along Alfv\'{e}n characteristics are then combined with quantities upwinded
along hydrodynamic fluid-flow characteristics in a self-consistent fashion.
We illustrate the procedure by outlining the calculation of $\epsilon 3$,
written schematically as
\begin{equation}
\epsilon 3 \ = \ \vonestr\btwostr \ - \ \vtwostr\bonestr.
\end{equation}
To aid the documentation, we introduce an ``ADV'' function in which
ADV$[b2,v1]$ denotes a mean value for b2 computed from a Van Leer average
upwinded according to the $v1$ velocity.  This functional notation will be
used to describe quantities upwinded along coordinate axes (fluid flow
characteristics) or Alfv\'{e}n characteristics.  Our method subdivides into
two stages.  In stage I, partial characteristics along the 2-axis are used
in the construction of values for $v1^*$ and $b1^*$ as follows:

\noindent Step Ia: upwind $b2$ and $v2$ along 1-axis:
\begin{eqnarray}
\btwobar\onedir & \equiv & {\rm ADV}[b2, v1]; \\
\vtwobar\onedir & \equiv & {\rm ADV}[v2, v1].
\end{eqnarray}
Step Ib: compute 2-characteristic Alfv\'{en} speeds:
\begin{eqnarray}
v\twochrp & = & \vtwobar\onedir \ - \ |\btwobar\onedir| / \srdp; \\
v\twochrm & = & \vtwobar\onedir \ + \ |\btwobar\onedir| / \srdm.
\end{eqnarray}
Step Ic: upwind $v1$ and $b1$ along the +/- characteristics:
\begin{eqnarray}
\vonebar\twochrp & = & {\rm ADV}\left[v1,v\twochrp\right]; \\
\vonebar\twochrm & = & {\rm ADV}\left[v1,v\twochrm\right]; \\
\bonebar\twochrp & = & {\rm ADV}\left[b1,v\twochrp\right]; \\
\bonebar\twochrm & = & {\rm ADV}\left[b1,v\twochrm\right].
\end{eqnarray}
Step Id: solve the characteristic equations (\ref{bchar} and~\ref{vchar}) for
$\bonestr$ and $\vonestr$:
\begin{eqnarray}
\bonestr & = & \left(\sqrt{\rho^+\rho^-}\over{\srdp + \srdm}\right)
               \left[{\bonebar\twochrp\over\srdp} + 
                     {\bonebar\twochrm\over\srdm} + 
                     \vonebar\twochrp - \vonebar\twochrm\right]; \\
\vonestr & = & \left(1\over{\srdp + \srdm}\right)
               \left[\vonebar\twochrp~\srdp \ + \ \vonebar\twochrm~\srdm
                     \ + \ \bonebar\twochrp \ - \ \bonebar\twochrm\right] \nonumber \\
         & + & \mathcal{S}\Delta t
\end{eqnarray}
Step Ie: compute and store the products $\vonestr~\btwobar\onedir$ and
$\vtwobar\onedir~\bonestr$.

Stage II is analogous to stage I, except that now we solve for $\vtwostr$ and
$\btwostr$ by examining partial characteristics in the 1 direction:

\noindent Step IIa:
\begin{eqnarray}
\bonebar\twodir & \equiv & {\rm ADV}[b1, v2]; \\
\vonebar\twodir & \equiv & {\rm ADV}[v1, v2].
\end{eqnarray}
Step IIb:
\begin{eqnarray}
v\onechrp & = & \vonebar\twodir \ - \ |\bonebar\twodir| / \srdp; \\
v\onechrm & = & \vonebar\twodir \ + \ |\bonebar\twodir| / \srdm.
\end{eqnarray}
Step IIc:
\begin{eqnarray}
\vtwobar\onechrp & = & {\rm ADV}\left[v2,v\onechrp\right]; \\
\vtwobar\onechrm & = & {\rm ADV}\left[v2,v\onechrm\right]; \\
\btwobar\onechrp & = & {\rm ADV}\left[b2,v\onechrp\right]; \\
\btwobar\onechrm & = & {\rm ADV}\left[b2,v\onechrm\right].
\end{eqnarray}
Step IId:
\begin{eqnarray}
\btwostr & = & \left(\sqrt{\rho^+\rho^-}\over{\srdp + \srdm}\right)
               \left[{\btwobar\onechrp\over\srdp} +
                     {\btwobar\onechrm\over\srdm} +
                     \vtwobar\onechrp - \vtwobar\onechrm\right]; \\
\vtwostr & = & \left(1\over{\srdp + \srdm}\right)
               \left[\vtwobar\onechrp~\srdp \ + \ \vtwobar\onechrm~\srdm
                     \ + \ \btwobar\onechrp \ - \ \btwobar\onechrm\right] \nonumber \\
         & + & \mathcal{S}\Delta t
\end{eqnarray}
Step IIe: compute and store the products $\vtwostr~\bonebar\twodir$ and
$\vonebar\twodir~\btwostr$.

With these two stages complete, we now write the 3-EMF as
\begin{equation}\label{emf3a}
\epsilon 3 \ = \ {{\vonestr\btwobar\onedir + \vonebar\twodir\btwostr}\over 2} - 
                 {{\vtwostr\bonebar\twodir + \vtwobar\onedir\bonestr}\over 2}.
\end{equation}
The 1-emf and 2-emf expressions are derived and expressed analogously as
\begin{equation}\label{emf1a}
\epsilon 1 \ = \ {{\vtwostr\bthrbar\twodir + \vtwobar\thrdir\bthrstr}\over 2} - 
                 {{\vthrstr\btwobar\thrdir + \vthrbar\twodir\btwostr}\over 2},
\end{equation}
and
\begin{equation}\label{emf2a}
\epsilon 2 \ = \ {{\vthrstr\bonebar\thrdir + \vthrbar\onedir\bonestr}\over 2} - 
                 {{\vonestr\bthrbar\onedir + \vonebar\thrdir\bthrstr}\over 2}.
\end{equation}
Because each component of the magnetic field (e.g. $b1$) depends upon EMF's
computed around both transverse axes (e.g. $\epsilon 2$ and $\epsilon 3$),
the evolution of each B-field component will depend upon the full set of characteristics.
This method is effectively a simple directional splitting of the full MOC algorithm.
As discussed in~\citet{hawley95}, each term in a given EMF expression is composed
of 1-D advection solutions in which hydrodynamic characteristics are mixed with
Alfv\'{e}n characteristics in a consistent fashion; i.e. in the leading term of
equation~\ref{emf3a}, $b2$ has been passively advected along the same coordinate
axis for which the characteristic velocity equation is solved.  This
consistency is maintained in all terms of the EMF equations.  Additional discussion
in~\citet{hawley95} notes that the practice of consistently mixing partial Alfv\'{e}n
characteristic solutions with hydrodynamic advection retains the relative
simplicity of 1-D upwinding yet is less prone to error in the presence of
strong magnetic discontinuities.

   \subsection{Lorentz Acceleration of Velocities}

The Lorentz accelerations are computed by a procedure analogous to the calculation
of the EMF's outlined above.  In what follows we make extensive use of the notation
introduced in the previous section.  We demonstrate the method in detail by writing
expressions for the 1-component of the Lorentz acceleration, which depends upon
information propagating along Alfv\'en characteristics in the 2- and 3-directions.
Stage I of the solution considers the Alfv\'en 2-characteristics:

\noindent Step Ia: define footpoint densities as
\begin{equation}
\srdp \ = \ \left(\rho\subijk\cdot\rho\subimjk\right)^{1/2};~~~
\srdm \ = \ \left(\rho\subijmk\cdot\rho\subimjmk\right)^{1/2}.
\end{equation}

\noindent Step Ib: define average of $\btwo$ and compute Alfv\'en speeds:
\begin{eqnarray}
\btwobar\subj & = & 0.5\left(b2\subijk + b2\subimjk\right); \\
\vatwop & = & -|\btwobar| \ / \ \srdp; \\
\vatwom & = &  |\btwobar| \ / \ \srdm.
\end{eqnarray}

\noindent Step Ic: upwind $\bone$ and $\vone$ along the (+) and (-) Alfv\'en
characteristics:
\begin{eqnarray}
b1\twochrp & = & {\rm ADV}\left[b1,\vatwop\right]; \\
b1\twochrm & = & {\rm ADV}\left[b1,\vatwom\right]; \\
v1\twochrp & = & {\rm ADV}\left[v1,\vatwop\right]; \\
v1\twochrm & = & {\rm ADV}\left[v1,\vatwom\right].
\end{eqnarray}

\noindent Step Id: solve characteristic equation for $b$:
\begin{equation}
\bonetwostr \ = \ \left\{
                         {\bonebar\twochrp\over\srdp} + 
                         {\bonebar\twochrm\over\srdm} + 
                         {\rm SGN}\left[1,\btwobar\right]\cdot
                         \left(\vonebar\twochrp - \vonebar\twochrm\right)
                  \right\} \left({1\over\srdp} + {1\over\srdm}\right),
\end{equation}
where ``${\rm SGN}\left[1,\btwobar\right]$'' is plus or minus 1 depending on the sign
of $\btwobar$.  Finally,

\noindent Step Ie: evaluate the first contribution to the Lorentz 1-acceleration:
\begin{equation}
\sone\lrnt \ = \ \left(\btwobar\subjp + \btwobar\subj\right)
                \left(\bonetwostr\subjp - \bonetwostr\subj\right) / 
                \left(g2a\subi~dx2a\subj\right).
\end{equation}
Stage II examines evolution along the Alfv\'en 3-characteristics as follows:

\noindent Step IIa: define footpoint densities as
\begin{equation}
\srdp \ = \ \left(\rho\subijk\cdot\rho\subijkm\right)^{1/2};~~~
\srdm \ = \ \left(\rho\subijkm\cdot\rho\subimjkm\right)^{1/2}.
\end{equation}

\noindent Step IIb: define average of $\bthr$ and compute Alfv\'en speeds:
\begin{eqnarray}
\bthrbar\subk & = & 0.5\left(b3\subijk + b2\subimjk\right); \\
\vathrp & = & -|\bthrbar| \ / \ \srdp; \\
\vathrm & = &  |\bthrbar| \ / \ \srdm.
\end{eqnarray}

\noindent Step IIc: upwind $\bone$ and $\vone$ along the (3+) and (3-) Alfv\'en
characteristics:
\begin{eqnarray}
b1\thrchrp & = & {\rm ADV}\left[b1,\vathrp\right]; \\
b1\thrchrm & = & {\rm ADV}\left[b1,\vathrm\right]; \\
v1\thrchrp & = & {\rm ADV}\left[v1,\vathrp\right]; \\
v1\thrchrm & = & {\rm ADV}\left[v1,\vathrm\right].
\end{eqnarray}

\noindent Step IId: solve characteristic equation for $b$:
\begin{equation}
\bonethrstr \ = \ \left\{
                         {\bonebar\thrchrp\over\srdp} +
                         {\bonebar\thrchrm\over\srdm} +
                         {\rm SGN}\left[1,\bthrbar\right]\cdot
                         \left(\vonebar\thrchrp - \vonebar\thrchrm\right)
                  \right\} \left({1\over\srdp} + {1\over\srdm}\right).
\end{equation}

\noindent Step IIe: add the second contribution to the Lorentz 1-acceleration to the
first:
\begin{eqnarray}
\sone\lrnt & = & \left(\btwobar\subjp + \btwobar\subj\right)
                 \left(\bonetwostr\subjp - \bonetwostr\subj\right) /
                 \left(g2a\subi~dx2a\subj\right) \nonumber \\
           & + & \left(\bthrbar\subkp + \bthrbar\subk\right)
                 \left(\bonethrstr\subkp - \bonethrstr\subk\right) /
                 \left(g31a\subi~g32b\subj~dx3a\subk\right). \label{sonelr}
\end{eqnarray}
The 2- and 3-components of $S\lrnt$ are similarly written as
\begin{eqnarray}
\stwo\lrnt & = & \left(\bthrbar\subkp + \bthrbar\subk\right)
                 \left(\btwothrstr\subjp - \btwothrstr\subj\right) /
                 \left(g31a\subi~g32b\subj~dx3a\subk\right) \nonumber \\
           & + & \left(\bonebar\subip + \bonebar\subi\right)
                 \left(\btwoonestr\subip - \btwoonestr\subi\right) /
                 \left(g2b\subi~dx1a\subi\right) \label{stwolr}; \\
\sthr\lrnt & = & \left(\bonebar\subip + \bonebar\subi\right)
                 \left(\bthronestr\subip - \bthronestr\subi\right) /
                 \left(g31b\subi~dx1a\subi\right) \nonumber \\
           & + & \left(\btwobar\subjp + \btwobar\subj\right)
                 \left(\bthrtwostr\subjp - \bthrtwostr\subj\right) /
                 \left(g2a\subi~g32b\subj~dx3a\subk\right). \label{sthrlr}
\end{eqnarray}
With the accelerations thus defined, the fluid velocities are accelerated 
according to
\begin{eqnarray}
\vone\supnc & = & \vone\supnb \ + \ {1\over 2}\dt\cdot\sone\lrnt~/~
                  \sqrt{\rho\subijk\cdot\rho\subimjk}; \label{vonelr} \\
\vtwo\supnc & = & \vtwo\supnb \ + \ {1\over 2}\dt\cdot\stwo\lrnt~/~
                  \sqrt{\rho\subijk\cdot\rho\subijmk}; \label{vtwolr} \\
\vthr\supnc & = & \vthr\supnb \ + \ {1\over 2}\dt\cdot\sthr\lrnt~/~
                  \sqrt{\rho\subijk\cdot\rho\subijkm}, \label{vthrlr}
\end{eqnarray}
where the $n+b$ superscript denotes velocities which have been updated in
the source step via local body forces (step ``a'') and artificial viscosity
(step ``b'').

   \subsection{Evolution of the Field Components}

With EMF's suitably computed on each edge of the 3-D grid cell, the three
magnetic field components are evolved from time level $n$ to level $n+1$ via
the 3-D CT formalism.  As with the gas hydrodynamic advection equations,
the HSMOCCT algorithm is formulated to account for grid motion along all three
coordinate axes, thus special care must be taken to time-center spatial
coordinate terms correctly.
The line integral equations describing the temporal evolution of the
magnetic fluxes through the $i$, $j$, and $k$ cell faces were introduced
in~\S\ref{mhdmeth}; for ease of reference we repeat them here:
\begin{equation}\label{bflux1a}
{{\phi 1\subijk\supnp - \phi 1\subijk\supn}\over \dt} \ = \
 \epsilon 2\subijk \Delta x2\subij + \epsilon 3\subijpk\Delta x3\subijpk
-\epsilon 2\subijkp\Delta x2\subij - \epsilon 3\subijk \Delta x3\subijk;
\end{equation}
\begin{equation}\label{bflux2a}
{{\phi 2\subijk\supnp - \phi 2\subijk\supn}\over \dt} \ = \
 \epsilon 1\subijkp\Delta x1\subi + \epsilon 3\subijpk\Delta x3\subijk
-\epsilon 1\subijk \Delta x1\subi - \epsilon 3\subipjk\Delta x3\subipjk;
\end{equation}
\begin{equation}\label{bflux3a}
{{\phi 3\subijk\supnp - \phi 3\subijk\supn}\over \dt} \ = \
 \epsilon 1\subijk \Delta x1\subi + \epsilon 2\subipjk\Delta x2\subipj
-\epsilon 1\subijpk\Delta x1\subi - \epsilon 2\subijk \Delta x2\subij.
\end{equation}
The magnetic fluxes are related to the field components and covariant
metric tensor coefficients through the following relations:
\begin{eqnarray}
\phi 1 & = & h_2~h_3~dx2~dx3~b1 \ \rightarrow \ 
                                   g2a\subi~g31a\subi~g32b\subj~dx2a\subj~dx3a\subk~
                                      \bone; \label{bftran1} \\
\phi 2 & = & h_1~h_3~dx1~dx3~b2 \ \rightarrow \ 
                                            g31b\subi~g32a\subj~dx1a\subi~dx3a\subk~
                                      \btwo; \label{bftran2} \\
\phi 3 & = & h_1~h_2~dx1~dx2~b3 \ \rightarrow \ 
                                   g2b\subi                    ~dx1a\subi~dx2a\subj~
                                      \bthr. \label{bftran3}
\end{eqnarray}
Cell edge line elements are transformed to covariant coordinates via
\begin{eqnarray}
\Delta x1 & \rightarrow &                     dx1a\subi; \label{ltran1} \\
\Delta x2 & \rightarrow &  g2a\subi~          dx2a\subj; \label{ltran2} \\
\Delta x3 & \rightarrow & g31a\subi~g32a\subj~dx3a\subk. \label{ltran3}
\end{eqnarray}
The evolution equations for $\bone$, $\btwo$, and $\bthr$ are then written as:
\begin{equation}
\left({\cal A}1\subijk\right)\supnp\bone\supnp \ = \
\left({\cal A}1\subijk\right)\supn \bone\supn 
 + \dt \left[
                 {\cal E}2\subijk  + {\cal E}3\subijpk 
                -{\cal E}2\subijkp - {\cal E}3\subijk  \right]; \label{b1evol}
\end{equation}
\begin{equation}
\left({\cal A}2\subijk\right)\supnp\btwo\supnp \ = \
\left({\cal A}2\subijk\right)\supn \btwo\supn 
 + \dt \left[
                 {\cal E}3\subijk  + {\cal E}1\subijkp 
                -{\cal E}3\subipjk - {\cal E}1\subijk  \right]; \label{b2evol}
\end{equation}
\begin{equation}
\left({\cal A}3\subijk\right)\supnp\bthr\supnp \ = \
\left({\cal A}3\subijk\right)\supn \bthr\supn 
 + \dt \left[
                 {\cal E}1\subijk  + {\cal E}2\subipjk 
                -{\cal E}1\subijpk - {\cal E}2\subijk  \right]. \label{b3evol}
\end{equation}
Equations (\ref{b1evol}) - (\ref{b3evol}) make use of area factors, ${\cal A}1\subijk$,
${\cal A}2\subijk$, and ${\cal A}3\subijk$, which are simply the metric coefficients
multiplying the corresponding $b$ component in equations (\ref{bftran1}) -
(\ref{bftran3}), evaluated at time level $n$ or $n+1$ according to the associated
superscript.  In~\zmp, the ``emf[1,2,3](i,j,k)'' arrays store the ${\cal E}$ 
values indicated in
(\ref{b1evol}) - (\ref{b3evol}), and are given by the true EMF components multiplied
by the appropriate time-centered line element:
\begin{eqnarray}
{\cal E}1\subijk & = & \epsilon 1\subijk\left(dx1a\subi\right)\supnh; \label{zemf1} \\
{\cal E}2\subijk & = & \epsilon 2\subijk\left(g2a\subi~dx2a\subj\right)\supnh; 
                       \label{zemf2} \\
{\cal E}3\subijk & = & \epsilon 3\subijk\left(g31a\subi~g32a\subj~dx3a\subk\right)\supnh.
                       \label{zemf3}
\end{eqnarray}

\section{The 3-D Discrete Radiation Diffusion Matrix}\label{fldapp}

   \subsection{Flux Limiters}

We write the three components of radiation flux as
\begin{eqnarray}
\F1 & = & -D1\left(\grad\erad\right)^1, \label{rflux1} \\
\F2 & = & -D2\left(\grad\erad\right)^2, \label{rflux2} \\
\F3 & = & -D3\left(\grad\erad\right)^3, \label{rflux3}
\end{eqnarray}
where the quantities $\left(D1,D2,D3\right)$ represent flux-limited
diffusion coefficients computed independently along each axis, and the
superscripts on $\grad\erad$ indicate the appropriate component of the
gradient operator.  Recall from equation~\ref{radm} that each diffusion
coefficient takes the following form:
\begin{equation}
D \ = \ \left(c\flim\over\chif\right).
\end{equation}
\zmp~currently implements two forms of the flux-limiter, $\flim$.  The
first is due to~\citet{levermore81}, (c.f. equation 28 of their paper):
\begin{equation}\label{lpflim}
\flim({\rm LP}) \ = \ {{2 + R}\over {6 + 3R + R^2}},
\end{equation}
where $R$ is given by
\begin{equation}
R \equiv \|\grad\erad\| / \erad.  
\end{equation}
The second option is a construction
derived by~\citet{minerbo78}:
\begin{equation}\label{miflim}
\flim({\rm Mi}) \ = \ 
 \left\{
  \begin{array}{lr}
   {2 \over{3 + \sqrt{9 + 12R^2}}}, & R \leq 1.5; \\
   {1 \over{1 + R + \sqrt{1 + 2R}}}, & R > 1.5;
  \end{array}
 \right.
\end{equation}
where $R$ is as defined previously. An important feature of the implementation
is that the numerical value of $R$ is {\it lagged} in time because it is
evaluated with converged values of E from the previous time step:
\begin{equation}
R\subijk \ = \ {\|\grad\erad\subijk\supn\| \over \erad\supn\subijk}.
\end{equation}
This choice preserves the linearity of our discrete solution for $\erad\subijk\supnp$.

Because $\divf$ must be defined at cell centers for consistency with $\erad$
in (\ref{rade}), the flux components are considered to be centered on cell
faces.  This introduces an additional subtlety in the computation of diffusion
coefficients, as the opacities ($\chif$) and $R$ values (and hence $\grad\erad\supn$)
must be colocated with $\F$.  Thus, while $R$ is manifestly a scalar quantity,
the face-centered opacity must be computed from an average of neighboring
cell-centered values whose spatial relationship depends upon the cell face in
question.  Face-centered gradients in $\erad$ are subject to a similar constraint.
At a given cell, each component of flux acquires a (generally) unique value
of the E-dependent flux-limiter, which further underscores the simplification
gained by time-lagging the evaluation of $R$ as a function of E.

   \subsection{The Matrix}

Recall from~\S\ref{fldmeth} that the discrete radiation and gas energy
equations solved in the ZEUS source step are written as
\begin{eqnarray}
f^{(1)}\subijk & = & \erad\supnp\subijk \ - \ \erad\supn\subijk
  \ - \ \dt\left[4\pi\kp\pfunc - c\ke\erad\supnp\subijk
  - \divf\subijk\supnp - \gradvp\subijk\supnp\right]; \label{apprad}\\
f^{(2)}\subijk & = & \egas\supnp\subijk \ - \ \egas\supn\subijk
  \ - \ \dt\left[-4\pi\kp\pfunc + c\ke\erad\supnp\subijk
  - \pg\divv\right]. \label{appgas}
\end{eqnarray}
Our derivation of the FLD matrix proceeds by first differentiating
equations (\ref{apprad}) and (\ref{appgas}) with respect to $\egas\subijk$
and $\erad\subijk$.  Considering first the radiation energy equation, we
note that $f^{(1)}\subijk$ depends on the value of $\egas\subijk$ through
the evaluation of $\pfunc$, which requires an (in general) energy-dependent
material temperature.  The dependence of $f^{(1)}\subijk$ on E is more complex,
owing to the flux-divergence term.  As will be documented below, $\divf\subijk$
is written as a 7-point function in E coupling $\erad\subijk$ to nearest-neighbor
values along all 3 coordinate axes.  Evaluating the Jacobian for the radiation
energy equation will yield a system of the following form:
\begin{equation}\label{radjac}
\begin{array}{lclclclcc}
{\cal A}\subijk\delta\erad\subijkm & + & {\cal B}\subijk\delta\erad\subijmk & + &
{\cal C}\subijk\delta\erad\subimjk & + & {\cal D}\subijk\delta\erad\subijk & + &
 \\
{\cal E}\subijk\delta\erad\subipjk & + & {\cal F}\subijk\delta\erad\subijpk & + &
{\cal G}\subijk\delta\erad\subijkp & + & {\cal H}\subijk\delta\egas\subijk & = &
-f^{(1)}\subijk,
\end{array}
\end{equation}
where ${\cal A}\subijk$, through ${\cal H}\subijk$ are given by:
\begin{equation}
\begin{array}{ccl}
{\cal A}\subijk \ = \ {\partial f^{(1)}\subijk\over\partial\erad\supnp\subijkm};
   &   &
{\cal B}\subijk \ = \ {\partial f^{(1)}\subijk\over\partial\erad\supnp\subijmk}; \\
{\cal C}\subijk \ = \ {\partial f^{(1)}\subijk\over\partial\erad\supnp\subimjk};
   &   &
{\cal D}\subijk \ = \ {\partial f^{(1)}\subijk\over\partial\erad\supnp\subijk}; \\
{\cal E}\subijk \ = \ {\partial f^{(1)}\subijk\over\partial\erad\supnp\subipjk};
   &   &
{\cal F}\subijk \ = \ {\partial f^{(1)}\subijk\over\partial\erad\supnp\subijpk}; \\
{\cal G}\subijk \ = \ {\partial f^{(1)}\subijk\over\partial\erad\supnp\subijkp};
   &   &
{\cal H}\subijk \ = \ {\partial f^{(1)}\subijk\over\partial\egas\supnp\subijk}.
\end{array} \nonumber
\end{equation}
Because the gas energy equation involves no space derivatives in the solution
variables, the Jacobian expression is considerably simpler:
\begin{equation}\label{gasjac}
{\partial f^{(2)}\subijk\over\partial\egas\subijk }\delta\egas\subijk  \ + \
{\partial f^{(2)}\subijk\over\partial\erad\subijk }\delta\erad\subijk  \ = \ 
-f^{(2)}\subijk.
\end{equation}
The fact that $f^{(2)}\subijk$ depends on the gas energy only through $\egas\supnp\subijk$
allows $\delta\egas\subijk$ to be written algebraically as
\begin{equation}\label{dele}
\delta\egas\subijk \ = \ -{f^{(2)}\subijk + \left(\partial f^{(2)}\subijk / 
                       \partial\erad\subijk\supnp
\right)\delta\erad\subijk \over {\partial f^{(2)}\subijk / 
                       \partial\egas\subijk\supnp}}.
\end{equation}
Substitution of (\ref{dele}) into (\ref{radjac}) eliminates the explicit
dependence of the radiation energy Jacobian on $\delta\egas\subijk$,
resulting in a reduced linear system for the radiation
energy corrections:
\begin{equation}\label{redjac}
\begin{array}{lclll}
   &   & {\cal A}\subijk \ \delta\erad\subijkm   & + &   \\
   &   & {\cal B}\subijk \ \delta\erad\subijmk   & + &   \\
{\cal C}\subijk \ \delta\erad\subimjk  & + &
{\cal D}'\subijk \ \delta\erad\subijk  & + &
{\cal E}\subijk \ \delta\erad\subipjk \ + \   \\ 
   &   & {\cal F}\subijk \ \delta\erad\subijpk  & + &  \\
   &   & {\cal G}\subijk \ \delta\erad\subijkp  &   &  \\
   &   &   & = &  {\cal I}\subijk f^{(2)}\subijk \ - \ f^{(1)}\subijk,
\end{array}
\end{equation}
where
\begin{eqnarray}
{\cal D}' & = & \left\{{\partial f^{(1)}\subijk\over\partial\erad\subijk\supnp}
 \ - \ {\partial f^{(1)}\subijk\over\partial\egas\subijk\supnp}
 \left({\partial f^{(2)}\subijk / \partial\erad\supnp\subijk}\over
       {\partial f^{(2)}\subijk / \partial\egas\supnp\subijk}\right)\right\}; \\
{\cal I}  & = & \left({\partial f^{(1)}\subijk / \partial\egas\supnp\subijk}\over
                      {\partial f^{(2)}\subijk / \partial\egas\supnp\subijk}\right).
\label{cali}
\end{eqnarray}
The ${\cal H}$ coefficient of $\delta\egas\subijk$ has been absorbed into
${\cal D}'$; coefficients ${\cal B}$ and ${\cal D}$ through ${\cal H}$ remain
unchanged.  The terms on the LHS of (\ref{redjac}) have been been arranged along
multiple lines in a manner illustrating the band structure of the resulting
matrix, which is described by a tridiagonal structure coupling points (i-1,j,k),
(i,j,k), and (i+1,j,k), accompanied by subdiagonals coupling points (i,j-1,k)
and (i,j,k-1) and superdiagonals coupling points (i,j+1,k) and (i,j,k+1).

Equation (\ref{redjac}) is equivalent to a matrix equation of the form
${\cal M}\vec{\delta\erad} \ = \ \vec{\cal R}$, where ${\cal M}$ is a 7-banded
matrix whose diagonals are specified by the values of ${\cal A}$ through
${\cal G}$.  As with the covariant form of the Poisson equation matrix
(appendix~\ref{poissonapp}), ${\cal M}$ may be symmetrized by multiplying
each row by a total volume element for zone (i,j,k): 
$\Delta V1a\subi\Delta V2a\subj\Delta V3a\subk$.  Written in this way, it is
necessary to evaluate (and document) only the five bands ${\cal D}'$
through ${\cal G}$, and the RHS vector, $\vec{\cal R}$.

The main diagonal of the symmetrized matrix is given by
\begin{equation}
{\cal D}^{sym}\subijk \ = \ \left\{{\partial f^{(1)}\subijk\over\partial\erad\subijk\supnp}
 \ - \ {\partial f^{(1)}\subijk\over\partial\egas\subijk\supnp}
 \left({\partial f^{(2)}\subijk / \partial\erad\supnp\subijk}\over
       {\partial f^{(2)}\subijk / \partial\egas\supnp\subijk}\right)\right\}
 \times \Delta V1a\subi\Delta V2a\subj\Delta V3a\subk.
\end{equation}
We evaluate the four required function derivatives as a function of a 
time-centering parameter, $\theta$, such that $\theta = 1$ gives
fully implicit time differencing.  (The time step, $\dt$, is by definition
time centered.)  We present the derivatives in order of increasing
complexity, thus:
\begin{equation}
{\partial f^{(2)}\over\partial\erad\supnp\subijk} \ = \ -\theta\dt c \ke\supnt\subijk;
\end{equation}
\begin{eqnarray}\label{df1degas}
{\partial f^{(1)}\over\partial\egas\supnp\subijk} & = &
-\theta\dt\left[4\pi\pfunc\subijk\supnt\left(\dkpde\right)\subijk\supnt \ + \
               \kp\subijk\supnt\left(\dbbde\right)\subijk\supnt
         \right] 
\nonumber \\
       &  & +\theta\dt~c\erad\subijk\supnt\left(\dkede\right)\supnt\subijk;
\end{eqnarray}
\begin{eqnarray}\label{df2degas}
{\partial f^{(2)}\over\partial\egas\supnp\subijk} \ = \ 1 & + & 
\theta\dt\left[4\pi\pfunc\subijk\supnt\left(\dkpde\right)\subijk\supnt \ + \
               \kp\subijk\supnt\left(\dbbde\right)\subijk\supnt
         \right] 
\nonumber \\
 & + & 
\theta\dt\left[-c\erad\subijk\supnt\left(\dkede\right)\supnt\subijk \ + \ 
                p\supnt\subijk\left(\divv\right)\subijk
         \right].
\end{eqnarray}
The final derivative expression is written schematically as
\begin{equation}\label{df1der}
{\partial f^{(1)}\over\partial\erad\supnp\subijk} \ = \ 1 \ + \ \theta\dt
\left[c\ke\supnt\subijk \ + \ {\partial\over\partial\erad\supnp\subijk}
\left(\divf\subijk\right) \ + \ {\partial\over\partial\erad\supnp\subijk}
\left(\gradvp\right)\subijk
\right].
\end{equation}
Because we assume that $\p = \f\erad$, the final term in (\ref{df1der})
is simply $\grad\vel :\f$, where $\f$ is assumed known and held fixed
during the N-R iteration.  To evaluate $\divf$, we assume that the three
components of $\F$ are given by (\ref{rflux1}) - (\ref{rflux3}), and we
express the divergence operator in covariant coordinates
using equation (116) of~\citet{stone92a}:
\begin{eqnarray}
\divf & = & {1\over h_{1}h_{2}h_{3}}\left[
{\partial\over\partial x_{1}}\left(h_2h_3\F1\right) \ + \
{\partial\over\partial x_{2}}\left(h_1h_3\F2\right) \ + \
{\partial\over\partial x_{3}}\left(h_1h_2\F3\right)
                                           \right]  \nonumber \\
 & \equiv & 
{\partial\left(g_2g_{31}\F1\right)\over\partial V_{1}} \ + \
{1\over g_2}{\partial\left(g_{32}\F2\right)\over\partial V_{2}} \ + \
{1\over g_{31}g_{32}}{\partial\left(\F3\right)\over\partial V_{3}},
\label{codiv}
\end{eqnarray}
where in the latter expression we have transformed spatial derivatives
into volume derivatives.  A similar operation is performed on each 
component of the gradient operator:
\begin{eqnarray}
\grad\erad & = & \left(
{1\over h_1}{\partial\erad\over\partial x_1},
{1\over h_2}{\partial\erad\over\partial x_2},
{1\over h_3}{\partial\erad\over\partial x_3} \right) \nonumber \\
 & \equiv & \left(
g_2g_{31}{\partial\erad\over\partial V_1},
{g_{32}\over g_2}{\partial\erad\over\partial V_2},
{1\over g_{31}g_{32}}{\partial\erad\over\partial V_3}
            \right).\label{cograd}
\end{eqnarray}
With (\ref{codiv}) and (\ref{cograd}) in hand, we may construct a discrete
form of $\divf$ explicitly in terms of the 7-point stencil in $\erad\subijk$,
from which derivatives of $\divf\subijk$ with respect to the appropriate
E variables may be read by inspection.  We provide the result here:
\begin{eqnarray}
\divf\subijk & = & 
-{\left(g2a\subip g31a\subip\right)^2 D1\subipjk\over\Delta V1a\subi}
\left({\erad\subipjk - \erad\subijk}\over\Delta V1b\subip\right) \nonumber \\
& + & {\left(g2a\subi g31a\subi\right)^2 D1\subijk\over\Delta V1a\subi}
\left({\erad\subijk - \erad\subimjk}\over\Delta V1b\subi\right) \nonumber \\
& - & {\left(g32a\subjp\right)^2 D2\subijpk\over g2b^2\subi\Delta V2a\subj}
\left({\erad\subijpk - \erad\subijk}\over\Delta V2b\subjp\right) \nonumber \\
& + & {\left(g32a\subj\right)^2 D2\subijk\over g2b^2\subi\Delta V2a\subj}
\left({\erad\subijk - \erad\subijmk}\over\Delta V2b\subj\right) \nonumber \\
& - & {D3\subijkp\over \left(g31b\subi g32b\subj\right)^2\Delta V3a\subk}
\left({\erad\subijkp - \erad\subijk}\over\Delta V3b\subkp\right) \nonumber \\
& + & {D3\subijk\over \left(g31b\subi g32b\subj\right)^2\Delta V3a\subk}
\left({\erad\subijk - \erad\subijkm}\over\Delta V3b\subk\right).
\end{eqnarray}
The middle term in (\ref{df1der}) then follows at once:
\begin{eqnarray}
{\partial\divf\subijk\over\partial\erad\subijk} & = & 
{\left(g2a\subip g31a\subip\right)^2 D1\subipjk\over\Delta V1a\subi\Delta V1b\subip}
\ + \ 
{\left(g2a\subi g31a\subi\right)^2 D1\subijk\over\Delta V1a\subi\Delta V1b\subi}
\nonumber \\
& + & 
{\left(g32a\subjp\right)^2 D2\subijpk\over g2b^2\subi\Delta V2a\subj\Delta V2b\subjp}
\ + \ 
{\left(g32a\subj\right)^2 D2\subijk\over g2b^2\subi\Delta V2a\subj\Delta V2b\subj}
\nonumber \\
& + & 
{D3\subijkp\over \left(g31b\subi g32b\subj\right)^2\Delta V3a\subk\Delta V3b\subkp}
\ + \ 
{D3\subijk\over \left(g31b\subi g32b\subj\right)^2\Delta V3a\subk\Delta V3b\subk}.
\label{ddivfder}
\end{eqnarray}

The three super-diagonal bands of the symmetric matrix, ${\cal E}\subijk$,
${\cal F}\subijk$, and ${\cal G}\subijk$, originate in the derivatives of
$\divf\subijk$ with respect to $\erad\subipjk$, $\erad\subijpk$, and
$\erad\subijkp$, respectively.  We therefore have:
\begin{eqnarray}
{\cal E}^{sym}\subijk & \equiv & {\partial f^{(1)}\subijk\over\partial\erad\subipjk}
\times \Delta V1a\subi\Delta V2a\subj\Delta V3a\subk \nonumber \\
 & = & -\theta\dt~
{\left(g2a\subip g31a\subip\right)^2 D1\subipjk\over\Delta V1b\subip}
\times \Delta V2a\subj\Delta V3a\subk; \\
{\cal F}^{sym}\subijk & \equiv & {\partial f^{(1)}\subijk\over\partial\erad\subijpk}
\times \Delta V1a\subi\Delta V2a\subj\Delta V3a\subk \nonumber \\
 & = & -\theta\dt~
{\left(g32a\subjp\right)^2 D2\subijpk\over g2b^2\subi\Delta V2b\subjp}
\times \Delta V1a\subi\Delta V3a\subk; \\
{\cal G}^{sym}\subijk & \equiv & {\partial f^{(1)}\subijk\over\partial\erad\subijkp}
\times \Delta V1a\subi\Delta V2a\subj\Delta V3a\subk \nonumber \\
 & = & -\theta\dt~
{D3\subijkp\over \left(g31b\subi g32b\subj\right)^2\Delta V3b\subkp}
\times \Delta V1a\subi\Delta V2a\subj.
\end{eqnarray}
Finally, the RHS of the symmetrized linear system is evaluated as
\begin{equation}
{\cal R}\subijk \ = \
\left\{
\left({\partial f^{(1)}\subijk / \partial\egas\supnp\subijk}\over
                      {\partial f^{(2)}\subijk / \partial\egas\supnp\subijk}\right)
f^{(2)}\subijk \ - \ f^{(1)}\subijk
\right\}
\times\Delta V1a\subi \Delta V2a\subj \Delta V3a\subk,
\end{equation}
with $\partial f^{(1)}\subijk / \partial\egas\supnp\subijk$ and
$\partial f^{(2)}\subijk / \partial\egas\supnp\subijk$ given by equations
(\ref{df1degas}) and (\ref{df2degas}), respectively.

\section{The 3-D Discrete Poisson Equation Matrix}\label{poissonapp}

The 2-D form of Poisson's equation was written (although not formally
derived) in~\citet{stone92a}; here we extend the discrete expression to
3-D and explicitly derive and document the matrix elements.
Following~\citet{stone92a}, we write the general tensor form of the
Laplacian operating on a scalar function, $\Phi$, as
\begin{equation}\label{phiofh}
\nabla^2\Phi \ = \ {1\over h_{1}h_{2}h_{3}}\left[
{\partial\over\partial x_{1}}\left({h_2h_3\over h_1}{\partial\Phi\over
                                                     \partial x_1}\right) \ + \ 
{\partial\over\partial x_{2}}\left({h_1h_3\over h_2}{\partial\Phi\over
                                                     \partial x_2}\right) \ + \ 
{\partial\over\partial x_{3}}\left({h_1h_2\over h_3}{\partial\Phi\over
                                                     \partial x_3}\right)
                                           \right]
\end{equation}
The inner partial derivatives of $\Phi$ are rewritten as functions of ZEUS
metric coefficients:
\begin{equation}
q_1 \ \equiv \ g_2g_{31}g_{32}{\partial\Phi\over\partial x_1}; \ \ \
q_2 \ \equiv \ {g_{31}g_{32}\over g_2}{\partial\Phi\over\partial x_2}; \ \ \
q_3 \ \equiv \ {g_2\over g_{31}g_{32}}{\partial\Phi\over\partial x_3};
\end{equation}
and the outer derivatives over the ``q'' functions so defined are transformed
into volume derivatives and written in discrete form as
\begin{eqnarray}
{1\over g_{32}}{\partial q_1\over\partial V_1} & = & 
{1\over g32b_j}\left({q1\subip - q1\subi}\over\Delta V1a\subi \right), \label{dq1}\\
{1\over g_2g_{31}}{\partial q_2\over\partial V_2} & = &
{1\over g2b\subi g31b\subi}\left({q2\subjp - q2\subj}\over\Delta V2a\subj \right),
\label{dq2} \\
{1\over g_2g_{31}g_{32}}{\partial q_3\over\partial V_3} & = &
{1\over g2b\subi g31b\subi g32b\subj}\left({q3\subkp - q3\subk}\over\Delta V3a\subk 
\right).\label{dq3}
\end{eqnarray}
The derivatives inside of the $q$ functions are left as discrete coordinate
differences:
\begin{eqnarray}
q1\subi  & = & g2a\subi g31a\subi g32b\subj           
\left(\Phi\subijk - \Phi\subimjk\right)\ / \ \Delta x1b\subi,  \label{q1}\\
q2\subj  & = & \left(g31b\subi g32a\subj / g2b\subi\right)
\left(\Phi\subijk - \Phi\subijmk\right) \ / \ \Delta x2b\subj, \label{q2}\\
q3\subk  & = & \left(g2b\subi / g31b\subi g32b\subj\right)
\left(\Phi\subijk - \Phi\subijkm\right)\ / \ \Delta x3b\subk . \label{q3}
\end{eqnarray}
Leaving the inner derivatives as functions of coordinate
differences was done for consistency with the formulation in the public
\ztwd~code.  We have also formulated the linear system for the case in which
the inner derivatives are also transformed into volume differences.  We have
not discovered an application in which this distinction has a measurable
effect.  We therefore adopt the former approach for the purposes of this
document.
Evaluating (\ref{dq1}) - (\ref{dq3}) with (\ref{q1}) - (\ref{q3}) yields
\begin{eqnarray}
{1\over g_{32}}{\partial q_1\over\partial V_1} & \rightarrow &
{1\over\Delta V1a\subi} \ \times \ \nonumber \\
  &   &                \left\{
                              {\cal P}1\Phi\subipjk \ + \
                              {\cal D}1\Phi\subijk \ + \
                              {\cal M}1\Phi\subimjk
                       \right\}, \label{dv1}\\
{1\over g_2g_{31}}{\partial q_2\over\partial V_2} & \rightarrow &
{1\over \left(g2b\subi\right)^2\Delta V2a\subj} \ \times \nonumber \\
  &   &                \left\{
                              {\cal P}2\Phi\subijpk \ + \
                              {\cal D}2\Phi\subijk \ + \
                              {\cal M}2\Phi\subijmk
                       \right\}, \label{dv2}\\
{1\over g_2g_{31}g_{32}}{\partial q_3\over\partial V_3} & \rightarrow &
\left(1\over g31b\subi g32b\subj\right)^2
\left(1\over \Delta V3a\subk\right) \ \times \nonumber \\
  &   &                \left\{
                              {\cal P}3\Phi\subijkp \ + \
                              {\cal D}3\Phi\subijk \ + \
                              {\cal M}3\Phi\subijkp
                       \right\}. \label{dv3}
\end{eqnarray}
In (\ref{dv1}) - (\ref{dv3}), the ${\cal P}$, ${\cal D}$, and ${\cal M}$
functions are written as
\begin{eqnarray}
{\cal P}1 & = & \left(g2a~g31a\over\Delta x1b\right)\subip, \\
{\cal D}1 & = & -\left[\left(g2a~g31a\over\Delta x1b\right)\subip \ + \
                 \left(g2a~g31a\over\Delta x1b\right)\subi\right], \\
{\cal M}1 & = & \left(g2a~g31a\over\Delta x1b\right)\subi; \\
{\cal P}2 & = & \left(g32a\over\Delta x2b\right)\subjp, \\
{\cal D}2 & = & -\left[\left(g32a\over\Delta x2b\right)\subjp \ + \
                 \left(g32a\over\Delta x2b\right)\subj\right], \\
{\cal M}2 & = & \left(g32a\over\Delta x2b\right)\subj; \\
{\cal P}3 & = & \left(1\over\Delta x3b\right)\subkp, \\
{\cal D}3 & = & -\left[\left(1\over\Delta x3b\right)\subkp \ + \
                 \left(1\over\Delta x3b\right)\subk\right], \\
{\cal M}3 & = & \left(1\over\Delta x3b\right)\subk.
\end{eqnarray}

The LHS of the discrete Poisson equation may be constructed by a direct
summation of expressions (\ref{dv1}) - (\ref{dv3}).  Such action results
in a 7-banded sparse matrix (cf. equation~\ref{discpoi})
in which elements along the main diagonal are
given by the sum of the 3 ${\cal D}$ expressions listed above, multiplied
by the inverse volume factor $(1/\Delta V1a)\subi$.  Similarly, the first
superdiagonal band (coupling $\Phi\subijk$ to $\Phi\subipjk$) is given by 
the ${\cal P}1$ expression multiplied by the associated volume factor
in (\ref{dv2}).  The remaining two superdiagonals and the three subdiagonal
bands are derived in analogous fashion.  The matrix may be symmetrized, however,
if expressions (\ref{dv1}) - (\ref{dv3}) are first multiplied by a total volume 
element $\Delta V \ \equiv \ \Delta V1a\subi\Delta V2a\subj\Delta V3a\subk$.
The resulting transpose symmetry allows explicit calculation, storage, and
operation upon the three subdiagonals to be avoided.  The symmetric linear
system may be written symbolically (compare with equation~\ref{discpoi}) as
\begin{eqnarray}
s_4\Phi\subipjk \ + \ s_5\Phi\subijpk \ + \ s_6\Phi\subijkp & + & \nonumber \\
s_7\Phi\subijk & = & 4\pi G\left(\Delta V1a\subi\Delta V2a\subj\Delta V3a\subk\right)
                     \rho\subijk,
\label{sympoi}
\end{eqnarray}
with
\begin{eqnarray}
s_4 & = & \Delta V2a\subj\Delta V3a\subk \ \times \ {\cal D}1 \ + \
          {\Delta V1a\subi\Delta V3a\subk\over (g2b\subi)^2} \ \times \ {\cal D}2
   \ + \ \nonumber \\ 
    &   & {\Delta V1a\subi\Delta V2a\subj \over(g31b\subi g32b\subj)^2} \times 
       {\cal D}3, \\
s_5 & = & \Delta V2a\subj\Delta V3a\subk \ \times \ {\cal P}1, \\
s_6 & = & {\Delta V1a\subi\Delta V3a\subk\over (g2b\subi)^2} \ \times \ {\cal P}2, \\
s_7 & = & {\Delta V1a\subi\Delta V2a\subj \over(g31b\subi g32b\subj)^2} \times 
          {\cal P}3.
\end{eqnarray}

\section{Implementation Techniques and Strategies}\label{impapp}

The ZEUS algorithm solves the partial differential equations describing astrophysical 
fluid flows by means of an ``operator-split'' finite difference scheme. The field 
variables are advanced in time through a series of substeps corresponding to each 
operator (physical process) contributing to the full evolution equations. Whether the 
field variables are updated in an explicit or implicit manner, they use values of 
quantities computed during the previous substep. Therefore, a parallel algorithm in 
which multiple substeps are executed concurrently is not feasible. Instead, our 
parallelization strategy is based on domain decomposition, in which the spatial mesh is 
divided into ``tiles'' and the field variables are updated in each tile concurrently. Each
substep in the time-stepping scheme is completed in all tiles before moving on to the 
next substep, so that the time levels of all variables remain synchronized between tiles. 

Gradients and other spatial derivatives appearing in the evolution equations are 
approximated by linear combinations of field variable values evaluated at discrete 
points in a set of several neighboring mesh zones comprising the ``stencils'' of the 
difference operators. Evaluating spatial derivatives in mesh zones near tile boundaries 
requires values of some quantities at locations in zones belonging to neighboring tiles. 
Therefore, before we can update the field variables in zones near the boundaries of a 
tile, we must receive some data from neighboring tiles as required by the stencils. We 
perform the required exchange of data between tiles by means of ``message passing'', using
the MPI library. MPI enables the code to execute efficiently on many types of parallel 
architectures, from systems with globally shared memory to clusters of workstations.

Optimal paralled efficiency is achieved by minimizing the ratio of communication overhead
to computational work (updating field variables). The amount of data that needs to be 
exchanged between tiles is proportional to the number of zones near tile boundaries 
(not physical boundaries, unless periodic boundary conditions are prescribed there). 
We therefore minimize the ratio of the number of zones near tile surfaces to zones in 
tile interiors by decomposing the domain along each active spatial dimension. We balance 
the load by assigning nearly the same number of zones to each tile.

Communication overhead involves more than merely the transit time for the messages 
(which is proportional to message size, i.e., the number of array elements). It also 
includes network latency (same for any message size), time for the CPUs to 
copy/pack/unpack the data to be passed, and context switching delays as the CPUs 
alternate between updating variables and passing messages. Fortunately much of the 
communication overhead is comprised of idle cycles, some of which can be spent doing 
other useful work, provided one makes use of the ``non-blocking'' communications 
operations in MPI.

\begin{figure*}
\plotone{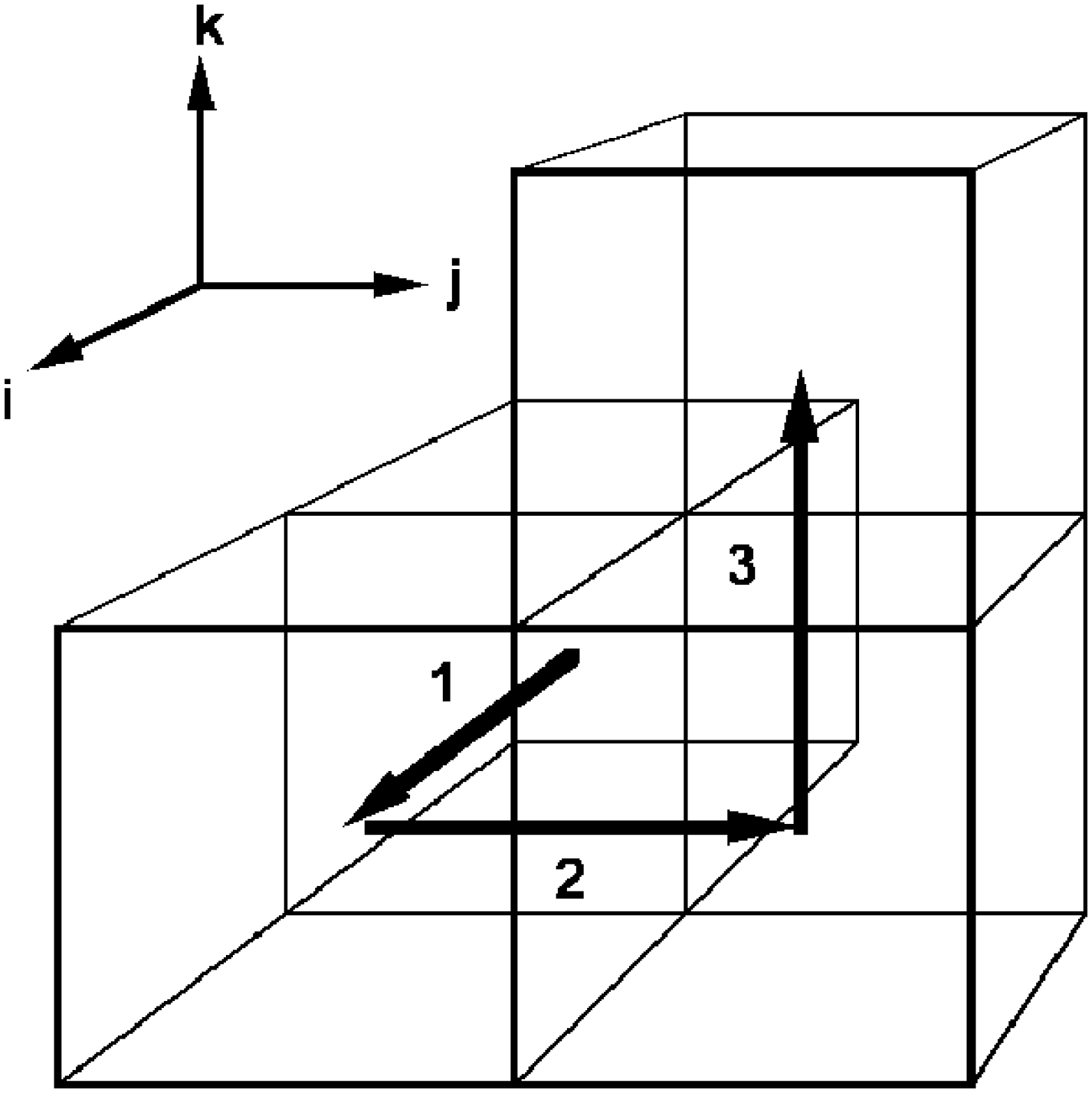}
\caption{MPI communication flow in~\zmp.}\label{fraf01}
\end{figure*}

One way to reduce communication overhead is to minimize the number of mesages that are 
sent. Of particular concern is the exchange of data between tiles that share only one 
corner point. Only a few zones near a corner require any data from tiles sharing only 
that corner, but each tile has 8 corners, each of which are shared by 7 neighboring 
tiles. Each tile also has 8 edges which are shared by 3 neighboring tiles. In contrast, 
each of the 6 tile faces has at most just 1 neighboring face. We can avoid passing a 
large number of small messages by exchanging messages across tile faces in 3 stages, 
sending and receiving messages along just one dimension per stage (see 
Figure~\ref{fraf01}). We begin the next communication stage only after the previous 
stage is completed. Data from neighboring tiles is stored in the 2 layers of ``ghost'' 
zones on the surfaces of each tile.  This ghost cell data is included in all messages 
and automatically carries edge and corner cell values to the tiles that share only those 
edges and corners. 

In some of the substeps in the ZEUS algorithm, such as advection along the ``i'' direction
for pure hydrodynamics, updating the field variables involves relatively little 
computational work. In such cases, we employ a more agressive strategy to overlap more 
communication operations with computations (at the expense of a more complicated code). 
We subdivide the zones in each tile into 3 roughly equal groups, so that one third of 
the interior zones can be updated while the messages for each communication stage are in 
transit. After the messages for a given stage are received, we update field variables in 
zones near tile boundaries for which all data is available. The precise procedure is as 
follows:
\begin{enumerate}
  \item Boundary data is exchanged with neighbors along the i-axis while 
     updates are performed for the 1st third of interior zones.
  \item Boundary data is exchanged along the j-axis while updates are 
     performed
     (a) for the values of i skipped in step (1) in the 1st third of
        interior zones, and
     (b) for the 2nd third of interior zones (all i). 
  \item Boundary data is exchanged along the k-axis while updates are
     performed 
     (a) for the values of j skipped in step (2) in the first two
        thirds of interior zones, and
     (b) for the 3rd third of interior zones (all i and j).
  \item Updates are performed for the values of k skipped in the previous
     steps (all i and j). 
\end{enumerate}

The procedure outlined above is adopted for several stages in the source step
portion of the hydrodynamics update, including the artificial viscosity, body
forces, compressional heating ($pdV$), and EOS updates.  Because the radiation
module employs an implicit solution which updates $\erad\subijk$ and $\egas\subijk$
at all mesh points simultaneously, partial mesh updates are not possible.  A related
procedure {\it is} employed, however, in both the subroutines which compute matrix
elements and within the CG linear solver routine which returns
corrections for $\erad\subijk$ during a Newton-Raphson iteration.  By construction,
the FLD stencil never accesses data which lies outside of a tile boundary along more
than one axis; i.e. ``ghost corner'' zones are never accessed.  Because of this,
the asynchronous MPI exchanges of tile faces may be initiated for all three axes
simultaneously, since only ghost corner cells depend upon the ordering of face
updates.  Within the FLD module, therefore, we exploit a simplified procedure
in which updates are performed along tile faces, MPI ``ISEND'' and ``IRECV''
calls are posted for face data along all axes, interior updates are performed,
and an MPI ``WAIT'' operation is performed to ensure that message-passing has
completed before proceeding.  This allows considerable opportunity for overlapping
communication with computation both in the evaluation of matrix elements
and the processing of matrix data during the CG linear solution step.

At the very end of a time step, the size of the next explicit time step is computed from 
the minimum value of the Courant limit over all zones.  
MPI and other message-passing libraries provide routines
to perform such global reduction operations efficiently in parallel.  
The time spent waiting for this operation to complete 
comprises only a fraction of the total communication time.

\end{appendix}

%%\clearpage

\end{document}